\documentclass[twocolumn,aps,prb,groupedaddress]{revtex4}
\usepackage[T1]{fontenc}
\usepackage[utf8]{inputenc}
\usepackage{color}
\definecolor{note_fontcolor}{rgb}{1, 0, 1}
\usepackage{amsmath}
\usepackage{amssymb}
\usepackage{graphicx}

\makeatletter

\providecommand{\tabularnewline}{\\}

\usepackage{etoolbox}
\usepackage{booktabs}

\makeatother

\begin{document}

\title{Non-ergodic delocalized phase in Anderson model on Bethe lattice
and regular graph.}

\author{V.~E.~Kravtsov }

\affiliation{Abdus Salam International Center for Theoretical Physics, Strada
Costiera 11, 34151 Trieste, Italy}

\affiliation{ L. D. Landau Institute for Theoretical Physics, Chernogolovka, 142432,
Moscow region, Russia}

\author{ B.~L.~Altshuler }

\affiliation{Physics Department, Columbia University, 538 West 120th Street, New
York, New York 10027, USA }

\author{L.~B.~Ioffe }

\affiliation{LPTHE - CNRS - UPMC, 4 place Jussieu Paris, 75252, France}

\affiliation{National Research University Higher School of Economics, Moscow,
Russia}

\affiliation{ L. D. Landau Institute for Theoretical Physics, Chernogolovka, 142432,
Moscow region, Russia}
\begin{abstract}
We develop a novel analytical approach to the problem of single particle
localization in infinite dimensional spaces such as Bethe lattice
and random regular graph models. The key ingredient of the approach
is the notion of the inverted order thermodynamic limit (IOTL) in
which the coupling to the environment goes to zero before the system
size goes to infinity. Using IOTL and Replica Symmetry Breaking (RSB)
formalism we derive analytical expressions for the fractal dimension
$D_{1}$ that distinguishes between the extended ergodic, $D_{1}=1$,
and extended non-ergodic (multifractal), $0<D_{1}<1$ states on the
Bethe lattice and random regular graphs with the branching number
$K$. We also employ RSB formalism to derive the analytical expression
$\ln\mathfrak{S}_{\text{typ}}^{-1}=-\langle\ln\mathfrak{S}\rangle\sim(W_{c}-W)^{-1}$
for the typical imaginary part of self-energy $\mathfrak{S}_{\text{typ}}$
in the non-ergodic phase close to the Anderson transition in the conventional
thermodynamic limit. We prove the existence of an extended non-ergodic
phase in a broad range of disorder strength and energy and establish
the phase diagrams of the models as a function of disorder and energy.
The results of the analytical theory are compared with large-scale
population dynamics and with the exact diagonalization of Anderson
model on random regular graphs. We discuss the consequences of these
results for the many body localization.
\end{abstract}
\maketitle
\tableofcontents{}

\section{Introduction}

Recent progress in understanding the dynamical processes of mesoscopic
and macroscopic isolated disordered quantum many-body systems is based
on the concept of Many-Body Localization (MBL) \cite{BAA,Ogan-Huse,Ogan-Huse1}:
the many-body eigenstates can be localized in the Hilbert space in
a way similar to the conventional real space Anderson localization
\cite{Anderson} of a single quantum particle in a quenched disorder.
Depending on the temperature (total energy) or other tunable parameters
the system can find itself either in many-body localized or in the
many-body extended phase. In the former case the system cannot be
described in terms of conventional Statistical Mechanics: the notion
of the thermal equilibrium loses its meaning, as not all positions
in the Hilbert space are reachable.

There are reasons to believe that the violation of the conventional
thermodynamics does not disappear with the Anderson transition from
the localized to the extended state \cite{LevPino,PinoKrav}: in a
finite range of the parameters one expects the appearance of a non-ergodic
extended phase for which the conventional theory is inapplicable.

\begin{figure*}[t!]
\centering{ \includegraphics[width=0.45\linewidth]{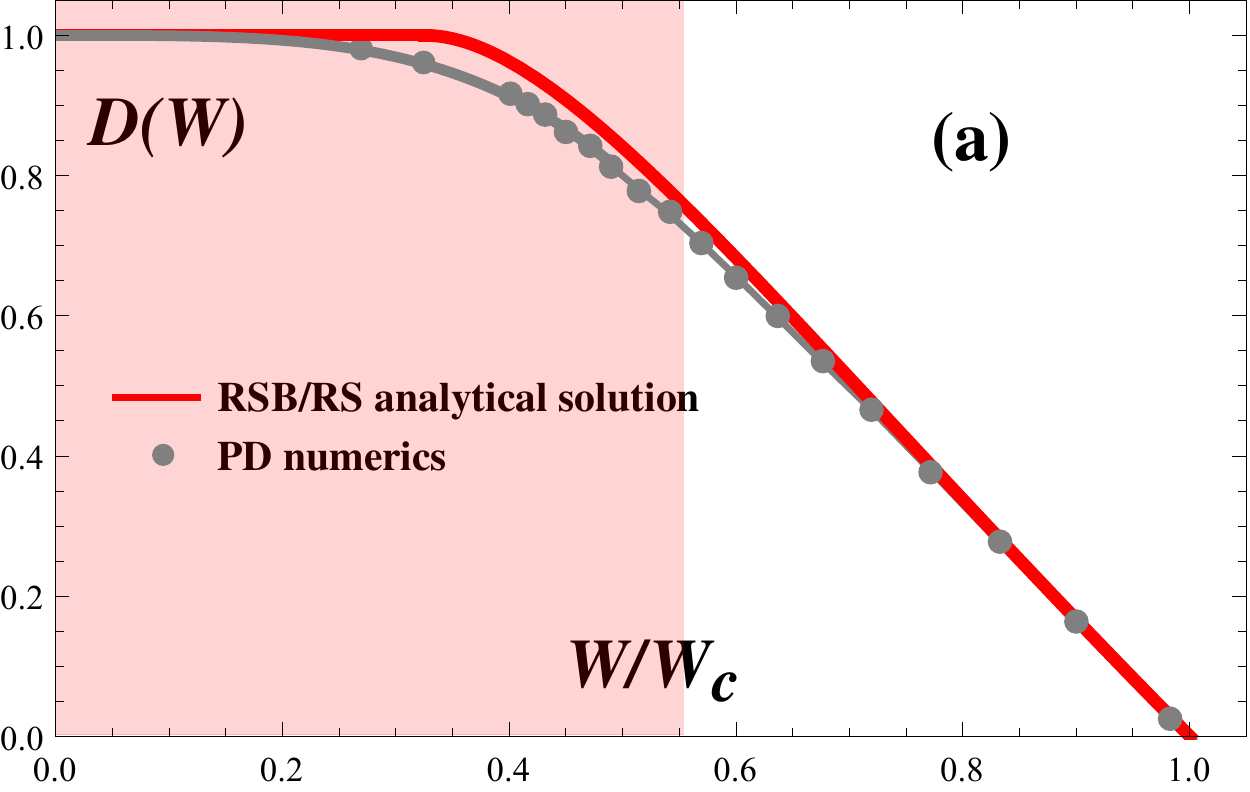}
\includegraphics[width=0.45\linewidth]{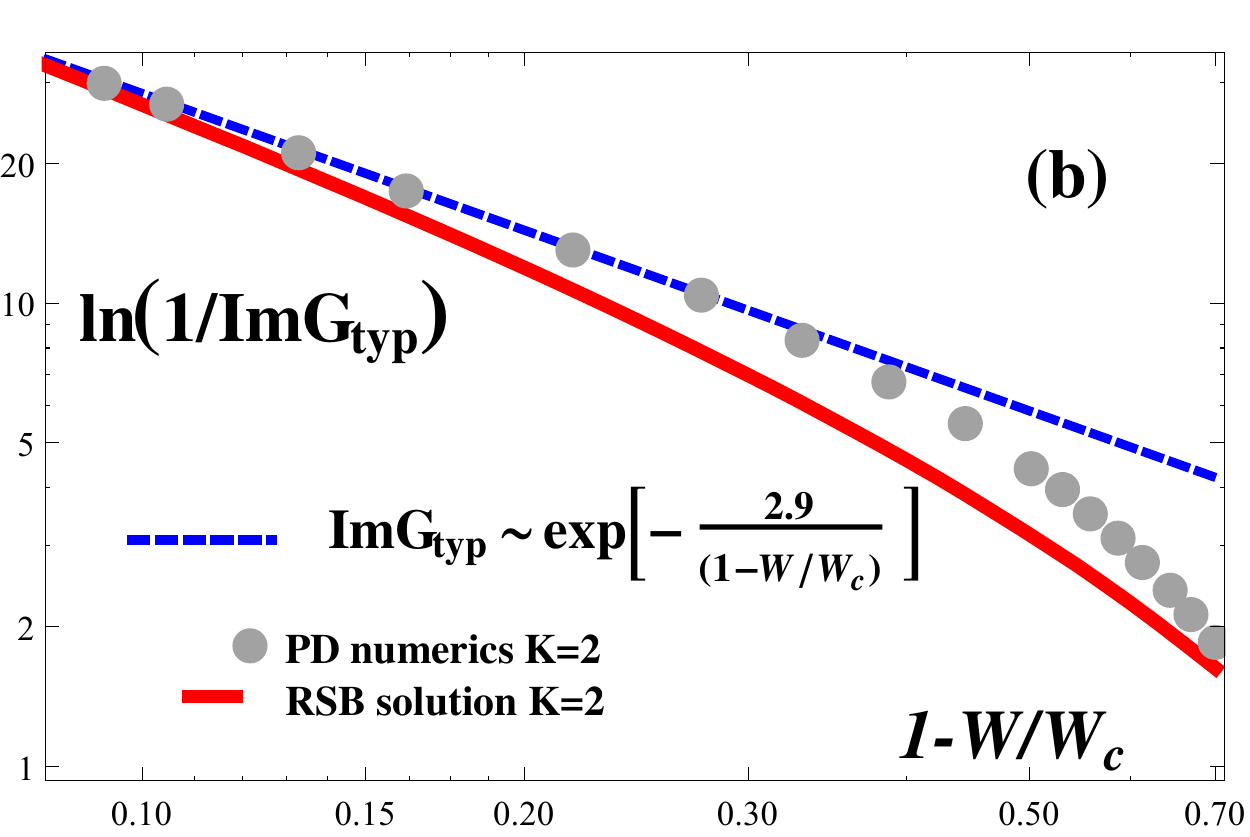} }
\caption{(Color online) The results of one-step Replica Symmetry Breaking (RSB)
analytical treatment (thick red lines) and Population Dynamics (PD)
(gray symbols) for (a) fractal dimension $D_{1}$ of wave functions
in the multifractal phase and (b) for typical imaginary part of Green's
function in the Anderson's thermodynamic limit (ATL) on the Bethe
lattice (BL) as a function of disorder strength $W$. The one-step
RSB predicts a continuous ergodic transition, while PD shows a power-law
decreases of $1-D_{1}\propto W^{4}$ down to the lowest disorder $W\rightarrow0$.
The one-step RSB analytical and PD numerical results coincide and
confirm existence of non-ergodic phase on both BL and RRG unless disorder
is not very small. For small disorder where the typical Lyapunov exponent
$\lambda_{{\rm typ}}<\ln K$ (rose area), the behavior on BL represented
by PD and that on a Random Regular Graph (RRG) may differ significantly
due to multiple connectivity on RRG, so that the ergodic transition
on RRG is not excluded in this area. \label{Fig:Intro} }
\end{figure*}

In a many body problem the number of the states connected with a given
one in the $n$-th order of the perturbation theory in the interaction
increases exponentially or faster with $n$.\cite{AGKL} Similar situation
takes place in the problem of single particle localization on hierarchical
lattices such as the Bethe lattice (BL) or random regular graphs of
connectivity $K$ where the number of sites at a given distance increases
exponentially with distance, $N(\ell)=K^{\ell}$. One thus believes
that these problems might be viewed as toy models of the many body
localization. The rapid growth of the number of sites with distance
from a given site is very important feature of these problems that
distinguishes them from the one-particle Anderson problem in a finite-dimensional
space in which the number of sites growth as a power-law. In particular,
the slow growth of the number of sites with distance implies that
it cannot compensate the exponential decay of the tunneling amplitude
with distance. Thus, in this case the resonances either appear at
short distances or not at all. This is the reason why in finite dimensional
localization all extended quantum states are ergodic and the ergodicity
is violated only at the critical point of Anderson transition, which
is manifested by the multifractality of the critical quantum states
\cite{Mirlin-rev}.

Recent numerical studies \cite{Biroli,Our-BL,AltshulerCuevasIoffeKravtsov2016}
of the Anderson problem on a random regular graph (RRG), which is
known to be almost indistinguishable from the Bethe lattice at short
length scales, brought up strong evidence in favor of the existence
of the non-ergodic phase: the eigenfunctions were found to be multifractal
with the fractal dimensions depending on disorder. It was also suggested
that the transition (referred to below as \textit{ergodic transition})
from the extended ergodic (EE) to the non-ergodic extended (NEE) phases
is a true transition as evidenced by the jump in the fractal dimensions
rather than a crossover \cite{AltshulerCuevasIoffeKravtsov2016}.

Existence of NEE phase and the transition from NEE to EE states has
been recently proven \cite{KravtsovKhaymovichCuevas2015,RP-Bir} for
an apparently different model, the random matrix theory with the special
diagonal, suggested in 1960 by Rosenzweig and Porter (RP) \cite{RPort}
and generalized in Ref.\cite{KravtsovKhaymovichCuevas2015}. As we
explain below, the property that unifies both models is the self-consistent
equations for the Green's function suggested for the Bethe lattice
by Abou-Chakra, Thouless and Anderson \cite{AbouChacAnd}. These equation
are valid for the Bethe lattice with any connectivity $K$ due to
its loopless, tree structure. However, being a kind of self-consistent
theory, these equations are also valid exactly for the RP model due
to its infinite connectivity in the thermodynamic limit.

An important boost for the interest to single-particle localization
problem on BL comes from the recent work \cite{Biroli-time-dep} that
proposes a mapping of dynamical correlations function in the full
many body problem onto single particle correlation functions on RRG
and studied them numerically. The results can be applied to the spin
correlation function \cite{Alet} and the time-dependent even-odd
site imbalance \cite{Bloch,Bloch-Science}. The power-law time dependence
of the single particle correlation functions on RRG with the exponents that continuously
depend on disorder implies similar power-law time-evolution of the
corresponding correlation functions in the many body problems \cite{Alet,Bloch-Science,Bloch}.
This behavior is  indeed observed \cite{Bloch}
in a system of interacting cold atoms in a disordered optical lattice.

In this paper we develop an analytical approach to the non-ergodic
phase of the Anderson model on the large-connectivity Bethe lattice
and random regular graphs which is based on the replica symmetry breaking,
the preliminary short version of this paper can be found in\cite{ArXiv16}.

The main result of the work is the behavior of the fractal dimension
of the wave functions summarized in Fig.~\ref{Fig:Intro}. Fig.~\ref{Fig:Intro}a
shows the analytical (RSB) and population dynamics results for the
fractal dimension $D_{1}$ for the Bethe lattice. The former predicts
that the dimension $D_{1}$ is a smooth function which varies from
$0$ to $1$ as $W$ varies in the range of $W_{E}<W<W_{c}$ proving
existence of \textit{non-ergodic extended phase} (NEE). This phase
terminates at the Anderson localization transition $W=W_{c}$ at large
disorder and at the \textit{ergodic transition} $W=W_{E}$ at low
disorder. The results of population dynamics coincide with RSB theory
for $W\gtrsim8$, and corroborate the existence of the non-ergodic
extended phase. However, at small $W$ the population dynamics predicts
gradual crossover to $D=1$ as $W\rightarrow0$ whilst RSB predicts
the transition at $W_{E}$. We discuss the origin of the discrepancy
and the region of the validity of these results in section \ref{sec:Population-dynamics-Lyapunov}.
Briefly, we expect
that {\it in the bulk} of a large Bethe lattice $D_{1}(W)$ obtained in population dynamics
is valid for all $W$ in the limit of infinite size $N\rightarrow\infty$, so that the non-ergodic phase survives to the
lowest $W\rightarrow0$.
For RRG the \textit{ergodic}
transition from NEE to EE phase might happen at sufficiently small
disorder $W\leq W_{0}\approx10$ at which RRG and BL are no longer
equivalent.

Fig.~\ref{Fig:Intro}b shows the dependence of the typical imaginary
part $\rho_{\text{typ}}$ of a single-site Green's function $G_{i}(E)$
at the band center $E=0$ in the Anderson thermodynamic limit \cite{AbouChacAnd}
as a function of disorder. Everywhere in the domain of NEE phase it
is smaller than $\Im G_{i}(E)$ averaged over disorder $\langle\rho\rangle$,
it approaches it only at small $W$, in ergodic or almost ergodic
phase. Furthermore, it becomes exponentially small near the Anderson
localization transition.

To verify the results of the analytical theory developed in this paper,
we further develop the population dynamics (PD) method by exploiting,
in addition to the standard \textit{equilibrium} PD, a new \emph{inflationary}
PD formalism introduced previously in \cite{AltshulerCuevasIoffeKravtsov2016}.
This formalism corresponds to the unusual (\char`\"{}inverted-order\char`\"{})
thermodynamic limit (IOTL) in which the bare energy level width $\eta\rightarrow0$
\textit{prior} to the system size $N\rightarrow\infty$, which allows
to compute the fractal dimension of a single wave function. The agreement
with analytical result appears to be very good for $W$ which is not
very small: $W\gtrsim8$, see Fig.~\ref{Fig:Intro}a. In a separate
computation we verified the results of the analytical theory for the
critical behavior of $\rho_{\text{typ}}(W)$ in the conventional thermodynamic
limit ($N\rightarrow\infty$ first) in the vicinity of Anderson transition.
The results of population dynamics and analytical theory are shown
in Fig.~\ref{Fig:Intro}b. They unambiguously show that $\ln1/\rho_{\text{typ}}\propto|W-W_{c}|^{-1}$
as $W$ approach $W_{c}$ from the \textit{metallic} side of the Anderson
transition.

The plan of the remainder of the paper is the following. In section
\ref{sec:Support-set-of} we define the support set of random wave
functions and give a definition of the NEE phase in terms of the scaling
of the support set volume with the total volume. In section \ref{sec:Distribution-of-LDoS}
we review the behavior of the typical local density of states in the
conventional Anderson Thermodynamic Limit (ATL) and in the Inverted
Order Thermodynamic Limit (IOTL) which allows to distinguish between
the EE and the NEE phases. Section \ref{sec:The-model} formulates
the models while section \ref{sec:Abou-Chakra-Thouless-Anderson-eq}
gives the basic equations for the Green's functions in these models.
In section \ref{sec:Inflationary-population-dynamics} we describe
the new method of Inflationary Population Dynamics and derive a relationship
between the increment $\Lambda$ of exponential inflation of the typical
imaginary part of Green's function and the fractal dimension $D_{1}$.
We derive the basic equations for the one-step Replica Symmetry Breaking
in sections \ref{sec:Large-connectivity-approximation}-\ref{sec:Minimal-account-for-real-part}.
In section \ref{sec:Improved-large-K-approximation} we use the basic
symmetry of the problem to derive a new algebraic equation for the
critical disorder $W_{c}$ at the localization transition on the Bethe
lattice and Random Regular Graphs. This simple equation considerably
improves the accuracy of $W_{c}$ compared to the classical result
of Ref.\cite{AbouChacAnd} for the small branching numbers $K\gtrsim2$.
In section \ref{sec:Analytical-results-for-D(W)} we derive, within
the one-step replica symmetry breaking method, the analytical results
for the fractal dimension $D_{1}$ as a function of disorder at the
branching number $K=2$ and compare them with the results of inflationary
population dynamics and exact diagonalization on random regular graphs.
In section \ref{sec:Application-to-Rosenzweig-Porter} we apply the
large-$K$ approximate solution for $D_{1}$ to the Generalized Rosenzweig-Porter
random matrix ensemble and re-derive the dependence of $D_{1}$ on
the control parameter $\gamma$ in the NEE phase that shows a continuous
transition to the EE phase discovered earlier in \cite{KravtsovKhaymovichCuevas2015}.
In section \ref{sec:RSB-results-for-rho_typ} we derive, in the framework
of one-step replica symmetry breaking, the dependence of the typical
imaginary part of Green's function $\rho_{\text{typ}}$ on disorder
strength $W$ in the Anderson thermodynamic limit and compare it with
equilibrium population dynamics numerics in section \ref{sec:Population-dynamics-for-ImG}.
In addition, in section \ref{sec:Population-dynamics-for-ImG} we
present the results of population dynamics numerics for the correlation
function $K(\omega)$ of $\Im G$ at different energies and relate
it with the $1/f$ noise in interacting spin systems. In section \ref{sec:Analytical-results-for-Lyapunov} we reformulate
the condition for the localization and ergodic transitions in terms
of the Lyapunov exponents and find the corresponding expressions within
the one-step RSB. These analytical results are compared with the population
dynamics for the Lyapunov exponents in section \ref{sec:Population-dynamics-Lyapunov}.
The obtained behavior of Lyapunov exponent allows us to estimate the
contribution of large loops present in RRG and obtain the upper bound
for the applicability of the analytical and population dynamic results
to RRG in section \ref{sec:Population-dynamics-Lyapunov}. The phase
diagram in the energy-disorder plane for the Bethe lattice and random
regular graphs is presented and discussed in section \ref{sec:Phase-diagram}.
Section \ref{sec:Discussion} compares the main results of this paper
with the results of other workers on the critical behavior of $\rho_{\text{typ}}(W)$
and on the existence of non-ergodic phases in finite lattices and
many body systems. The main results of the paper are summarized in
Conclusion, section \ref{sec:Conclusion}.

The paper contains six Appendices which provide the details of the
computations and proofs.

\section{Support set of random wave functions\label{sec:Support-set-of}}

A fundamental concept that distinguishes non-ergodic extended states
from the ergodic ones is \textit{the support set} of wave functions.
Suppose that wave function amplitudes $|\psi_{a}(i)|^{2}\equiv|\langle a|i\rangle|^{2}$
are ordered:
\[
|\psi_{a}(1)|^{2}\geq|\psi_{a}(2)|^{2}\geq...|\psi_{a}(N)|^{2}
\]
and obey the normalization condition:
\begin{equation}
\sum_{i=1}^{N}|\psi_{a}(i)|^{2}=1.\label{norm}
\end{equation}
In order to define the support set we introduce the integer valued
function $M_{\epsilon}$ that gives the number of sites needed to
get the normalization condition with a prescribed accuracy $\epsilon\ll1$:
\begin{eqnarray}
 & \sum_{i=1}^{M_{\epsilon}}|\psi_{a}(i)|^{2} & \leq1-\epsilon\label{sset1}\\
{\rm \text{but}}\nonumber \\
 & \sum_{i=1}^{M_{\epsilon}+1}|\psi_{a}(i)|^{2} & >1-\epsilon.\label{sset2}
\end{eqnarray}
The manifold of sites $i$ contributing to the sum in the left hand
side of (\ref{sset1}) constitutes a \textit{support set} of the wave
function and the number $M_{\epsilon}$ is the \textit{support set
volume} \cite{Our-sset}. The wave function is localized if $M_{\epsilon}$
is finite for any fixed $\epsilon>0$\textit{ }\textit{\emph{in}}
the limit $N\rightarrow\infty$. It is extended and ergodic (EE) if $M_{\epsilon}/N$ is finite in
this limit, and it is extended, \textit{non-ergodic} if $M_{\epsilon}\rightarrow\infty$
while $M_{\epsilon}/N\rightarrow0$.

A special class of extended non-ergodic states are \textit{multifractal}
states for which $M_{\epsilon}=A(\epsilon)\,N^{D}$, where $0<D<1$.
Because the probability to find the particle in a given state on a
site of the support set is almost unity, the typical value of the
wave function on the support set sites is $|\psi_{a}(i)|_{\text{sup}}^{2}\sim N^{-D}$.
In contrast, the typical value of the wave function on a generic site
is much smaller; it is controlled by an exponent $\alpha_{0}>1$
: $|\psi(i)|_{\text{typ}}^{2}\propto N^{-\alpha_{0}}\ll N^{-1}\ll|\psi_{a}(i)|_{\text{sup}}^{2}$.
Qualitatively, the sites of the support set are in resonance with
each other while the sites outside the support set have very different
energies and are connected to the support set only in high orders
of perturbation theory which makes their wave function very small.

One can show \cite{Our-sset} that the exponent $D$ coincides with
the fractal dimension $D_{1}$. The latter is determined by the \char`\"{}Shannon
entropy\char`\"{} which leading term in the limit $N\rightarrow\infty$
is $D_{1}\,\ln N$:
\begin{equation}
\left\langle \sum_{i}|\psi_{a}(i)|^{2}\,\ln(|\psi_{a}(i)|^{-2})\right\rangle =D_{1}\,\ln N.\label{D1}
\end{equation}

\section{Distribution of LDoS\label{sec:Distribution-of-LDoS}}

\begin{figure}[h]
\center{ \includegraphics[width=1\linewidth]{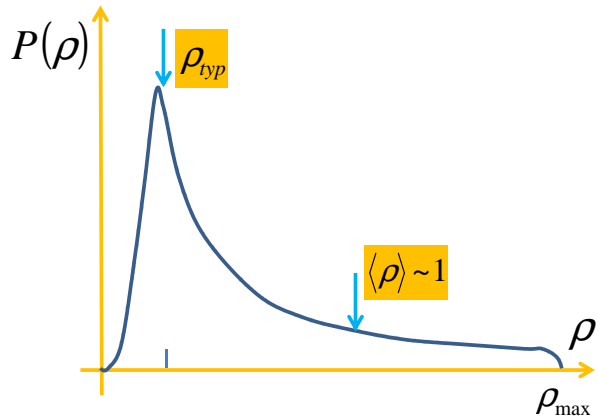}}
\caption{(Color online) Distribution function of local density of states $P_{0}(\rho)$
for the non-ergodic extended states in the limit $\eta\rightarrow0$
taken after the limit $N\rightarrow\infty$ (solid curve). The typical
$\rho_{\text{typ}}$ is much smaller than the average $\langle\rho\rangle$
and depends critically on disorder close to the Anderson transition.\label{Fig:PDF} }
\end{figure}

As argued in \cite{AbouChacAnd} the information on the character
of wave functions $\psi_{a}(i)=\langle i|a\rangle$ can be extracted
from the probability distribution function (PDF) $P_{\eta}(\rho)$
of the generalized local density of states (LDoS) $\rho_{i}\equiv(1/\pi)\,\Im G_{i}$:
\begin{equation}
\rho_{i}(E)=\frac{1}{\pi}\sum_{a}|\langle i|a\rangle|^{2}\,\frac{\eta}{(E-E_{a})^{2}+\eta^{2}},\label{LDoS}
\end{equation}
where $G_{i}$ is the Green's function and $\eta$ is the broadening
of energy levels.

For localized wave functions $P_{\eta\rightarrow0}(\rho)=\delta(\rho)$
is singular: for all $\rho>0$ it vanishes in the limit $\eta\rightarrow0$.
In contrast, in case of the extended ergodic wave functions the result
for the distribution $P_{0}(\rho)$ depends on the order of limits.
If limit $N\rightarrow\infty$ is taken first, before $\eta\rightarrow0$,
one gets a stable non-singular $P_{0}(\rho)$ \cite{AbouChacAnd}
with the typical $\rho_{\text{typ}}$ of the order of the averaged
value $\langle\rho\rangle$:
\begin{equation}
\rho_{\text{typ}}\equiv{\rm exp}[\langle\ln\rho\rangle]\sim\langle\rho\rangle.\label{typ-av}
\end{equation}
In the following we shall refer to this order of limits as Anderson
thermodynamic limit (ATL).

In this paper we show that the non-ergodic extended states on BL are
characterized by a non-singular but extremely broad distribution of
$\rho$, such that $\rho_{\text{typ}}$ is \textit{non-zero} but \textit{parametrically
smaller} than $\langle\rho\rangle$ in ATL:
\begin{equation}
0<\rho_{\text{typ}}\ll\langle\rho\rangle.\label{NEE-rho-distr}
\end{equation}
It becomes exponentially small
\begin{equation}
\rho_{\text{typ}}\sim{\rm exp}[-a/(1-W/W_{c})]\label{rho_at_AT}
\end{equation}
at the disorder strength $W\rightarrow W_{c}$ approaching the Anderson
localization transition. Notice that (\ref{NEE-rho-distr}) does not
by itself prove the existence of a distinct non-ergodic phase because
$\rho_{\text{typ}}$ smoothly changes as $W$ is decreased and becomes
$\sim\langle\rho\rangle$ at $W\ll W_{c}$.

The way to unambiguously characterize the type of extended wave functions
is to consider the \textit{opposite} order of limits when $N\rightarrow\infty$
after $\eta\rightarrow0$ (inverted-order thermodynamic limit, IOTL).
In this limit $\eta$ becomes smaller than the mean level spacing
$\delta\sim1/N$, so that the typical $\rho_{\text{typ}}$ is dominated
by the state closest in energy to the observation energy $E$. In
this regime $\rho_{\text{typ}}$ can be estimated from (\ref{LDoS}):
\begin{equation}
\rho_{\text{typ}}\sim\frac{\eta}{\delta^{2}}\,|\psi|_{\text{typ}}^{2}.\label{rho-typ-lin}
\end{equation}
For ergodic states $|\psi|_{\text{typ}}^{2}\propto N^{-1}$, so that
one gets
\begin{equation}
\rho_{\text{typ}}^{({\rm erg})}\sim\eta\,N.\label{erg-typ}
\end{equation}
For multifractal wave functions the 'main body' of the wave function
is located on its fractal support set, so its typical amplitude $|\psi(i)|_{\text{typ}}^{2}\propto N^{-\alpha_{0}}\ll N^{-1}$
is very small; it is characterized by a non-trivial exponent $\alpha_{0}>1$.
For this class of states we obtain:
\begin{equation}
\rho_{\text{typ}}^{({\rm mf})}\sim\eta\,N^{2-\alpha_{0}},\label{str-ne}
\end{equation}
where the exponent $0<2-\alpha_{0}<1$.

The exponent $2-\alpha_{0}$ is in fact equal to the anomalous dimension,
$D$. There are two ways to prove it. One is to use the physical arguments
of Ref.\cite{AltshulerCuevasIoffeKravtsov2016} that discussed the
crossover from the linear behavior of $\rho_{\text{typ}}(\eta)$ at
small $\eta$ to $\eta-$independent $\rho_{\text{typ}}(\eta)$ at
large $\eta$ and argued that it should occur when $\eta\sim N^{-D}$
which is the distance between the levels in the support set. Because
the $\rho_{\text{typ}}(\eta)$ crosses over from linear function (\ref{str-ne})
to the constant at $\rho\sim O(N^{0})$ these exponents should be
equal. Another argument relies on Mirlin-Fyodorov symmetry of multifractal
spectrum \cite{MF-sym} which gives (see Appendix A):

\begin{equation}
2-\alpha_{0}=D_{1}=D,\label{MF-alpha-D}
\end{equation}
and consequently:
\begin{equation}
\lim_{\eta\rightarrow0}\frac{\rho_{\text{typ}}}{\eta}\sim N^{D}.\label{basic-rho-typ}
\end{equation}
We conclude that behavior of $\rho_{\text{typ}}$ in the IOTL gives
directly the support set dimension $D$.

The scaling behavior $|\psi(i)|_{\text{typ}}^{2}\propto N^{-\alpha_{0}}$
of the \emph{typical }value of the wave function amplitude is much
easier to determine numerically in the exact diagonalization of finite
graphs than the values of the wave function dimensions $D_{q}$. In
particular, its extrapolation to the graphs of infinite size is much
more reliable than that of $D_{q}$ for $q\geq1$. The reason for
that is that for broad distributions the average of $|\psi|^{2q}$
are controlled by the distribution tail which is more sensitive to
the finite size effects and insufficient statitstics than the distribution
main body.

The behavior of the distribution function of $\rho$ at $\rho\gg\rho_{\text{typ}}$
in the IOTL does not contain useful information. The reason for it
is the presence of the $(E-E_{a})^{-2}$ factor in the definition
of $\rho$ that implies that $P(\rho)\propto1/\rho^{3/2}$ at large
$\rho$. This power law dependence simply reflects the probability
to find the state very close to the given energy. It serves, however,
a useful check on the consistency of the analytic approximations developed
below.

\section{The model\label{sec:The-model}}

We are considering the Anderson model on a graph with \textit{$N\gg1$}
sites:

\begin{equation}
\hat{H}=\sum_{i}^{N}\varepsilon_{i}\left\vert i\right\rangle \left\langle i\right\vert +\sum_{i,j=1}^{N}t_{ij}\left\vert i\right\rangle \left\langle j\right\vert \label{eq:H}
\end{equation}
Here \textit{$i=1,2,...,N$ } labels sites of the graph and \textit{$t_{ij}$
}is connectivity matrix of this graph: \textit{$t_{ij}$} equals to
$1$ if the sites \textit{$i$} and \textit{$j$} are connected,$\ $otherwise
\textit{$t_{ij}=0$}. This class of models is characterized by the
on-site disorder: $\epsilon_{i}$ are random on-site energies uniformly
distributed in the interval \textit{$(-W/2,W/2)$. } For the random
regular graph (RRG) all $N$ sites are statistically equivalent and
each of them has $K+1$ neighbors, while the Cayley tree is a \textit{directed,
hierarchical} graph: each site is connected to \textit{$K$ }neighbors
of the previous generation and to one site on the next generation.
The common feature of both graphs is the local tree structure. The
difference is that the Cayley tree is loop-less, while RRG contains
loops. Whilst the number of small loops is only $O(1)$ for the whole
graph, so they cannot have an effect on its properties, the long loops
might be more dangerous. A typical random path starting from a site
$i$ comes back to this site in $\ln N$ steps, so a typical loop
has the length equal to the graph diameter $d_{RRG}\approx\ln N/\ln K$.
Another important feature is that the finite Cayley tree is statistically
inhomogeneous: it has a root and a boundary where a finite fraction
of states is located.

\section{Abou-Chakra-Thouless-Anderson equations\label{sec:Abou-Chakra-Thouless-Anderson-eq}}

For a general lattice one can write self-consistent equations for
the \emph{two point} Green's functions $G_{ij}=\left\langle i\right\vert \left(E-H\right)^{-1}\left\vert j\right\rangle $.
In the absence of the loops, it is possible to derive self-consistent
equations for the \emph{single site} Green's functions, $G_{i}\equiv G_{ii}$
and $G_{i\rightarrow j}$ where the latter denotes single site Green's
function with the bond $i\rightarrow j$ removed:
\begin{align}
G_{i\rightarrow k} & =\frac{1}{E-\epsilon_{i}-\sum_{j\neq k}G_{j\rightarrow i}}\label{eq:Recursion}\\
G_{i} & =\frac{1}{E-\epsilon_{i}-\sum_{j}G_{j\rightarrow i}}
\end{align}

For the Bethe lattice one can introduce the notion of generations:
each site of a given generation, $\ell$, is connected to $K$ ancestors
(generation $\ell-1$) and $1$ descendant (generation $\ell+1$)
and focus only on the Green's functions $G_{i\rightarrow k}$ in which
$k$ is descendant of $i$: $G_{i\rightarrow m}=G_{i}^{(\ell)}$ where
\begin{equation}
G_{i}^{(\ell+1)}(E)=\frac{1}{E-\epsilon_{i}-\sum_{j(i)}G_{j}^{(\ell)}(E)}\label{eq:G_i^(l+1)}
\end{equation}
where $j(i)$ are ancestors of $i$.

The equations Eq.\eqref{eq:G_i^(l+1)} are under-determined: the pole-like
singularities in the right hand side of this equation might be regularized
by adding an infinitesimal imaginary part $\eta$ to $E\rightarrow E+i\eta$
similar to Eq.(\ref{LDoS}). In what follows we will mostly assume
the imaginary part $i\eta$ always added to $E$ in Eq.(\ref{eq:G_i^(l+1)}).

At sufficiently small $W$ the recursion Eq.\eqref{eq:G_i^(l+1)}
might become unstable with respect to addition of $i\eta$. This instability
signals the Anderson transition (AT) point and persists everywhere
in the extended (both EE and NEE) phases.

One can use the recursive relation Eq.(\ref{eq:G_i^(l+1)}) to find
the stationary distribution of $G$. This approach was first employed
by the authors of the seminal paper \cite{AbouChacAnd} who used it
to prove the existence of the localized phase on Bethe lattice and
to determine the critical disorder $W_{c}$ of the AT. Recently we
have generalized it to identify the non-ergodic phase on Bethe lattice
\cite{AltshulerCuevasIoffeKravtsov2016}.

\section{Inflationary population dynamics and fractal dimension $D_{1}$\label{sec:Inflationary-population-dynamics}}

The recursive procedure Eq.(\ref{eq:G_i^(l+1)}) is the basis for
the recursive algorithm known as \textit{population dynamics} (PD)
\cite{pop-dyn} that can be used in two versions. In the linear version
we assume infinitesimally small $\eta>0$ and $W<W_{c}$ . In this
regime the typical imaginary part $(\Im G)_{\text{typ}}=\pi\,\rho_{\text{typ}}$
increases exponentially with the number of recursion steps $\ell$
in Eq.(\ref{eq:G_i^(l+1)}) but remains proportional to $\eta$:
\begin{equation}
\rho_{\text{typ}}(\ell)\propto\eta\,e^{\Lambda\,\ell},\label{Lambda}
\end{equation}
where $\Lambda$ is the corresponding increment.

This \textit{non-stationary} (\char`\"{}inflationary\char`\"{}) regime
of PD holds at the initial stage of iteration for any sufficiently
small $\eta$. It should be contrasted with the conventional \textit{stationary}
regime when the number of iterations is sufficiently large for the
non-linear in $\Im G$ terms to become relevant, and the distributions
of $\Im G$ and $\Re G$ reach their equilibrium which is independent
of $\eta$ as $\eta\rightarrow0$. This regime exactly corresponds
to the Anderson thermodynamic limit.

At an infinitesimal $\eta$ the stationary regime is reached only
in the $\ell\rightarrow\infty$ limit. At the same time, the number
of generations in a finite graph is limited by the size of the graph.
For instance it is equal to $\ln N/\ln K$ for a finite Cayley tree
with $N$ sites. A part of RRG corresponding to $\ell$ generation
with a common ancestor is equivalent to the tree for $\ell<d_{\text{RRG}}\approx\ln N/\ln K$.
Thus, the exponential growth should persist up to $d\approx d_{\text{RRG}}$.
At larger distances the loops have to be taken into account. Consider
an iteration of the equations along the loop that corresponds to the
positive exponent $\Lambda$: $\Im G\rightarrow e^{\Lambda}\,\Im G$.
The recursion in this case would predict the infinite growth of the
$\Im G$ which is impossible for a single finite loop. Clearly, in
order to get the correct result one has to stop the recursion after
one turn. For short loops the number of loops of length $\ell$ is
\cite{Bollobas} $K^{\ell}/(2\ell)$, so that the probability that
a site belongs to a loop of length $\ell$ is $P_{\ell}\approx\exp(\ell\ln K)/2N$.
Thus the loops with $\ell<\ln N/\ln K$ have vanishingly small probability.
A typical loop has the length $\ell\approx d_{RRG}$, where $d_{\text{RRG}}\approx\ln N/\ln K$
is the graph diameter \cite{Bollobas}. Therefore in order to avoid
spurious feedback onto itself, the recursion typically has to stop
after a number of steps equal to the graph diameter $d$:
\begin{equation}
\ell_{t}(N)=d\approx\ln N/\ln K.\label{ell-t}
\end{equation}

The reasoning above neglects the constructive interference between
different paths leading to the same point, we shall discuss the consistency
of this assumption below in Section \ref{sec:Population-dynamics-Lyapunov}.
Here we notice that the prescription Eq.(\ref{ell-t}) gives a correct
result for the fractal dimensions in the Rosenzweig-Porter random
matrix ensemble earlier obtained in \cite{KravtsovKhaymovichCuevas2015}.
Note that this random matrix theory can be mapped on a graph where
every site is connected to any other site directly and thus the corresponding
graph diameter is $d_{\text{RP}}=1$, despite $N\rightarrow\infty$,
so that there are plenty of loops on this graph!

Combining (\ref{Lambda}) and (\ref{ell-t}) one obtains for a finite
RRG:
\begin{equation}
\rho_{\text{typ}}\sim\eta\,N^{\Lambda/\ln K}.\label{typ-term}
\end{equation}
Now, comparing Eq.(\ref{typ-term}) with Eq.(\ref{str-ne}) we see
that the inflationary PD corresponds to the inverted-order thermodynamic
limit (IOTL), and using Eq.(\ref{MF-alpha-D}) we obtain \cite{AltshulerCuevasIoffeKravtsov2016}
for the fractal dimension $D_{1}$:
\begin{equation}
D_{1}=\frac{\Lambda}{\ln K}.\label{Lambda-D}
\end{equation}

\section{Large connectivity approximation\label{sec:Large-connectivity-approximation}}

The increment $\Lambda$ in Eq.(\ref{Lambda}) was computed numerically
for $K=2$ in \cite{AltshulerCuevasIoffeKravtsov2016} using a specially
designed \textit{non-stationary} population dynamics algorithm. The
results unambiguously show that $\Lambda$ is a continuous function
of disorder $W$ that vanishes and changes sign at the AT point $W=W_{c}$
and grows almost linearly as $W$ decreases below $W_{c}$. At low
$W$ the function $\Lambda(\omega)$ flattens out, so that $\Lambda$
never exceeds $\ln K$ . This is the simplest test for consistency
of Eq.(\ref{Lambda-D}), as $D_{1}$ cannot exceed 1. We discuss the
behavior at small $W$ in more detail in section \ref{sec:Analytical-results-for-D(W)}.

In this section we derive an analytic expression for $\Lambda(W)$
and $D_{1}(W)$ in the case of large connectivity $\ln K\gg1$. Linearizing
the right hand side of (\ref{eq:G_i^(l+1)}) in $\Im G$ we obtain:
\begin{equation}
\Im G_{i}^{(\ell)}(E)=\frac{\sum_{j(i)}\Im G_{j}^{(\ell-1)}}{\left(E-\epsilon_{i}-\Re\Sigma_{i}^{(\ell)}\right)^{2}}\label{eq:ImG_lin}
\end{equation}
The general method for the solution of the equations of this type
was developed in \cite{Derrida} that employed mapping to traveling
wave problem. More compact solution uses replica approach and one
step replica symmetry breaking (see e.g. \cite{IoffeMezard,FeigelmanIoffeMezard}).

We begin with the expression $\Lambda(E)=\overline{\ln Z(E)}/\ell$
for $E$- (and $W$-) dependent $\Lambda$, where
\begin{equation}
Z(E)=\sum_{P}\prod_{k=1}^{\ell}\frac{1}{\left(E-\epsilon_{P}^{(k)}\right)^{2}}.\label{eq:f_i}
\end{equation}
In Eq.(\ref{eq:f_i}) $P$ determines a path that goes from an initial
point in generation 1 to a point in generation $\ell$ ($k=1...\ell$),
and $\epsilon_{P}^{(k)}=\varepsilon_{P}^{(k)}+\Re\Sigma_{P}^{(k)}$
is the random on-site energy renormalized by the real part of self
energy $\Re\Sigma$ on this path in the $k$-th generation.

The function $\Lambda(E)$ has a meaning of the free energy of a polymer
on the Bethe lattice \cite{Derrida}with unusual disordered site energies
$\beta E_{j}=\ln\left(E-\epsilon_{j}\right)^{2}$. In order to compute
it we use the replica method:
\begin{eqnarray*}
\Lambda(E) & =\lim_{n\rightarrow0} & \frac{1}{n\ell}\left(\overline{Z^{n}}-1\right).
\end{eqnarray*}
Where $\overline{Z^{n}}$ can be written as:
\begin{equation}
\overline{Z^{n}}=\overline{\sum_{P_{1},...P_{n}}\prod_{a=1}^{n}\prod_{k=1}^{\ell}\frac{1}{(E-\epsilon_{P_{a}}^{(k)})^{2}}}.\label{all-pathes}
\end{equation}
\begin{figure}[h]
\center{ \includegraphics[width=0.9\linewidth]{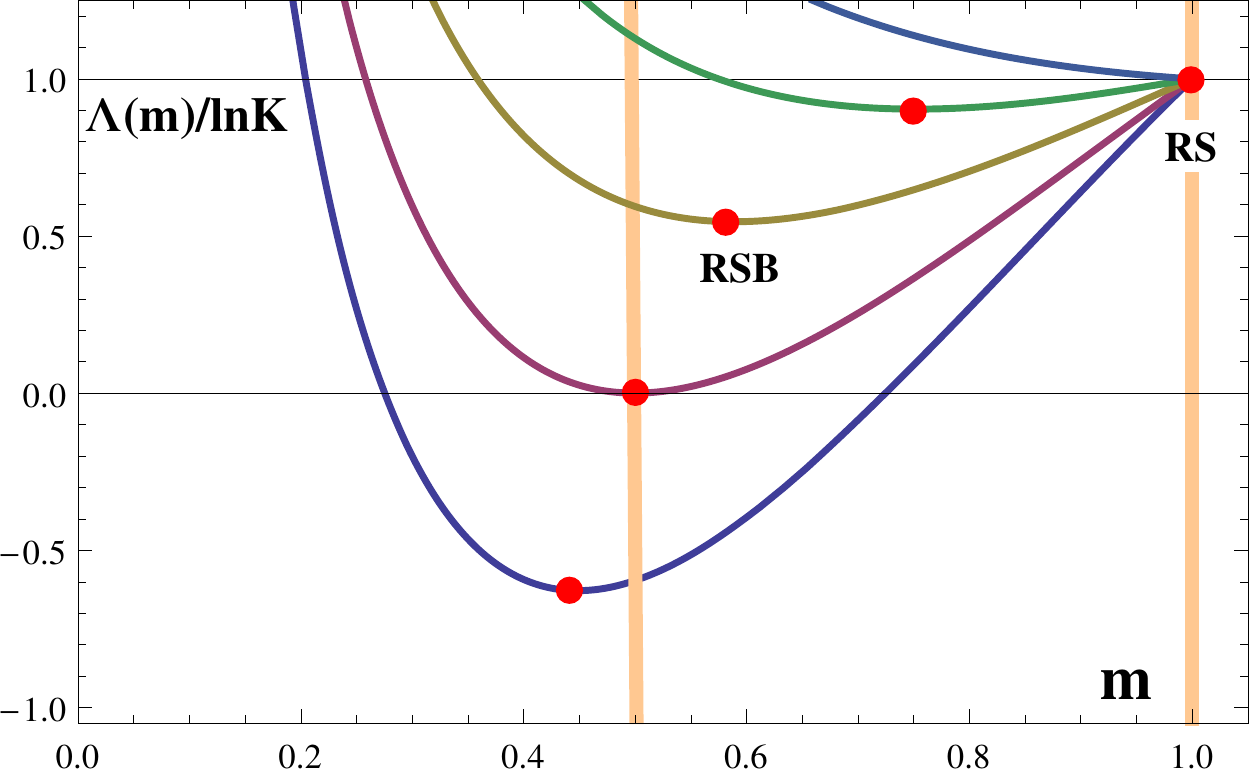}
} \caption{(Color online) Plots of $\Lambda(m)/\ln K$ at $E=0$ and $K=2$ for
$W=25$, $W=W_{c}=17.6$, $W=12$ $W=8$ and $W=5$ in the approximation
Eq.(\ref{low-cutoff}). The one-step RSB solution for the increment
$\Lambda$ corresponds to the minimum on each curve. It exists only
for $W>W_{E}\approx5.74$. The replica symmetric solution always exists
and corresponds to $m=1$, $\Lambda=\ln K$. However, for $W>W_{E}$
the RSB solution gives smaller $\Lambda<\ln K$ and thus is more favorable.
\label{Fig:minimization} }
\end{figure}

We refer to $(E-\epsilon_{P_{a}}^{(k)})^{-2}$ as \textit{entries}
and the product of entries in (\ref{all-pathes}) as a \textit{path}.
Without replica symmetry breaking (RSB) there would be $K^{\ell n}$
pathes contributing to $\Lambda(E)$, each path containing $\ell\,n$
entries. The \textit{one step replica symmetry breaking} implies that
the main contribution is given by paths where these $\ell\,n$ entries
are grouped into $\ell\,n/m$ groups of $m$ \textit{identical} entries
each, considering the contribution of different groups as statistically
independent (the \char`\"{}RSB ansatz\char`\"{}). The RSB solution
for the increment $\Lambda$ is obtained by minimization of $\Lambda(E,m)$
with respect to $m$:
\begin{equation}
\Lambda(E)={\rm min}_{m}\Lambda(E,m)\equiv\Lambda(E,m_{0}).\label{min}
\end{equation}
One step RSB gives exact result for the free energy, $F=\ln Z$, (\ref{all-pathes})
for any distribution function of entries $(E-\epsilon_{P_{a}}^{(k)})^{-2}$
provided that these entries are not correlated between different sites.
In order to understand the reason for this we define the 'free energy'
of individual path by
\begin{equation}
\mathfrak{f}_{P}=\frac{1}{\ell}\sum_{k=1}^{\ell}\ln(E-\epsilon_{P_{a}}^{(k)})^{2}\label{eq:f_P}
\end{equation}
The partition sum, $Z$ is controlled by the path that corresponds
to the minimal energy, $\mathfrak{f}_{P}=\mathfrak{f}_{0}$. The number
of paths with energies larger than $\mathfrak{f}_{0}$ grows exponentially,
$S_{\text{conf}}(\mathfrak{f})=\ln\mathcal{N}=m\ell(\mathfrak{f}-\mathfrak{f}_{0})$,
so that main contribution to $Z^{(m)}=\sum_{P}\exp(-m\ell\mathfrak{f}_{P})$
comes from the thermodynamically large number of paths with $\mathfrak{f}_{P}\approx\mathfrak{f}_{0}$.
The large number of paths implies that $Z^{(m)}$ is self averaging
and thus can be computed straightforwardly. On the other hand, all
these paths have the same $\mathfrak{f}_{P}$ as the optimal one in
thermodynamic limit which allows one to extract the value of $\mathfrak{f}_{0}$
from this computation. These arguments implicitly assume that the
two paths are either fully correlated (if they pass through the same
point) or completely uncorrelated (if they pass through different
points). In this situation, the one step RSB approximation gives exact
results to this problem (see also \cite{IoffeMezard,FeigelmanIoffeMezard}).

In the presence of non-local correlations between entries caused by
the correlations in $\Re\Sigma$, the configurational entropy acquires
a non-trivial dependence on the distance between paths. In this case
one probably needs to introduce re-weighting factors in order to get
the self averaging partition function, which is equivalent to the
full RSB scheme instead of the one-step RSB ansatz. The development
of such scheme is beyond the scope of this paper and our abilities.

Formally, the next step is averaging with respect to random on-site
energies:
\begin{eqnarray}
\Lambda(E,m) & = & \lim_{n\rightarrow0}\frac{1}{n}\left[\left(K\int F(\epsilon)\,\frac{d\epsilon}{\left|E-\epsilon\right|^{2m}}\right)^{n/m}-1\right]\nonumber \\
 & = & \frac{1}{m}\ln\left(K\tilde{I}_{m}\right),\label{eq:Lambda-averaging}
\end{eqnarray}
where
\begin{equation}
\tilde{I}_{m}=\int F(\epsilon)\,\frac{d\epsilon}{\left|E-\epsilon\right|^{2m}}\label{Imm}
\end{equation}
In this equation $F(\epsilon)=(1/W)\,\theta(W/2-|\epsilon|)$ is the
box-shaped on-site energy distribution function, and the contribution
of $\Re\Sigma$ is \textit{completely neglected} which can be justified
at very large $K$. Indeed, as it has been shown in \cite{AbouChacAnd},
in the limit $\ln K\gg1$ the relevant range of disorder potential
corresponds to $\ln |\epsilon_{i}|\sim \ln W\sim \ln K\gg1$ so, the real part of the self-energy in
the denominator of Eq.(\ref{eq:ImG_lin}) can be neglected.

Then the increment $\Lambda(E)$ found from Eq.(\ref{eq:Lambda-averaging})
takes the form:
\begin{equation}
\Lambda=2\,\ln\left(\frac{W_{c}(E)}{W}\right).\label{Lambda-fin-large-K}
\end{equation}

In Eq.(\ref{Lambda-fin-large-K}) the critical disorder $W_{c}=W_{c}(0)$
close to the middle of the band is defined as:
\begin{equation}
\ln\frac{W_{c}}{2}=\frac{1}{2m_{0}}\ln\frac{K}{1-2m_{0}}.\label{eq:lnWc/2}
\end{equation}
where $m_{0}$ is found from the minimization condition Eq.(\ref{min}):
\begin{equation}
\frac{2m_{0}}{1-2m_{0}}=\ln\frac{K}{1-2m_{0}}\label{mstar}
\end{equation}
Combining (\ref{eq:lnWc/2},\ref{mstar}) to exclude $1/(1-2m_{0})$
we get an equation for $W_{c}$:
\begin{equation}
K\,\ln\left(\frac{W_{c}}{2}\right)=\frac{W_{c}}{2e},\label{upper-limit}
\end{equation}
which is exactly the \char`\"{}upper bound\char`\"{} equation for
$W_{c}$ of Ref.\cite{AbouChacAnd}. At large $\ln K\gg1$ one obtains
with logarithmic accuracy
\begin{equation}
W_{c}\approx2eK\ln(eK)\label{lKWc}
\end{equation}
in agreement with \cite{AbouChacAnd}.

\section{RSB parameter $m_{0}$ and Abou-Chakra-Thouless-Anderson exponent
$\beta$\label{sec:RSB-parameter-m_0}}

The parameter $m_{0}$ found from Eq.(\ref{mstar}) (which is independent
of $W$ in the leading approximation) has a special physical meaning,
it is related to the power-law dependence of the distribution function.
To establish this correspondence consider the behavior of the moments
$\overline{(\Im G)^{q}}$ in IOTL.

The value of $\Lambda$ computed in section \ref{sec:Large-connectivity-approximation}
describes the exponential growth of the typical value of $\Im G$.
One can use the same method to determine $\Lambda_{q}=\ln\overline{(\Im G)^{q}}/\ell$
that governs the growth of the moments $\overline{(\Im G)^{q}}$.
Repeating the arguments of section \ref{sec:Large-connectivity-approximation}
we get $\Lambda_{q}=q\Lambda$ provided that $q$ is sufficiently
small, $q<m_{0}$, so that analytic continuation to this value of
$q$ has the same structure as continuation to $n=0$ employed in
section \ref{sec:Large-connectivity-approximation} . For $q>m_{0}$
the solution disappears which describes the fact that higher moments
of $\Im G$ grow faster with $\ell$ than its typical value or even
diverge indicating the absence of linear response ($\Im G\propto\eta$).
This behavior of the moments implies that the distribution function
of $y=\Im G/\rho_{\text{typ}}$ acquires a stationary form at $\ell\rightarrow\infty$
with the power law tail $P(y)\propto1/y^{1+m_{0}}$ with the lower
cutoff $\sim1$ and the upper cutoff that grows with $\ell$. Indeed,
for this distribution all moments $\left\langle y^{q}\right\rangle $
are finite for $q<m_{0}$ and diverge for $q>m_{0}$. The same conclusion
can be obtained by solving the equation for the evolution of the distribution
function directly (see \cite{FeigelmanIoffeMezard} and Appendix B).

In particular, at the Anderson transition $m_{0}$ determines the
power of $\rho$ (or $N|\psi|^{2}$) in the power-law distribution
function $P(\rho)$ in both IOTL and ATL limits (see Appendix A):
\begin{equation}
P(\rho)\propto\frac{1}{\rho^{1+m_{0}}},\label{PDF}
\end{equation}

Thus the RSB parameter $m_{0}$ at $W=W_{c}$ is \textit{identical}
to the exponent $\beta$ introduced in Ref.\cite{AbouChacAnd}. The
work \cite{AbouChacAnd} also shows that in agreement with the general
arguments of section \ref{sec:Distribution-of-LDoS} the exponent
at the transition is
\begin{equation}
\beta=1/2\label{beta-c}
\end{equation}
The same result, Eq.(\ref{beta-c}), follows from the duality (\ref{eq:duality}),(\ref{MF-sym})
for a linear $f(\alpha)$.

On the other hand, one obtains from Eq.(\ref{mstar}):
\begin{equation}
m_{0}\approx1/2-1/(2\ln K).\label{m_0}
\end{equation}
which coincides with the exact result within the accuracy of the approximation
that neglected the effects of the real part of the Green's function.
In the next section we discuss how one can take into account these
effects and develop better approximation.

\section{Minimal account for the real part of self-energy\label{sec:Minimal-account-for-real-part}}

Fairly large errors of the ``upper-limit'' value of $W_{c}$ (\ref{upper-limit})
can be traced back to the inaccurate value of $m_{0}$ (\ref{m_0})
at $W=W_{c}$ that differs from the exact result, $1/2$, by $m_{0}-1/2\sim1/\ln K.$
This difference is due to the complete neglect of the real part of
the self energy in the denominator of (\ref{eq:ImG_lin}) and the
resulting logarithmic divergence of the average $\int_{-W/2}^{W/2}\frac{d\epsilon}{W}\,\epsilon^{-2m}$
at $m=1/2$. In order to improve the accuracy of the analytic theory
we introduce the \textit{effective} distribution function $F_{{\rm eff}}(\epsilon)$
of the real part of $\varepsilon_{i}+\sum_{j(i)}G_{i}=E-G_{i}^{-1}$
instead of the distribution $F(\epsilon)$ of the on-site energies
$\varepsilon_{i}$. Because in the resulting model the entries in
(\ref{all-pathes}) remain uncorrelated, for this effective distribution
the one-step RSB ansatz is treated exactly and leads to (\ref{eq:Lambda-averaging}).
This will allow us to restore the exact value of $m_{0}=1/2$ at the
transition point and dramatically reduce the error in the value of
$W_{c}.$

We emphasize that function $F_{{\rm eff}}(\epsilon)$ is an \textit{effective
distribution} which takes into account correlations of different entries
caused by correlations of $\Re G_{i}$ in a given path in Eq.(\ref{eq:f_i})
when $\Re\Sigma$ is not neglected. Indeed, if $\Re G_{j\rightarrow i}$
is anomalously large, the Green's function at the affected sites $\Re G_{i}\sim[\Re G_{j\rightarrow i}]^{-1}$
should be anomalously small. This effect leads (see Appendix D for
detailed derivation) to the symmetry ${\cal P}_{\ell}(y)={\cal P_{\ell}}(1/y)$
of the PDF of the product $y=\prod_{k}|G_{i_{k}}|^{-1}$ along a path
$P$ of length $\ell\gg1$, which is equivalent to the symmetry of
$F_{{\rm eff}}(\epsilon)$:
\begin{equation}
F_{{\rm eff}}(\epsilon+E)=F_{{\rm eff}}(\epsilon^{-1}+E).\label{symmetry-omega}
\end{equation}
Notice that the introduction of $F_{{\rm eff}}(\epsilon)$ does not
solve the problem of non-local correlations between \textit{different}
paths which may invalidate the one-step RSB ansatz.

The simplest approximation for the effective distribution function
$F_{{\rm eff}}(\epsilon)$ that is close to the original distribution
but obeys the symmetry Eq.(\ref{symmetry-omega}) at $E=0$ is
\begin{equation}
F_{{\rm eff}}(\epsilon)=F_{{\rm eff}}(1/\epsilon)=\frac{\theta(|\epsilon|-2/W)\theta(W/2-|\epsilon|)}{W-4/W},\label{low-cutoff}
\end{equation}
Thus the minimal account of $\Re\Sigma$ is equivalent to imposing
the symmetry (\ref{symmetry-omega}) which eliminates small $|\epsilon|<2/W$.
Physically, it describes the level repulsion from the state at energy
$E$. We will see below that this is a crucial step with many implications.
For instance it allows for the \textit{replica-symmetric} solution
which corresponds to $D=1$. Notice that in the absence of the gap
at low $\epsilon$ implied by the distribution $F(\epsilon)$ (\ref{low-cutoff})
this solution did not exist as $\int F(\epsilon)\,|\epsilon|^{-2m}\,d\epsilon$
always diverges at $m=1$.

Now the critical disorder $W_{c}$ and $m_{0}$ are found from the
solution of the system of equations: \begin{subequations}

\begin{align}
\tilde{I}_{m} & =K^{-1},\\
\frac{\partial\,\tilde{I}_{m}}{\partial\,m} & =0,
\end{align}
\label{system}\end{subequations}where
\[
\tilde{I}_{m}=\int F_{{\rm eff}}(\epsilon+E)\,\frac{d\epsilon}{|\epsilon|^{2m}}.
\]
On can easily see that the symmetry Eq.(\ref{symmetry-omega}) results
in: \begin{subequations}
\begin{align}
\tilde{I}_{m} & =\tilde{I}_{1-m},\\
\partial\,\tilde{I}_{m} & =-\partial\,\tilde{I}_{1-m},
\end{align}
\label{sym-I_n}\end{subequations}where $\partial\,\tilde{I}_{m}=\partial\,\tilde{I}_{m}/\partial\,m$.
Eq.(\ref{sym-I_n}) implies that $\partial\tilde{I}_{1/2}=0$, so
that $m_{0}=1/2$ is an \textit{exact solution} to the equations (\ref{system}b).
The critical disorder is then found from the first equation
\begin{equation}
\tilde{I}_{1/2}(W_{c})=K^{-1}.\label{crit-exact}
\end{equation}

\section{Improved large-K approximation for $W_{c}$\label{sec:Improved-large-K-approximation}}

As we have seen, the Abou-Chakra-Thouless-Anderson \char`\"{}upper
bound\char`\"{} (\ref{upper-limit}) for $W_{c}$ has an accuracy
of $1/\ln K$. The symmetry (\ref{symmetry-omega}) allows one to
take into account all terms $\sim1/\ln K$ and obtain a new estimate
for the critical disorder, which accuracy is at least $1/\sqrt{K}$.

Computing $\tilde{I}_{m}(W)$ using Eq.(\ref{low-cutoff}) one reduces
Eq.(\ref{crit-exact}) to
\begin{equation}
2K\,\ln\left(\frac{W_{c}}{2}\right)=\frac{W_{c}}{2}-\frac{2}{W_{c}}.\label{Wc-K}
\end{equation}

The results of solution of this \textit{algebraic} equation for different
connectivity $K$ is summarized in Table I.

\begin{table}[h!]
\centering \caption{Comparison of values for $W_{c}(K)$ obtained from Eq.(\ref{Wc-K}),
from the \char`\"{}upper bound\char`\"{} of Ref.\cite{AbouChacAnd}
(Eq.(\ref{upper-limit}) of this paper) and from numerics of Ref.\cite{Biroli-Tarzia-rev-BL}.}
\label{tab:table1} %
\begin{tabular}{|c||c|c|c|c|c|c|c|}
\hline
K  & 2  & 3  & 4  & 5  & 6  & 7  & 8\tabularnewline
\hline
Eq.(\ref{Wc-K})  & 17.65  & 34.18  & 52.30  & 71.62  & 91.91  & 113.0  & 134.8 \tabularnewline
\hline
Ref.\cite{Biroli-Tarzia-rev-BL}  & 17.4  & 33.2  & 50.1  & 67.7  & 87.3  & 105  & 125.2 \tabularnewline
\hline
\char`\"{}upper bound\char`\"{}  & 29.1  & 53.6  & 80.3  & 108  & 138  & 169  & 200 \tabularnewline
\hline
\end{tabular}
\end{table}

One can notice an \textit{\emph{excellent}}\emph{ }agreement with
numerics even for the minimal $K=2$. The results for large $K>8$
are expected to be even more accurate.

We conclude that the correct symmetry improves at lot the large-K
approximation and leads to an extremely simple and powerful formula
for $W_{c}$ which accuracy exceeds by far any approximations to the
exact Abou-Chacra-Thouless-Anderson theory known so far.

\section{Analytical results for $D(W)$ and $m(W)$ at the band center $E=0$.\label{sec:Analytical-results-for-D(W)}}

\begin{figure}[h]
\includegraphics[width=0.9\linewidth]{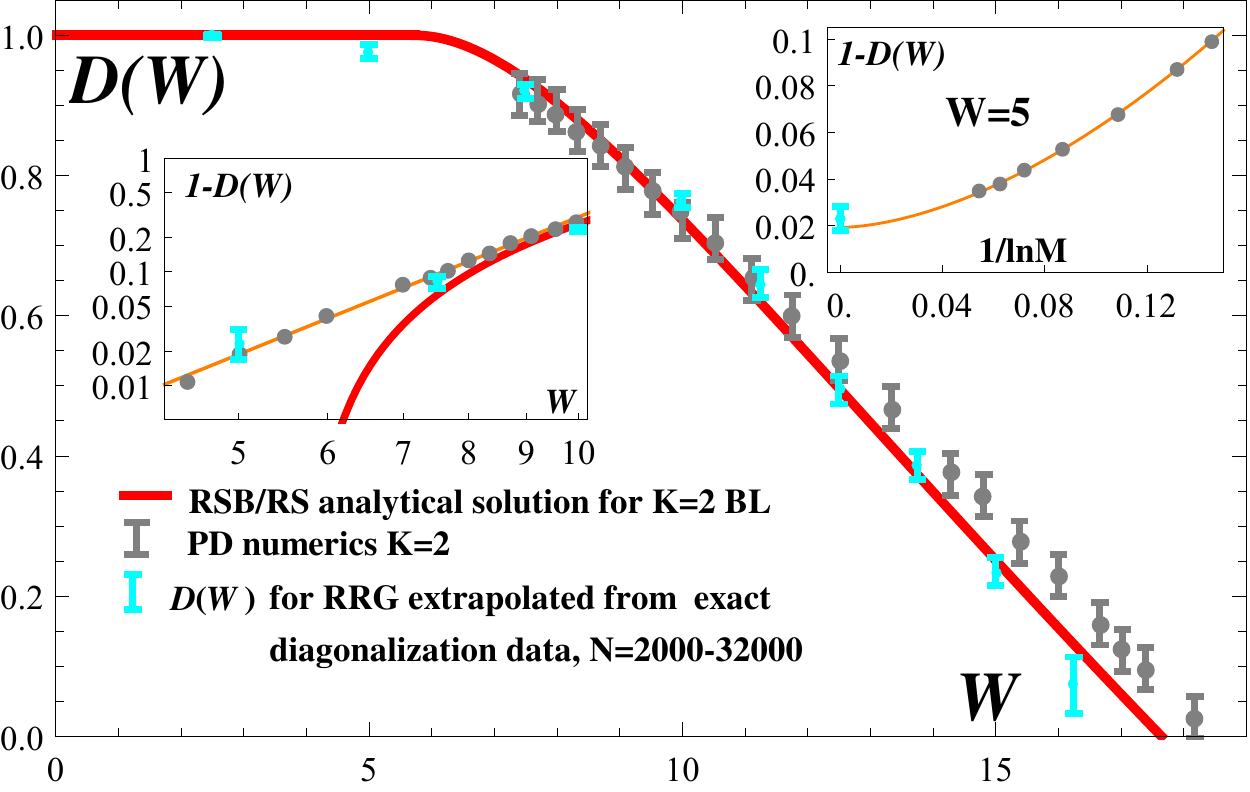}
\caption{(Color online) Fractal dimensions $D(W)$ at $E=0$ and $K=2$ determined
by different methods. The red solid curve gives the result of RSB
solution, (\ref{Lambda-D}),(\ref{min}), (\ref{eq:Lambda-averaging}),(\ref{low-cutoff}).
The grey error bars show the results of the inflationary population
dynamics. Cyan error bars display $D(W)$ obtained by extrapolation
of numerical diagonalization of finite RRG. The one-step RSB solution
of section \ref{sec:Minimal-account-for-real-part} gives two transitions,
from egrodic to non-ergodic at $W_{E}\approx5.74$ and from ergodic
to to completely localized states at $W_{c}\approx17.65$. The population
dynamics gives well defined transition to the localized state at $W_{c}\approx18.6\pm0.3$
but, in contrast to the RSB solution, it indicates to a gradual crossover
to completely ergodic state ( $D\equiv1$ ) as $W$ becomes smaller,
e.g. $D(W)\approx0.9808\pm0.0003<1$ even for $W=5$. For $W\lesssim12$
population dynamics results are in a perfect agreement with the extrapolation
to $N\rightarrow\infty$ of exact diagonalization data for modestly
large $N=2000-32000$, e.g. $D(W)=2-\alpha_{0}\approx0.977\pm0.005$
for $W=5$. Insets: the dependence of $1-D(W)=1-\Lambda/\ln K$ on
the logarithm of the population size $M$ obtained by inflationary
PD at $W=5$ and its extrapolation to $M\rightarrow\infty$; zoom
into $D(W)$ at low $W$ that displays $1-D\propto W^{\zeta}$ behavior
with $\zeta$ very close to 4. \label{Fig:D(W)} }
\end{figure}

The results of Section \ref{sec:Minimal-account-for-real-part} allows
one to get the analytical results for anomalous dimension $D(W)$.
Plugging (\ref{low-cutoff}) into (\ref{eq:Lambda-averaging}) we
compute the increment $\Lambda(W)$. Finally we use Eq.(\ref{Lambda-D})
to convert it into fractal dimension $D(W)$. The resulting prediction
of the RSB theory for the behavior of $D(W)$ at $K=2$ is displayed
in Fig.~\ref{Fig:D(W)}. In this figure we also compare the RSB result
for $D(W)$ with the results of the population dynamics and the results
of the direct numerical diagonalization for finite RRG. The latter
were obtained by using the data of Ref.\cite{Our-BL} for the distribution
of $|\psi|^{2}$ for RRG of moderate sizes $N=2000-32000$. In more
detail, we have computed the finite-size spectrum of fractal dimensions
defined by
\begin{align}
f(\alpha,N) & =\ln[N\,P(\ln|\psi_{\text{env}}|^{2})]/\ln N,\label{eq:f(alpha,N)}\\
\alpha & =-\ln|\psi_{\text{env}}|^{2}/\ln N\nonumber
\end{align}
where $\psi_{\text{env}}$ is wave function envelope and $P(\ln|\psi_{\text{env}}|^{2})$
is the corresponding probability density, extrapolated it to $N\rightarrow\infty$
and found $D_{1}=2-\alpha_{0}$ from the maximum point $\alpha_{0}$
of the extrapolated $f(\alpha,\infty)$. More details can be found
in Appendix \ref{app:Extraction-of-D(W)}. Here we only note that
the spectrum of fractal dimensions $f(\alpha)$ translates into
\begin{equation}
P(\ln Z)=A\exp\left[\ln N\,f\left(\frac{\ln Z}{\ln N}\right)\right]\label{eq:P(lnx)-general}
\end{equation}
where $\text{Z=\ensuremath{\ln\psi}}_{\text{env}}^{2}$. The form
(\ref{eq:P(lnx)-general}) of the distibution is a very general one
that corresponds to the fractality of the physical quantity described
by variable $Z.$ For linear function $f(x)$ it is reduced to the
power law. More generally it describes the crossover from the power
law to a Gaussian-like behavior. This distribution function appear
in a variety of problems, including classical ones\cite{KravtsovPekola2016}.

We notice excellent agreement between the results obtained by population
dynamics and the data of the direct diagonalization away from Anderson
transition, $W\lesssim12$. At larger $W$ population dynamics and
direct diagonalization results deviate from each other due to the
rapidly growing correlation volume as $W$ approaches $W_{c}.$ As
discussed in Section \ref{sec:RSB-results-for-rho_typ} the correlation
volume $N_{c}\sim2\times10^{4}$ at $W=12$, so the deviations of
the results of direct diagonalization from the infinite size limit
at $W\gtrsim12$ are only to be expected. The results of the RSB theory
and population dynamics are in a very reasonable quantitative agreement
with each other at $W\gtrsim7$.

\begin{figure}
\includegraphics[width=0.9\linewidth]{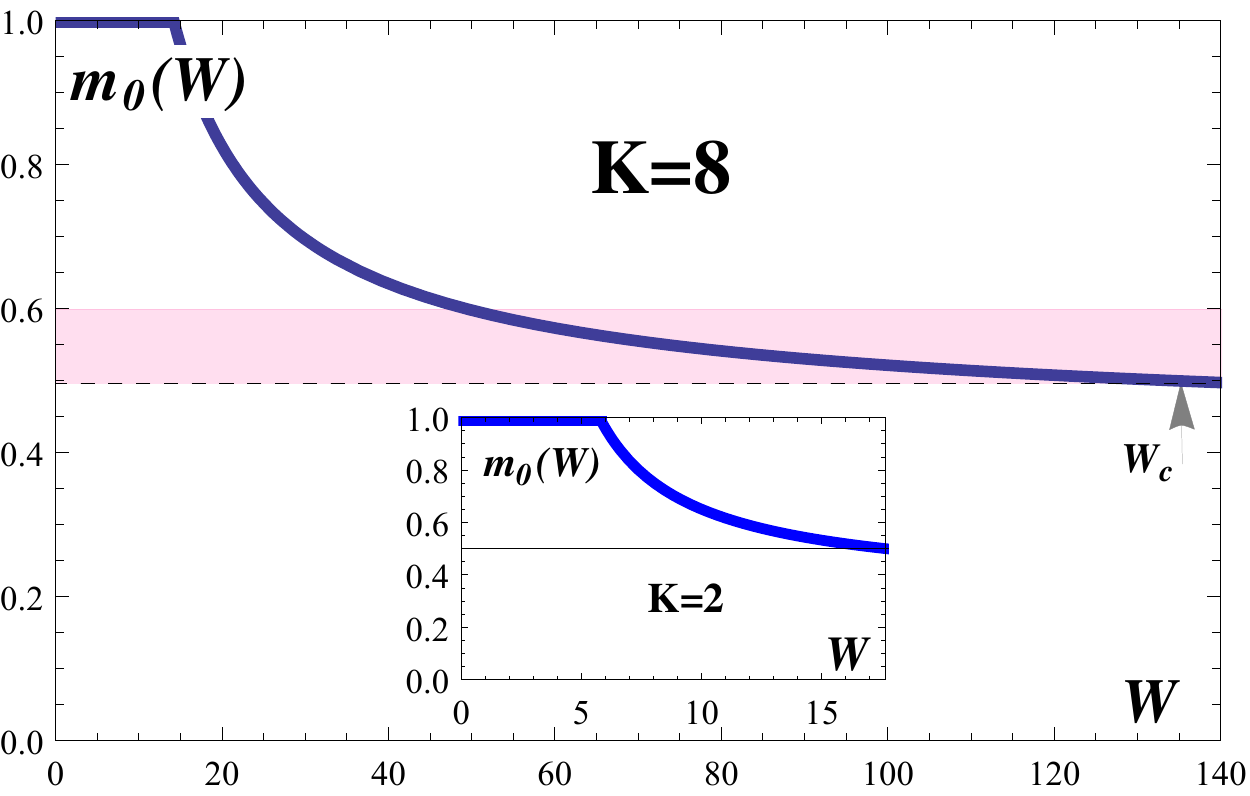}
\caption{Region of validity of RSB solution. The RSB prediction for the power
dependence of the distribution function $P(\rho)\propto\rho^{-(1+m_{0})}$
remains very close to the exact result ($m_{0}=1/2)$ in a wide region
of $W$ for $K=8$. For $K=2$ the value of $m_{0}$ deviates significantly
from $1/2$ at $W\lesssim8$ in accordance with the deviation of the
population dynamics and RSB results for anomalous dimension shown
in Fig.~\ref{Fig:D(W)}.\label{fig:m(W)}}
\end{figure}

At low $W\lesssim7$ the population dynamics and RSB give qualitatively
different predictions. The former predicts gradual crossover to $D=1$
that follows the scaling behavior $1-D\propto W^{\zeta}$ with $\zeta\approx4$
whilst RSB predicts the ergodic transition at $W_{E}\approx5.74$.
Both approaches correspond to infinite sizes, the difference between
them is due to incomplete account of the effects of the real part
of the Green's function in the analytic RSB solution. Generally, one
expects that analytic solution is exact at large $W$ at which all
effects of the real part of the Green's function are small. This can
be verified by computing the dependence of $m_{0}(W)$. Both the general
arguments of section \ref{sec:Distribution-of-LDoS} and population
dynamics (see Fig.\ref{fig:Distribution-functions-of-rho}) predicts the distribution function $P(\rho)\propto\rho^{-3/2}$
at $\rho\gg\rho_{\text{typ}}.$ This translates into $m_{0}=1/2$
for all $W<W_{c}$ (section \ref{sec:RSB-parameter-m_0}). For RSB
solution $m_{0}=1/2$ at $W=W_{c}$ but it deviates from it at $W<W_{c}$.
These deviations are small in a wide range of $W$:
\begin{equation}
m_{0}-\frac{1}{2}\approx\frac{3\ln\left(\frac{2K\ln\left(\frac{W}{2}\right)}
{\frac{W}{2}-\frac{2}{W}}\right)}{2\ln^{2}\left(\frac{W}{2}\right)}\label{eq:m_0-1/2}
\end{equation}
confirming the accuracy of RSB approach at large $W$. The Fig.~\ref{fig:m(W)}
displays $m_{0}(W)$ dependence that confirms that it stays close
to $1/2$ in a wide range of $W<W_{c}.$ Strong deviations of $m_{0}(W)$
from $1/2$ for $K=2$ at $W\lesssim8$ exactly correspond to the
deviations of the RSB and population dynamics results shown in Fig.~\ref{Fig:D(W)}.

\begin{figure}[h]
\includegraphics[width=0.9\linewidth]{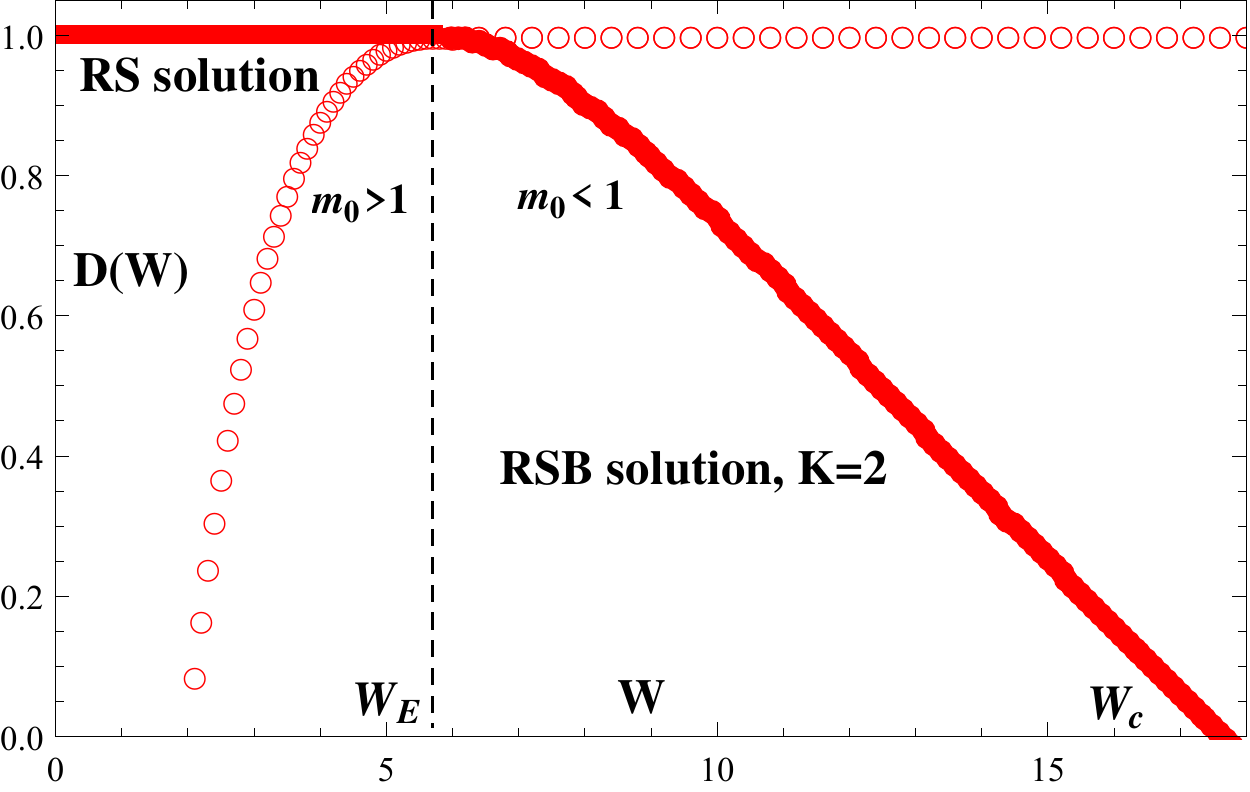}
\caption{(Color online) RSB and RS solutions to (\ref{Lambda-D}),(\ref{min}),(\ref{eq:Lambda-averaging}),(\ref{low-cutoff}).
The branch of the curve with $m_{0}>1$ (open circles) is unphysical.
For $W<W_{E}$ only the RS solution with $D_{RS}=1$ is valid. For
$W_{E}<W<W_{c}$ both solutions exist but only the one with the minimal
$\Lambda$ is realized in PD calculations. \label{Fig:D_RSB} }
\end{figure}

We now discuss in more detail the predictions of the RSB solution
for low $W$ where its accuracy is uncertain. As $W$ decreases below
$W_{c}$, $m_{0}$ increases monotonically from $m_{0}=1/2$ and it
reaches $m_{0}=1$ at
\begin{align}
(W_{E}/2)^{\frac{W_{E}^{2}+4}{W_{E}^{2}-4}} & =e\sqrt{K,}\label{eq:W_E}\\
W_{E} & \approx2e\sqrt{K},\ K\gg1
\end{align}
At this point the RSB solution terminates (see Fig.~\ref{Fig:D_RSB}),
because only $m<1$ are allowed in RSB solution. This proves, within
one-step RSB, existence of the \textit{ergodic transition} from the
non-ergodic extended (multifractal) phase described by the RSB solution
to the extended ergodic phase described by the \textit{replica symmetric}
(RS) solution with $m=1$. Existence of such a RS solution and the
fact that $D=1$ at $m=1$ is a consequence of the symmetry Eq.(\ref{low-cutoff}).
Indeed, at $m=1$ (and $E=0$) we have:
\begin{equation}
\Lambda(m=1)=\ln\left(K\int F_{{\rm eff}}(\epsilon)\,\frac{d\epsilon}{\epsilon^{2}}\right).\label{Lambda-RS}
\end{equation}
Because of the symmetry of $F_{{\rm eff}}(\epsilon)=F_{{\rm eff}}(1/\epsilon)$,
changing the variables of integration $\epsilon\rightarrow1/\epsilon$
converts the integral in Eq.(\ref{Lambda-RS}) into the normalization
integral for the effective distribution function $\int F_{{\rm eff}}(\epsilon)\,d\epsilon=1$.
Then we immediately obtain from Eq.(\ref{Lambda-D}) that the RS solution
corresponds to $D=1$, i.e. to the ergodic extended phase.

In Appendix \ref{app:Termination-point-of-RSB-solution} we prove
that the existence of termination point of the RSB solution at a non-zero
$W_{E}$ such that $m_{0}(W_{E})=1$, $D(W_{E})=1$, and $\partial_{W}D(W_{E})=0$
is a generic feature of the theory. It occurs at any function $F_{{\rm eff}}(\epsilon)$
obeying the symmetry Eq.(\ref{symmetry-omega}) and decreasing sufficiently
fast at large and small $\epsilon$ , e.g faster than $\epsilon^{-1}\,\ln^{-2}\epsilon$
at $\epsilon\rightarrow\infty$.

Neither population dynamics nor RSB theory precludes the 'first order'
jump in $D(W)$ for finite graphs in which loops become important
at large scales. In RSB theory $W_{E}$ gives a \textit{limit of stability}
of the non-ergodic extended phase. The actual ergodic transition may
occur \textit{before} this limit is reached, as the replica symmetric
solution exists in the entire region $W<W_{c}$. In this case it should
be a first order transition at $W=W_{E}$ similar to the one observed
in \cite{AltshulerCuevasIoffeKravtsov2016}. We estimate the effect
of the loops of large sizes in finite RRG in Section \ref{sec:Population-dynamics-Lyapunov}
and conclude that they might become relevant at $W\lesssim10$ in
agreement with the transition observed \cite{AltshulerCuevasIoffeKravtsov2016}
in the data for the largest graphs.

\section{Application to Rosenzweig-Porter model\label{sec:Application-to-Rosenzweig-Porter}}

The generalized Rosenzweig-Porter random matrix model (GRP) is probably
the simplest model in which both localization and ergodic transitions
happen, with the non-ergodic extended phase existing in between \cite{KravtsovKhaymovichCuevas2015}.
It plays the same role for the field of quantum non-ergodicity as
the random energy model for classical spin glasses.

The model is formally defined \cite{RPort,KravtsovKhaymovichCuevas2015}
as a Hermitian $N\times N$ matrix with random Gaussian entries $H_{ij}$
independently fluctuating about zero with the variance $\langle|H_{ii}|^{2}\rangle=1$,
and $\langle|H_{i\neq j}|^{2}\rangle=\lambda\,N^{-\gamma}$, where
$\lambda$ is an $N$-independent number. By changing the energy scale
one may define $h_{ij}$, where $\langle|h_{ii}|^{2}\rangle=\lambda^{-1}\,N^{\gamma}$
and $\langle|h_{i\neq j}|^{2}\rangle=1$. Thus the GRP model corresponds
to $W\sim N^{\gamma/2}$. The AT critical point in the limit $N\to\infty$
corresponds to $\gamma=2$ (see Ref.\cite{KravtsovKhaymovichCuevas2015}
and references therein) and thus $W_{c}\sim N$.

We apply Eq.(\ref{Lambda-fin-large-K}) to GRP, as it should be valid
for any graph with connectivity $K\rightarrow\infty$. The GRP can
be mapped on a graph where each site is connected with any other site
directly and thus in this model $K_{RP}=N$ and the graph diameter
$d_{RP}=1$. Thus we immediately obtain from Eq.(\ref{Lambda-fin-large-K}):
\begin{equation}
\Lambda=2\ln\left(\frac{N}{N^{\gamma/2}}\right)=(2-\gamma)\,\ln N.
\end{equation}
Now, Eq.(\ref{Lambda}) terminated at $\ell_{t}=d_{RP}=1$ and (\ref{str-ne},\ref{MF-alpha-D})
give
\begin{align}
\rho_{\text{typ}} & \sim\eta\,e^{\Lambda}\sim\eta\,N^{D_{1}}\label{RP-D}\\
D_{1}(\gamma) & =2-\gamma
\end{align}
This result coincides with the fractal dimensions $D_{q}=2-\gamma$
(valid for all $q>1/2$) for the GRP obtained in Ref.\cite{KravtsovKhaymovichCuevas2015}
from completely different arguments. Note that $m_{0}$ minimizing
$\Lambda(m)$ is $1/2$ in the entire region of non-ergodic extended
states $1<\gamma<2$ in the limit $\ln N\rightarrow\infty$. This
implies that the exponent in the power-law dependence (\ref{PDF})
is $3/2$ for all values of $\gamma>1$, in agreement with general
expectations and the results of works \cite{KravtsovKhaymovichCuevas2015,RP-Bir}.
We conclude that the exact result of the RSB theory in this case is
associated with the value of the exponent $m_{0}=1/2$ in the entire
region of non-ergodic states.

\section{RSB results for $\rho_{\text{typ}}$ \label{sec:RSB-results-for-rho_typ}}

The typical local density of states in Anderson thermodynamic limit,
$\rho_{\text{typ}},$ is an important characteristic of a strongly
disordered system. Physically, it characterizes the inverse escape
time, i.e. the time needed for a particle to leave a vicinity of a
given site. This time is finite in delocalized regime but becomes
infinite as $W\rightarrow W_{c}$. We discuss the relation between
$\rho_{\text{typ}}$and physical properties in the end of this section.

In this section we compute $\rho_{\text{typ}}$ as a function of disorder
using the one-step RSB and compare it with the results of the population
dynamics. Our goal is to obtain a \textit{stationary} distribution
of $\rho$ for RRG where all sites are statistically equivalent. Note
that stationarity of the \textit{probability distribution, }\textit{\emph{i.e.
its independence of}} $i$ and $\ell$ by no means implies homogeneity
of $\Im G_{i}^{(\ell)}$ for a particular realization of disorder.

It will be more convenient for us to solve the equations for the imaginary
part of the self-energy related to the imaginary part of the Green's
function by $\mathfrak{S}_{i}^{(\ell+1)}=\Im\Sigma_{i}^{(\ell+1)}=\sum_{j(i)}\Im G^{(\ell)}(j)$
. As we show below close to the critical point the typical $\mathfrak{S}$
becomes exponentially small in $1/(W_{c}-W)$. This strong dependence
on $W$ is the same for $\mathfrak{S}_{\text{typ}},$ $\Im G_{\text{typ}}$
and $\rho_{\text{typ}}$that differ from each other only by factors
$K$ and $\pi$. Below we shall focus on the strong exponential dependence
of these quantities and ignore the difference between them.

The recursion equation for $\mathfrak{S}_{j}^{(\ell)}$ follows directly
from (\ref{eq:G_i^(l+1)}):
\begin{equation}
\mathfrak{S}_{i}^{(\ell+1)}=\sum_{j(i)}\frac{\mathfrak{S}_{j}^{(\ell)}}{\epsilon_{j}^{2}+(\mathfrak{S}_{j}^{(\ell)})^{2}},\label{eq:ImG-recursion}
\end{equation}
where $\epsilon_{i}=\varepsilon_{i}+\Re\Sigma_{i}-E$, and $\mathfrak{S}_{i}$
is a sum of $K$ independent random $\Im G_{j(i)}$ on the ancestors
sites.

The power law distribution of $\rho$, $P\sim\rho^{-(1+m_{0})}$ which
is a general property one-step RSB solution (see Appendices \ref{app:m_0-m_1-and-the-power},\ref{app:power-law-distribution-function}),
implies that the contribution to the moment $\langle\mathfrak{S}{}^{m_{0}}\rangle$
comes from a wide region of $\mathfrak{S}$. Indeed, for this and
only this moment the integral $\int P(\mathfrak{S})d\mathfrak{S}$
is logarithmically divergent. The wide distribution of individual
terms in the sum (\ref{eq:ImG-recursion}) implies that in this sum
one term is much larger than others, so that the $m$-th power of
the sum is equal to sum of the powers. This allows us to write the
closed equation for $\langle\rho^{m_{0}}\rangle$ :
\begin{equation}
\langle\rho^{m_{0}}\rangle=K\,\left\langle \frac{\rho^{m_{0}}}{\left(\epsilon^{2}+\rho^{2}\right)^{m_{0}}}\right\rangle _{\epsilon},\label{RSB}
\end{equation}
where $\langle...\rangle_{\epsilon}$ denotes averaging with the distribution
function $F_{{\rm eff}}(\epsilon)$ approximated by Eq.(\ref{low-cutoff}).

Eq. (\ref{RSB}) can be rewritten in terms of the distribution function
$P_{0}(\rho)$
\begin{equation}
\langle\rho^{m}\rangle=K\int d\rho\,\rho^{m}\,P_{0}(\rho)\,\Xi(\rho;W,m),\label{RSB-distr}
\end{equation}
where
\begin{equation}
\Xi(\rho;W,m)=\int\frac{d\epsilon}{\left(\epsilon^{2}+\rho^{2}\right)^{m}}\,F_{{\rm eff}}(\epsilon).\label{Xi}
\end{equation}
The averaging over $\epsilon_{i}$ and $\rho_{i}$ in the same generation
are independent, because on the tree $\rho_{i}$ depends only on $\epsilon_{j}$
in the previous generations.

We now use the definition $\langle\rho^{m}\rangle=\int\rho^{m}\,P_{0}(\rho)\,d\rho$
valid for \textit{any} moment to arrive at the equation:
\begin{equation}
\int d\rho\;\rho^{m_{0}}\,P_{0}(\rho)\,\left[\Xi(\rho;W,m_{0}))-K^{-1}\right]=0,\label{eq:implicit-eq-rho_typ}
\end{equation}
where $m_{0}$ is taken from the solution of (\ref{min}).

Equality (\ref{eq:implicit-eq-rho_typ}) is an \textit{implicit} equation
for $\rho_{\text{typ}}$. In order to make it explicit we note that
$P_{0}(\rho)\propto\rho^{-(1+m_{0})}$ is the stationary distribution
function at $W=W_{c}.$ In the spirit of Ginzburg-Landau theory we
assume that this function does not change significantly when $\rho_{\text{typ}}$
appears below $W_{c}$ at large $\rho\gtrsim\rho_{\text{typ},}$ so
that
\begin{equation}
P_{0}(\rho)\propto1/\rho^{1+m_{0}(W)},\;\rho\gtrsim\rho_{\text{typ}}\label{P-power}
\end{equation}
One can also show that in the vicinity of the Anderson transition
point the curvature of $\ln P_{0}$ as a function of $\ln\rho$ is
very small: $\partial_{x}^{2}(\ln P_{0}(x=\ln\rho))|_{x=0}$ decreases
faster than $(1-W/W_{c})^{2}$ as $W\rightarrow W_{c}$ that colloborates
this assumption. We show in Appendix \ref{app:power-law-distribution-function}
that this fast decrease is sufficient to justify the usage of power
law $P_{0}(\rho)\propto\rho^{-(1+m_{0})}$ in (\ref{eq:implicit-eq-rho_typ})
at $W<W_{c}$ .

Plugging (\ref{P-power}) into (\ref{eq:implicit-eq-rho_typ}) we
obtain an explicit equation for $\rho_{\text{typ}}$:
\begin{equation}
\int_{\rho_{\text{typ}}}^{K\,W/4}\frac{d\rho}{\rho}\,\left[\Xi(\rho;W,m_{0})-K^{-1}\right]=0.\label{eq:rho_0}
\end{equation}
The value of $\rho_{\text{typ}}$ enters this equation as a \textit{hard
lower cutoff}. This is justified by the logarithmic divergence of
the integral (\ref{eq:rho_0}), so that any soft cutoff of the power-law
(\ref{P-power}) at small $\rho$ will only change a numerical prefactor
in the expression for $\rho_{\text{typ}}$ found from (\ref{eq:rho_0}).
The \textit{upper cutoff} emerges because the fraction in (\ref{eq:ImG-recursion})
is always smaller than $K/(2|\epsilon_{i}|)$ and $|\epsilon_{i}|$,
in its turn, is larger than $2/W$ due to the gap (or pseudo-gap)
in the effective distribution $F_{{\rm eff}}(\epsilon)$.

\begin{figure}[h]
\includegraphics[width=0.9\linewidth]{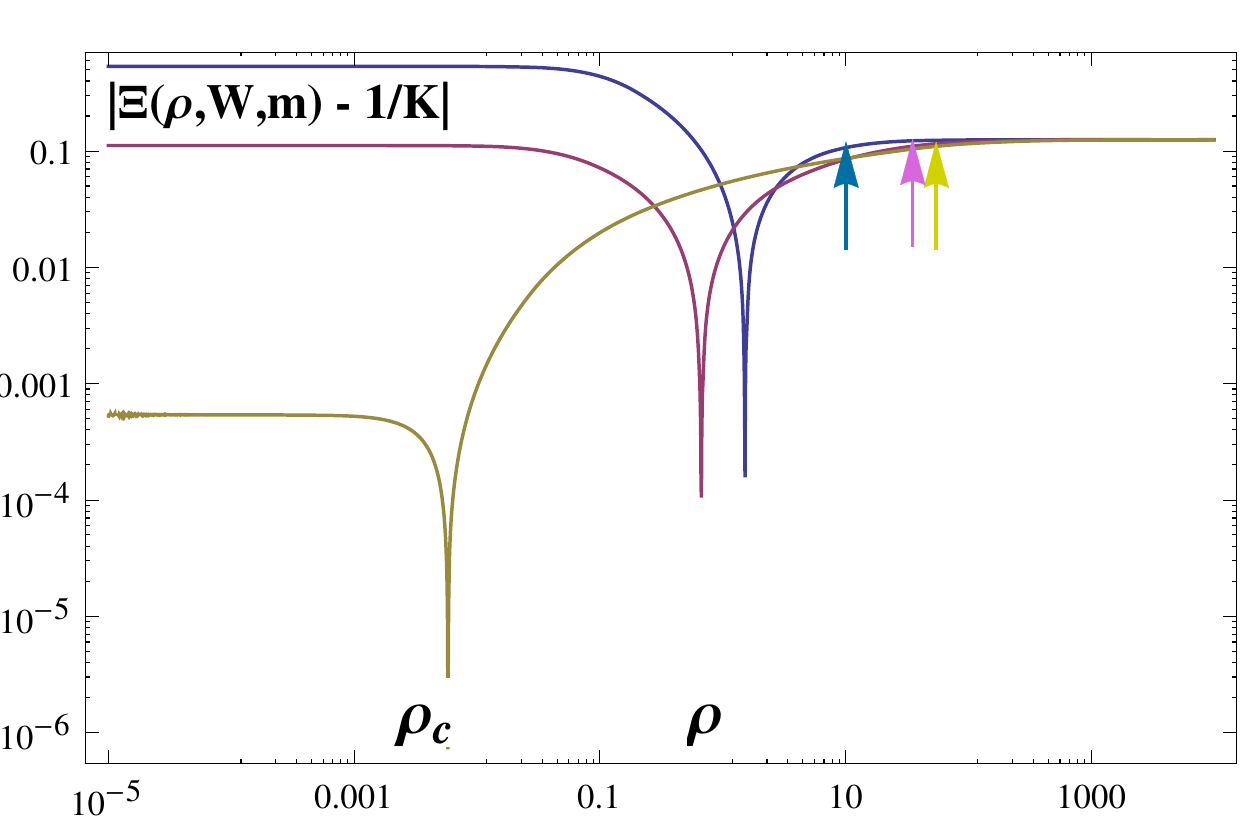}
\caption{(Color online) The function $|\Xi(\rho,W,m(W))-K^{-1}|$ for different
values of $W<W_{c}$. The cusp at $\rho=\rho_{c}(W)$ marks the point
where $\Xi(\rho,W,m(W))-K^{-1}$ changes sign from positive at $\rho<\rho_{c}$
to negative at $\rho>\rho_{c}$. In the limit $W\rightarrow W_{c}$
the positive plateau decreases proportional to $1-W/W_{c}$ and $\rho_{c}$
decreases as $\rho_{c}\propto\sqrt{1-W/W_{c}}$. The arrows show the
position of $\rho_{{\rm max}}=KW/4$.\label{Fig:plotXi} }
\end{figure}

Eq.(\ref{Xi}) for $\Xi(\rho;W,m)$ can be integrated exactly: \begin{widetext}
\begin{equation}
\Xi(\rho,W,m)=\frac{\rho^{-2m}}{W/2-2/W}\,\left[\frac{W}{2}\,_{2}F_{1}\left(\frac{1}{2},m,\frac{3}{2},-\frac{W^{2}}{4\rho^{2}}\right)-\frac{2}{W}\,_{2}F_{1}\left(\frac{1}{2},m,\frac{3}{2},-\frac{4}{W^{2}\rho^{2}}\right)\right].\label{Xi-exact}
\end{equation}
\end{widetext} The absolute value of $\Xi(\rho;W,m_{0})-K^{-1}$
is shown in Fig.~\ref{Fig:plotXi}. For $W<W_{c}$ it has a positive
plateau at small $\rho$ and a negative plateau $=-K^{-1}$ at large
$\rho$, the sign of $\Xi(\rho;W,m_{0})-K^{-1}$ changes at $\rho=\rho_{c}\propto\sqrt{1-W/W_{c}}$.
It follows from (\ref{crit-exact}) that the height of the positive
plateau $\approx\kappa\,(1-W/W_{c})$ vanishes as $W\rightarrow W_{c}$,
where $\kappa\approx4W_{c}^{-1}\,\ln(W_{c}/2e)$. For $W>W_{c}$ the
negative plateau disappears and a non-trivial solution of (\ref{eq:rho_0})
for $\rho_{\text{typ}}$ is no longer possible.

For $W<W_{c}$ the solution of (\ref{eq:rho_0}) always exists. To
find it one has to equate the contribution of the positive plateau
$\kappa\,(1-W/W_{c})\,\ln(\rho_{c}/\rho_{\text{typ}})$ to the contribution
(with the minus sign) of the negative domain $\rho_{c}<\rho<KW/2$
to the integral in Eq.(\ref{eq:rho_0}). The latter is independent
of $\rho_{\text{typ}}$ and determines the constant $a$ in the resulting
asymptotic expression for $\rho_{\text{typ}}$ at $W\rightarrow W_{c}$:
\begin{equation}
\rho_{\text{typ}}\sim\rho_{c}\,{\rm exp}\left[-\frac{a(K)}{(1-\frac{W}{W_{c}})}\right].\label{eq:rho_typ-final}
\end{equation}

Because the upper cutoff in Eq.(\ref{eq:rho_0}) is just at the onset
of the negative plateau (see Fig.~\ref{Fig:plotXi}) a good estimate
for $a(K)$ (which becomes asymptotically accurate at large $K$)
is
\begin{equation}
a(K)=\frac{\ln\left(\frac{K\,W_{c}}{4}\right)}{K\,\kappa}=\frac{W_{c}\ln(KW_{c}/4)}{4K\,\ln(W_{c}/2e)}\sim\ln K.\label{a-K}
\end{equation}
\begin{figure}[th]
\includegraphics[width=0.95\linewidth]{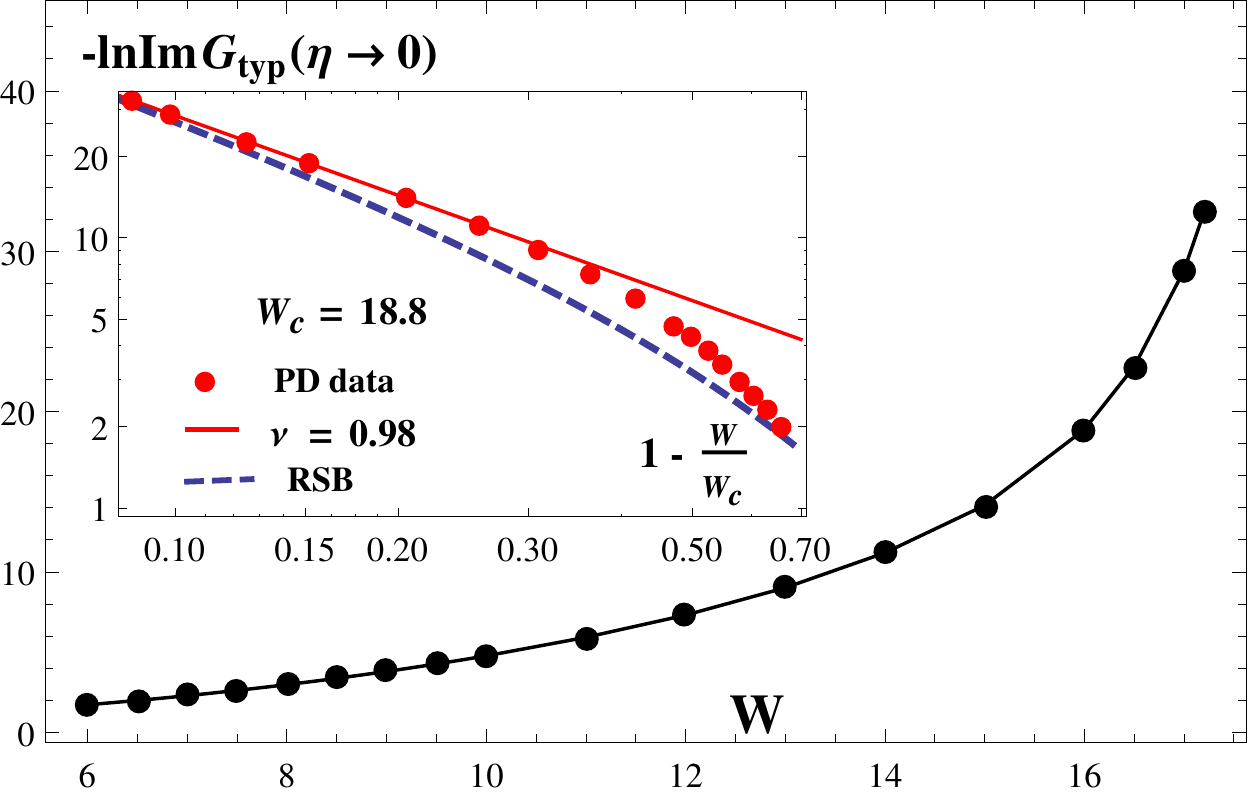}
\caption{(Color online) The plot of $-\ln\Im G_{\text{typ}}|_{\eta\rightarrow0}$
versus $W/W_{c}$ obtained by (equilibrium) population dynamics for
$K=2$ with the population size $\sim2\times10^{7}$ and the number
of generations $20000$. The bare level width $\eta$ changes from
$\eta=0.1$ down to $\eta=10^{-10}$. The saturation of $\Im G_{\text{typ}}$
as a function of $\eta$ with three digit accuracy has been observed
at sufficiently small $\eta$ for each value of $W$. The inset: The
log-log plot of $-\ln\Im G_{\text{typ}}|_{\eta\rightarrow0}$ vs.
$(1-W/W_{c})$ obtained by population dynamics (red points) and by
RSB analytical calculations \textit{with no fitting parameters} (blue
dashed line). Close to $W=W_{c}$ the deviation of PD data points
from a straight line is minimal at $W_{c}=18.8$, the exponent $\nu$
in $-\ln\Im G_{\text{typ}}\propto(1-W/W_{c})^{-\nu}$ (the slope of
red solid line) being $0.98$. Changing $W_{c}$ in the range $18.5-19.0$
results in variation of the slope in the range $0.85-1.03$. \label{Fig:exponent} }
\end{figure}

The result of numerical solution to (\ref{eq:rho_0}) is presented
in Fig.~\ref{Fig:exponent}. In this solution we have used for $\Xi(\rho;W,m)$
the formula (\ref{Xi-exact}) with $m=m_{0}$ given by the solution
of the second equation in (\ref{system}) at $E=0$ .

We now comment on the possible origin of the discrepancy between the
result (\ref{eq:rho_typ-final}) and the prediction of a weaker divergence
$\ln\rho_{\text{typ}}^{-1}\sim(1-W/W_{c})^{-1/2}$ in Ref. \cite{MirlinFyodorov1991}.
The result (\ref{eq:rho_typ-final}) requires a simultaneous account
for the symmetry (\ref{low-cutoff}) \textit{both} in the \textit{linear}
in $\mathfrak{S}_{i}$ regime (encoded in $m_{0}(W)$) and in the
\textit{non-linear} in $\mathfrak{S}_{i}$ regime (encoded in $\Xi(\rho;W,m)$).
If this symmetry is respected only in the linear regime (and thus
$m_{0}\approx1/2$ near AT point) but the effect of $\Re\Sigma$ is
disregarded in (\ref{Xi}) for $\Xi(\rho;W,m)$, the result for $\Xi(\rho;W,m)\propto\ln\rho$
would show logarithmic divergence at small $\rho$, so that the contribution
of the positive plateau at $\rho_{\text{typ}}<\rho<\rho_{c}$ to the
integral (\ref{eq:rho_0}) would be $\frac{1}{2}\kappa\,(1-W/W_{c})\,\ln^{2}\rho_{\text{typ}}$.
As the result we would get $\rho_{\text{typ}}\propto{\rm exp}[-\,c/\sqrt{1-W/W_{c}}]$
instead of (\ref{eq:rho_typ-final}). The numerics presented in the
next section rules out this possibility, while confirming (\ref{eq:rho_0})
with unexpected accuracy (see Fig.~\ref{Fig:exponent}).

Qualitatively, the parameter $\rho_{t\text{yp}}$ has the meaning
of the typical local level width, i.e. inverse escape time, it characterizes
the ``badness'' of the metal. In conventional localization it is
directly related to the diffusion coefficient and conductance. For
example, it was shown in Ref.\cite{AKL86} that for weak multifractality
the local density of states distribution function $P_{0}(\rho)$ takes
the form:
\begin{equation}
P_{0}(\rho)=\frac{1}{\rho\,\sqrt{4\pi\,u}}\;{\rm exp}\left[-\frac{\left(\ln\left(\frac{\rho}{\langle\rho\rangle}\right)+u\right)^{2}}{4u}\right],\label{P_0-AKL}
\end{equation}
where
\begin{equation}
u=\ln\left(\frac{\sigma_{0}}{\sigma}\right)=\ln\left(\frac{{\cal D}_{0}}{{\cal D}}\right).\label{u}
\end{equation}
Here $\sigma_{0}$ and ${\cal D}_{0}$ are Drude conductivity and
diffusion coefficient, respectively, while $\sigma$ and ${\cal D}$
are those with weak localization effects included.

It immediately follows from (\ref{P_0-AKL},\ref{u}) that
\begin{equation}
\frac{\rho_{\text{typ}}}{\langle\rho\rangle}=\frac{\sigma}{\sigma_{0}}=\frac{{\cal D}}{{\cal D}_{0}}.\label{rho_typ-sigma}
\end{equation}
We believe that Eq.(\ref{rho_typ-sigma}) holds well beyond the weak
localization condition of its derivation and it applies for generic
disordered system with multifractal statistics of eigenfunctions.

\section{Population dynamics for $\text{Im}G_{\text{typ}}$ and $K(\omega)$
\label{sec:Population-dynamics-for-ImG}}

\begin{figure}[h!]
\includegraphics[width=0.9\linewidth]{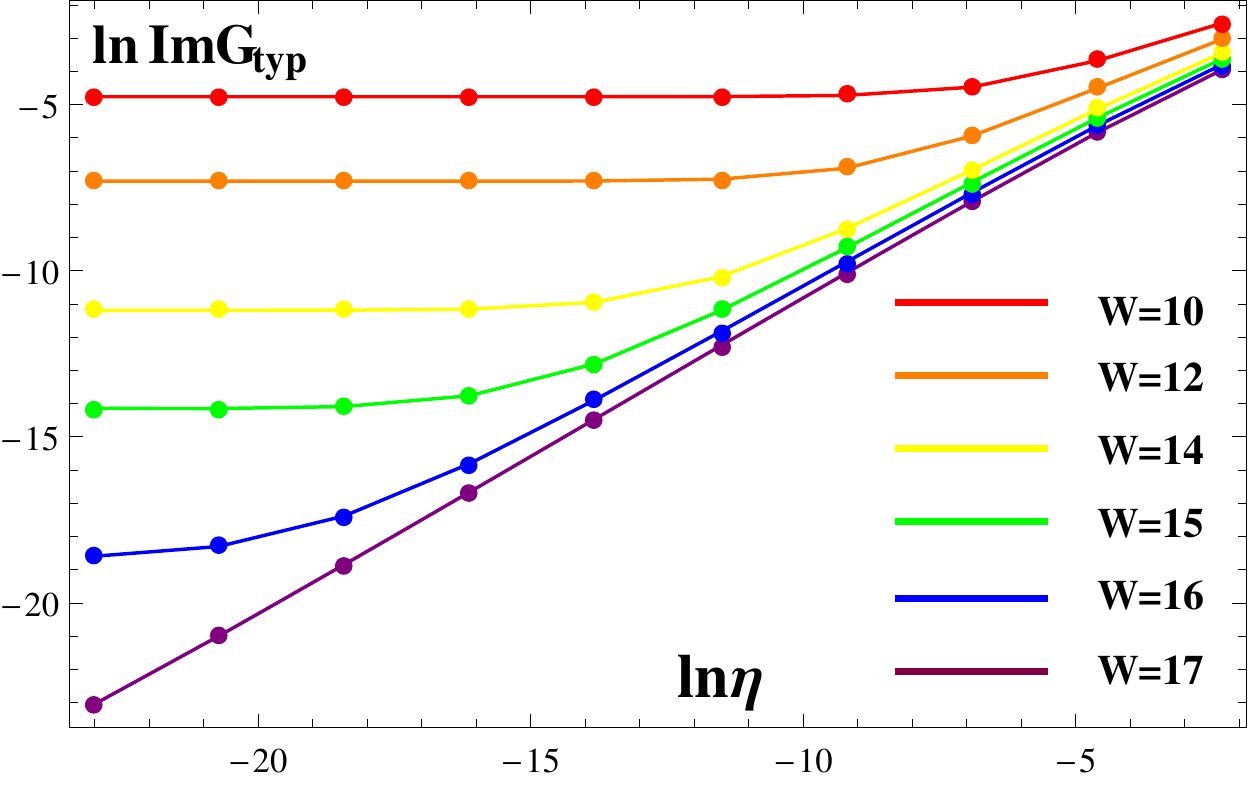}
\caption{(Color online) $\Im G_{\text{typ}}$ as a functions of imaginary part
of energy $\eta$ from PD numerics with the population size $10^{7}$
and number of generations 1000 to reach a stationary regime. At small
enough $\eta$ an $\eta$-independent $\Im G_{\text{typ}}$ is reached
which corresponds to ATL. In this limit $\Im G_{\text{typ}}$ is disorder-dependent
and decreases fast as $W\rightarrow W_{c}$. As $\eta$ increases
a crossover at $\eta=\eta_{{\rm cr}}(W)$ to the power-law-like behavior
$\Im G_{\text{typ}}\sim\eta^{\theta}$ is observed.\label{Fig:PD-ImG} }
\end{figure}

A plot of $\Im G_{\text{typ}}$ as a function $\eta$ obtained from
equilibrium PD is shown in Fig.~\ref{Fig:PD-ImG}. It demonstrates
how Anderson thermodynamic limit is reached when the imaginary part
of the energy $\eta$ in Eq.(\ref{eq:G_i^(l+1)}) decreases. At large
$\eta$ the imaginary part $\Im G_{\text{typ}}\sim\eta$, but as $\eta$
is decreased below the characteristic crossover value $\eta_{\text{cr}}$,
$\Im G_{\text{typ}}$ saturates at $\rho_{\text{typ}}$ discussed
in section \ref{sec:RSB-results-for-rho_typ}. The plots similar to
Fig.~\ref{Fig:PD-ImG} for different $W$ were used to extract the
limiting value of $\Im G_{\text{typ}}|_{\eta\rightarrow0}$ that is
plotted in Fig.~\ref{Fig:exponent} as a function of $W$. It shows
clearly a critical exponential growth of $1/\rho_{\text{typ}}\sim{\rm exp}[a/(1-W/W_{c})^{\nu}]$
as $W\rightarrow W_{c}$. In order to determine the exponent $\nu$
which controls this growth, we plotted in the inset of Fig.~\ref{Fig:exponent}
$\ln\ln(1/\rho_{\text{typ}})$ vs $\ln(1-W/W_{c})$. The best fit
corresponds to $W_{c}=18.8$ and $\nu=0.98_{-0.1}^{+0.05}$ which
completely excludes $\nu=1/2$ reported in Ref.\cite{MirlinFyodorov1991}.
The inset to Fig.~\ref{Fig:exponent} shows the analytical plot for
$\ln\ln1/\rho_{\text{typ}}$ vs. $\ln(1-W/W_{c})$ (blue dashed line)
that displays unexpectedly good agreement with the population dynamics
results, especially given that \textit{there is no fitting parameters}.
Notice that the value of the critical $W_{c}$ coincides with the
critical value obtained by inflationary population dynamics with the
accuracy of the population dynamics.

The Fig.~\ref{Fig:PD-ImG} also shows that the crossover energy (all
energies are measured in units of the band-width $\approx W/2$) $\eta_{\text{cr}}\sim\rho_{\text{typ }}$
as one should expect because it enters the recursion equations (\ref{eq:G_i^(l+1)})
as an addition to the imaginary part of the Green's functions. This
indicates that $\rho_{\text{typ }}$ gives the only frequency (time)
scale in this problem.

\begin{figure}[t]
\includegraphics[width=0.9\linewidth]{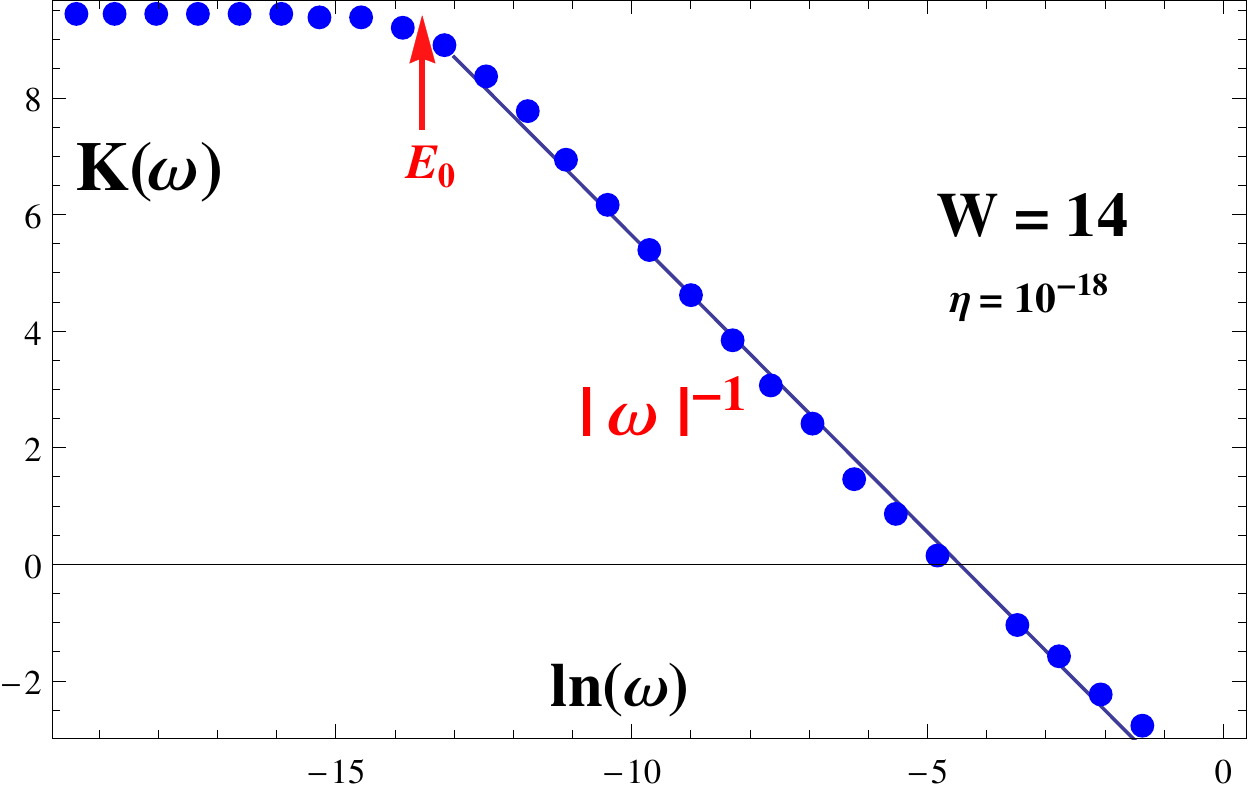}
\caption{(Color online) The correlation function $K(\omega)$ for $W=14$ obtained
from population dynamics in the limit of very small $\eta=10^{-18}$.
In the power-law region $K(\omega)\propto|\omega|^{-\mu}$ we obtain
$\mu=1$ in agreement with earlier result by exact diagonalization
of Anderson model on RRG \cite{KravtsovKhaymovichCuevas2015}. This
power-law is saturated at small $\omega<E_{0}$, where $E_{0}$ is
of the same order as $\eta_{{\rm cr}}$ where the saturation of $\rho_{\text{typ}}(\eta)$
dependence is reached. \label{Fig:K_w} }
\end{figure}

Finally, in Fig.~\ref{Fig:K_w} we present the results for the local
density of states correlation function:
\begin{equation}
K(\omega)=\lim_{\eta\rightarrow0}\langle\rho_{i}(E+\omega)\,\rho_{i}(E)\rangle,\label{K-omega}
\end{equation}
obtained by population dynamics. It shows a region of the power-law
behavior $K(\omega)\sim1/|\omega|^{\mu}$ typical of the multifractal
states \cite{Chalker-p,KrMut-p,CueKrav-orig}. However, the conventional
Chalker's relationship $\mu=1-D_{2}$ is violated, as $\mu=1$ almost
exactly. Close to Anderson transition, the height of the plateau at
small $\omega$ is related to the typical imaginary part $\rho_{\text{typ}}$by
\begin{equation}
\text{ln}K(0)+\ln(\rho_{\text{typ}})=c\approx-1.0\pm0.1\label{eq:lnK(0)}
\end{equation}
confirming again that $\rho_{\text{typ}}$ (inverse escape time) is
the main quantity that characterizes the system behavior. As one might
expect that the characteristic energy $E_{0}$ where saturation in
$K(\omega)$ at small $\omega$ occurs, is of the same order as the
characteristic $\eta_{\text{cr}}$ where $\Im G_{\text{typ}}(\eta)$
reaches the Anderson's thermodynamic limit (see Fig.~\ref{Fig:PD-ImG}).

We note that the dynamical correlation function of the the local density
of states on Bethe lattice is a proxy for the spin dynamic correlation
function in the many-body problem \cite{Biroli-time-dep}. Thus the
multifractality in the non-ergodic phase of many-body systems may
be a universal source of $1/f$ noise.

\section{Analytic results for Lyapunov exponents.\label{sec:Analytical-results-for-Lyapunov} }

At large distances the eigenfunctions on the Bethe lattice and RRG
decrease exponentially with the distance from their centers even in
the absence of disorder. The rate of this decrease turns out to be
a useful tool to characterize the disordered system as we show below.
As customary (see e.g. Ref.\cite{Warzel}) we define the ``Lyapunov
exponent'' by
\begin{equation}
\lambda_{\text{typ}}=\lim_{r\to\infty}r^{-1}\left\langle \ln\left|\frac{\psi^{(\ell+r)}}{\psi^{(\ell)}}\right|\right\rangle ,\label{L}
\end{equation}
where $r$ is the distance between the initial and the final point.
Note that this equation implies a non-normalizable solutions to the
Schroedinger equation (not eigenfunctions!) that \emph{increase} with
$r$. Comparing the Schroedinger equation for a tree,
\begin{equation}
\psi_{k(i)}^{(\ell+1)}+\sum_{j(i)=1}^{K}\psi_{j}^{(\ell-1)}=(E-\varepsilon_{i})\,\psi_{i}^{(\ell)},\label{Schroe}
\end{equation}
with equation for the one site Green's function (\ref{eq:G_i^(l+1)})
one finds a relationship between their solutions
\begin{equation}
G_{i}^{(\ell)}=\frac{\psi_{i}^{(\ell)}}{\psi_{k(i)}^{(\ell+1)}}.\label{G-psi}
\end{equation}
As the result, the Lyapunov exponent is expressed through the Green's
functions by
\begin{equation}
\lambda_{\text{typ}}=-\lim_{r\to\infty}r^{-1}\left\langle \ln\,\prod_{P}|G_{i_{P}}|\right\rangle ,\label{eq:lambda_typ-def-mf}
\end{equation}
where $P$ is the path that connects two points at a distance $r$.
We also define:
\begin{equation}
\lambda_{\text{av}}=-\lim_{r\to\infty}r^{-1}\,\ln\left\langle \prod_{P}|G_{i_{P}}|\right\rangle .\label{eq:lambda_av-def-mf}
\end{equation}

Alternatively, one can define the Lyapunov exponents as describing
the \emph{decrease} of the Green's function at large distances
\begin{align}
\lambda_{\text{typ}} & =-\lim_{r\to\infty}r^{-1}\left\langle \ln G_{i,i+r}\right\rangle \label{eq:lambda_typ-def}\\
\lambda_{\text{av}} & =-\lim_{r\to\infty}r^{-1}\ln\left\langle G_{i,i+r}\right\rangle \label{eq:lambda_av-def}
\end{align}

A typical decrease of the Green's function between the sites is due
to the typical increase of the wave functions, so the definitions
(\ref{eq:lambda_typ-def-mf}) and (\ref{eq:lambda_typ-def}) are equivalent.

One can establish a generic relationship between $\lambda_{\text{typ}}$
and $\lambda_{\text{av}}$ using the the Jensen inequality \cite{Jensen}.
It states that for any distribution of a random variable $Y$ the
following is valid:
\begin{equation}
\langle e^{Y}\rangle\geq e^{\langle Y\rangle}.\label{Jensen}
\end{equation}
Taking $Y=\ln|G|$ we arrive at:
\begin{equation}
\langle|G|\rangle=e^{-\lambda_{av}}\geq e^{-\lambda_{\text{typ}}},\;\;\Rightarrow\;\;\lambda_{av}\leq\lambda_{\text{typ}}.
\end{equation}

The limit of stability of Anderson insulator (AI) is naturally described
in terms of $\lambda_{av}$ \cite{Warzel}. Indeed, the equation for
$\Lambda(E,m)$ can be written as
\begin{equation}
\Lambda(E,m)=\lim_{\ell\rightarrow\infty}\frac{1}{m\ell}\ln\left[K^{\ell}\left\langle \prod_{P_{\ell}}|G_{i_{P}}|^{2m}\right\rangle \right]\label{eq:Lambda(E,m)-general}
\end{equation}
where $P_{\ell}$ is the path of length $\ell$ and $G_{i_{P}}$ denote
the Green's functions in the $\eta\rightarrow0$ limit. In contrast
to (\ref{eq:Lambda-averaging}) the equation (\ref{eq:Lambda(E,m)-general})
is very general, it does not make the assumption of independent $G$
at different sites. Exactly at Anderson transition $m_{0}=1/2$, so
\begin{equation}
\lambda_{\text{av}}=\ln K-\frac{1}{2}\Lambda\left(m=\frac{1}{2}\right).\label{eq:lambda_av-gen}
\end{equation}
Because the limit of stability of Anderson insulator corresponds to
$\Lambda(m_{0})=0$ on the entire upper arc of Fig.~\ref{Fig:phase-diagram}
we get the exact relation
\begin{equation}
\lambda_{\text{av}}(E,W_{c})=\ln K.\label{eq:lambda_av(E,W_c)}
\end{equation}
We shall use this equality to determine the phase diagram in the plane
$W-E$ in section \ref{sec:Phase-diagram}.

The transition from the non-ergodic to ergodic state is controlled
by a similar equation. Namely, at this transition two conditions are
satisfied
\begin{align}
\Lambda(E,m_{0} & )=\ln K\label{eq:Lambda(E,m_0)}\\
\frac{\partial\Lambda}{\partial m} & (E,m_{0})=0\label{eq:dLambda(E,m_0)dm}
\end{align}
for $m_{0}=1$. Introducing the distribution function, $\mathcal{P}_{\ell}(y)$
of
\begin{equation}
y=\prod_{P_{\ell}}|G_{i_{P}}|^{-1}\label{eq:y}
\end{equation}
and using the definition (\ref{eq:Lambda(E,m)-general}) we rewrite
the equations (\ref{eq:Lambda(E,m_0)},\ref{eq:dLambda(E,m_0)dm})
as
\begin{align*}
\int\mathcal{P}_{\ell}(y)\frac{dy}{y^{2}} & =1\\
\int\mathcal{P}_{\ell}(y)\frac{dy}{y^{2}} & (2\ln y+\ln K)=0
\end{align*}

Using the symmetry $\mathcal{P}_{\ell}(y)=\mathcal{P}_{\ell}(1/y)$
(see Appendix \ref{app:Proof-of-the-symmetry}) we get from the second
equation
\begin{equation}
\lambda_{\text{typ}}(E,W_{E})=\frac{1}{2}\ln K\label{eq:lambda_typ(E,W_E)}
\end{equation}
at the transition to a fully ergodic state.

The value of $\lambda_{\text{typ}}$ for the transition to ergodic
state coincides with the Lyapunov exponent on a clean Bethe lattice.
Indeed, the Green's function $G_{i}=G_{C}(E)$ for a clean tree is
a solution of the equation $G_{C}\,(E-K\,G_{C})=1$:
\begin{equation}
G_{C}(E)=\frac{e^{i\phi(E)}}{\sqrt{K}},\;\;\;\;E=2\sqrt{K}\,\cos\phi(E).\label{Laplace}
\end{equation}
Therefore inside the energy band $|E|\leq2\sqrt{K}$ of he Lyapunov
exponent is:
\begin{equation}
\lambda_{{\rm C}}=\ln|G_{C}|^{-1}=\frac{1}{2}\,\ln K.\label{L-clean}
\end{equation}

The importance of the condition (\ref{eq:lambda_typ(E,W_E)}) was
realized previously \cite{Warzel-pr}.

\section{Population dynamics results for Lyapunov exponents\label{sec:Population-dynamics-Lyapunov}}

\begin{figure}[h]
\includegraphics[width=0.9\columnwidth]{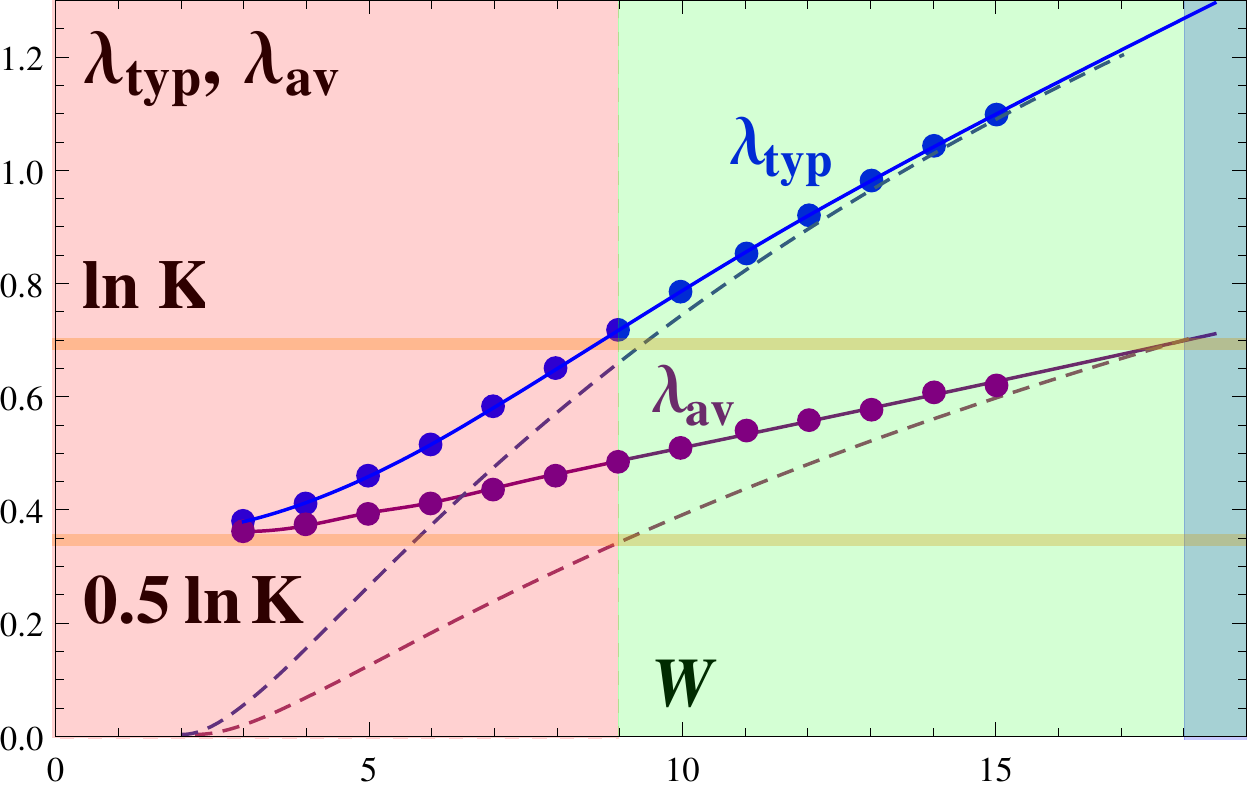}
\caption{(Color online) The Lyapunov exponents $\lambda_{\text{typ}}$ and
$\lambda_{\text{av}}$ for $K=2$ and $E=0$ from (\ref{eq:lambda_typ-def-mf},\ref{eq:lambda_av-def-mf})
evaluated by equilibrium population dynamics numerics (blue and purple
points) and obtained analytically within one-step RSB approach (dashed
lines) from (\ref{eq:lambda_typ^(RSB)},\ref{eq:lambda_av^(RSB)})
using approximation Eq.(\ref{low-cutoff}) for $F_{{\rm eff}}(\epsilon)$.
In the vicinity of AT (where $\Im G_{\text{typ}}\rightarrow0$) there
is a good agreement between the RSB analytical theory and equilibrium
population dynamics. However at small disorder there is an essential
difference related to a different order involved in thermodynamic
limits (ATL for equilibrium PD and IOTL for analytical calculations). }
\label{Fig:PD-Lyapunov}
\end{figure}

In order to compute the Green's function between different sites we
introduce the matrix Green's function between two \emph{descendents
}of a given site in generation $\ell$. The recursion equation for
this matrix Green's function, $\mathcal{G}_{ij}(E)$ is a straightforward
generalization of the equation for single site Green's function (\ref{eq:G_i^(l+1)}):
\begin{align}
\mathcal{G}_{ij}^{(\ell+1)}(E) & =\frac{1}{E\hat{1}-\hat{\Xi}_{ij}-\mathcal{G}_{ij}^{(\ell)}(E)},\label{eq:G_ij^(l+1)}\\
\hat{\Xi}_{ij} & =\left(\begin{array}{cc}
\varepsilon_{i}+\Sigma'_{i} & 0\\
0 & \varepsilon_{j}+\Sigma'_{j}
\end{array}\right)\nonumber
\end{align}
where $\varepsilon_{i}$ is a random on-site energy, $\Sigma'_{i}=\sum_{k\neq i}G_{k}(E)$
is the single site self energy with the site $i$ excluded. The off-diagonal
part of the $2\times2$ matrix Green's function $\mathcal{G}_{ij}^{(\ell)}(E)$
gives the Green's function $\mathfrak{g}_{2\ell}(E)$ between sites
at distance $2\ell$ from each other ($\equiv G_{ij}$ at $\left\Vert i-j\right\Vert =2\ell$).

The Lyapunov exponents are defined as the limiting behavior of $\lambda=-\ln\mathfrak{g}_{2\ell}(E)/2\ell$
at large distances. As usual one should distinguish the typical Green's
function decrease (cf. (\ref{eq:lambda_typ-def}))
\begin{equation}
\lambda_{\text{typ}}=-\lim_{\ell\rightarrow\infty}\left\langle \ln\left[\mathfrak{g}_{2\ell}(E)\right]/2\ell\right\rangle \label{eq:lambda_typ}
\end{equation}
and the average one (cf. (\ref{eq:lambda_av-def}))
\begin{equation}
\lambda_{\text{av}}=-\lim_{\ell\rightarrow\infty}\ln\left\langle \mathfrak{g}_{2\ell}(E)\right\rangle /2\ell.\label{eq:lambda_av}
\end{equation}
The average of the Green's function is dominated by rare events, so
$\lambda_{\text{av}}<\lambda_{\text{typ}}$. These exponents were
computed by applying population dynamics to the equation (\ref{eq:G_ij^(l+1)}).
The results are shown in Fig.~\ref{Fig:PD-Lyapunov} and compared
with those obtained analytically in section \ref{sec:Phase-diagram}
in the framework of the linear RSB theory developed in section \ref{sec:Minimal-account-for-real-part}.

\begin{figure*}[t]
\includegraphics[width=0.3\linewidth]{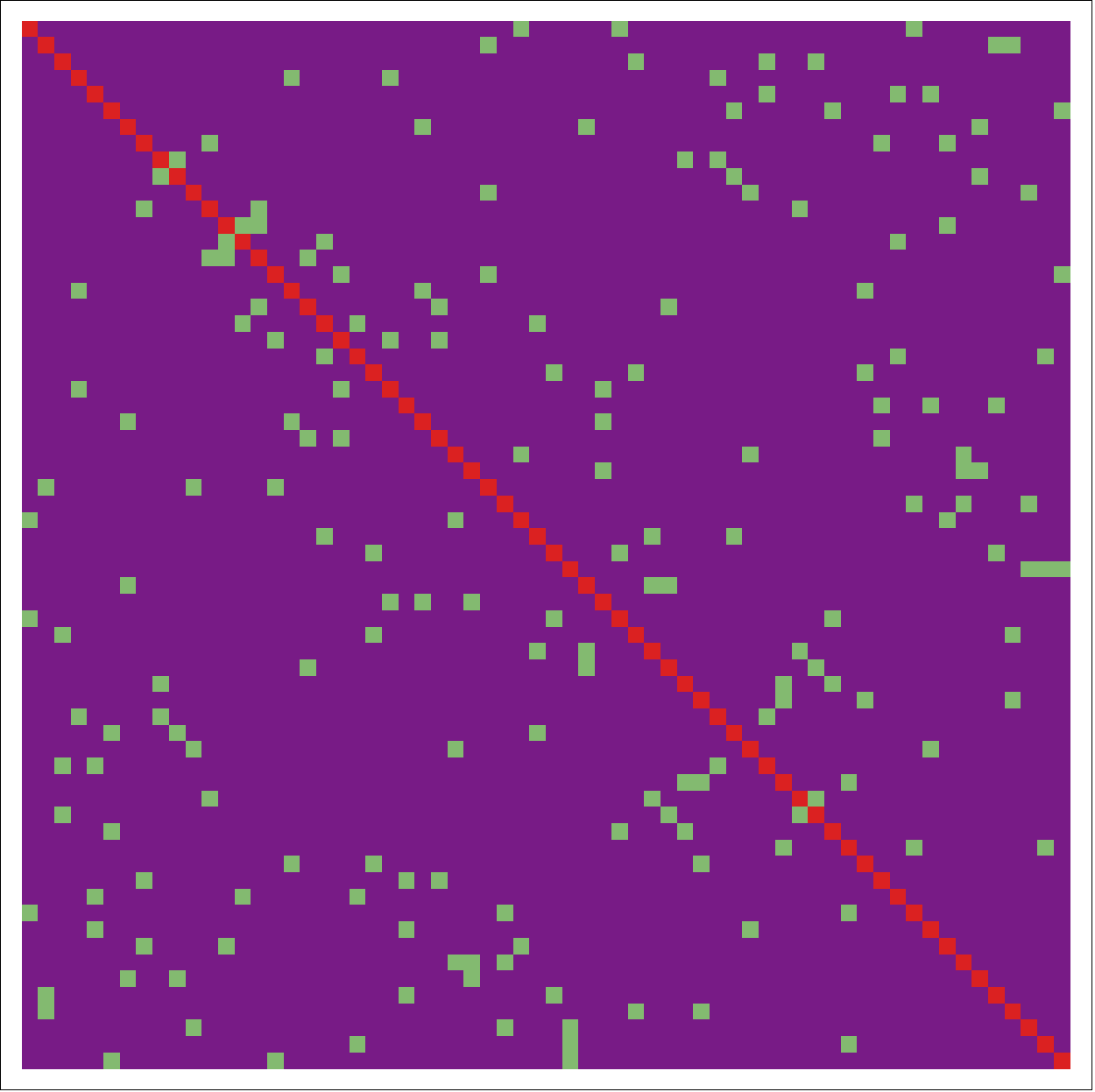}
\includegraphics[width=0.3\linewidth]{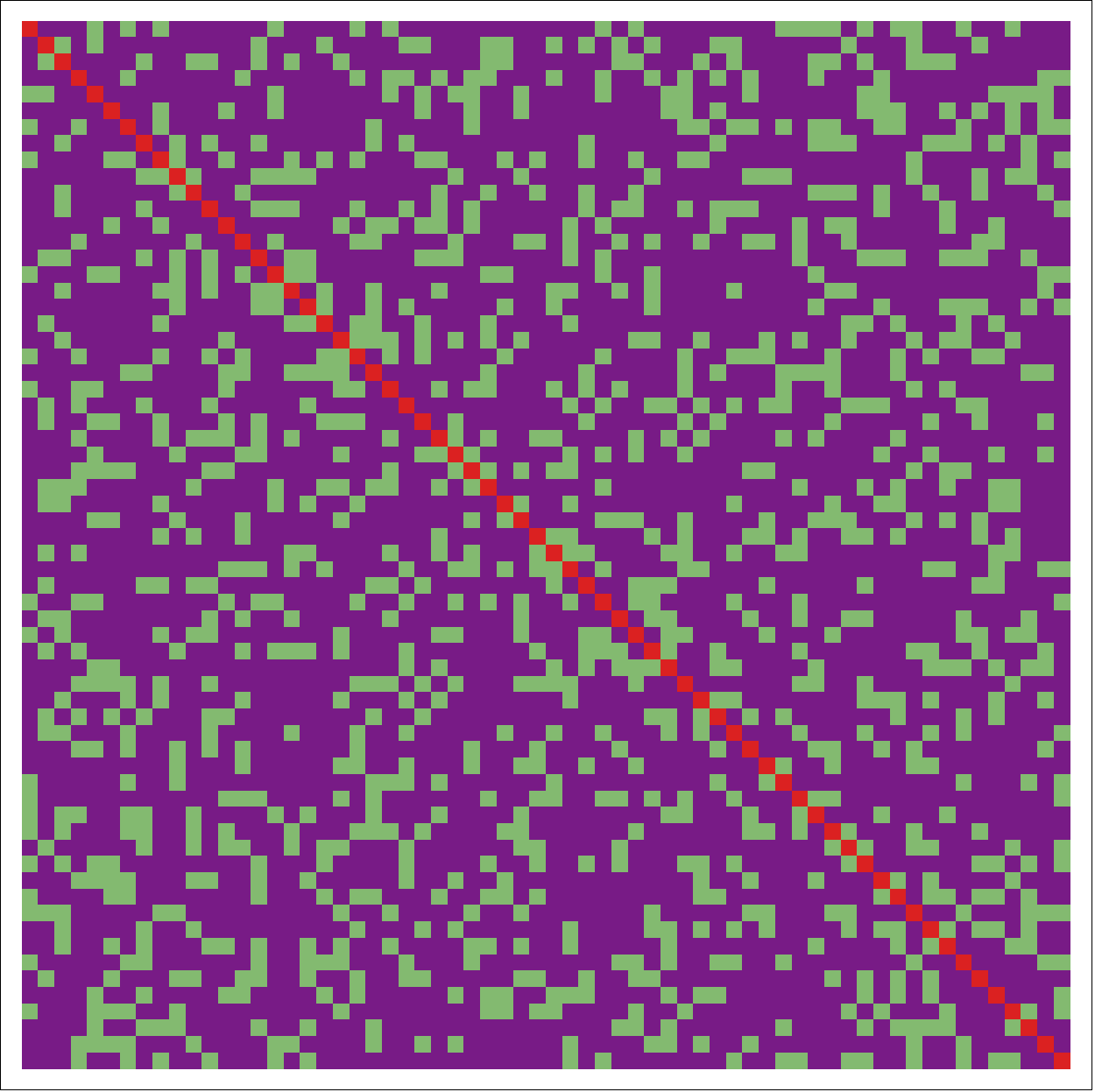}
\includegraphics[width=0.3\linewidth]{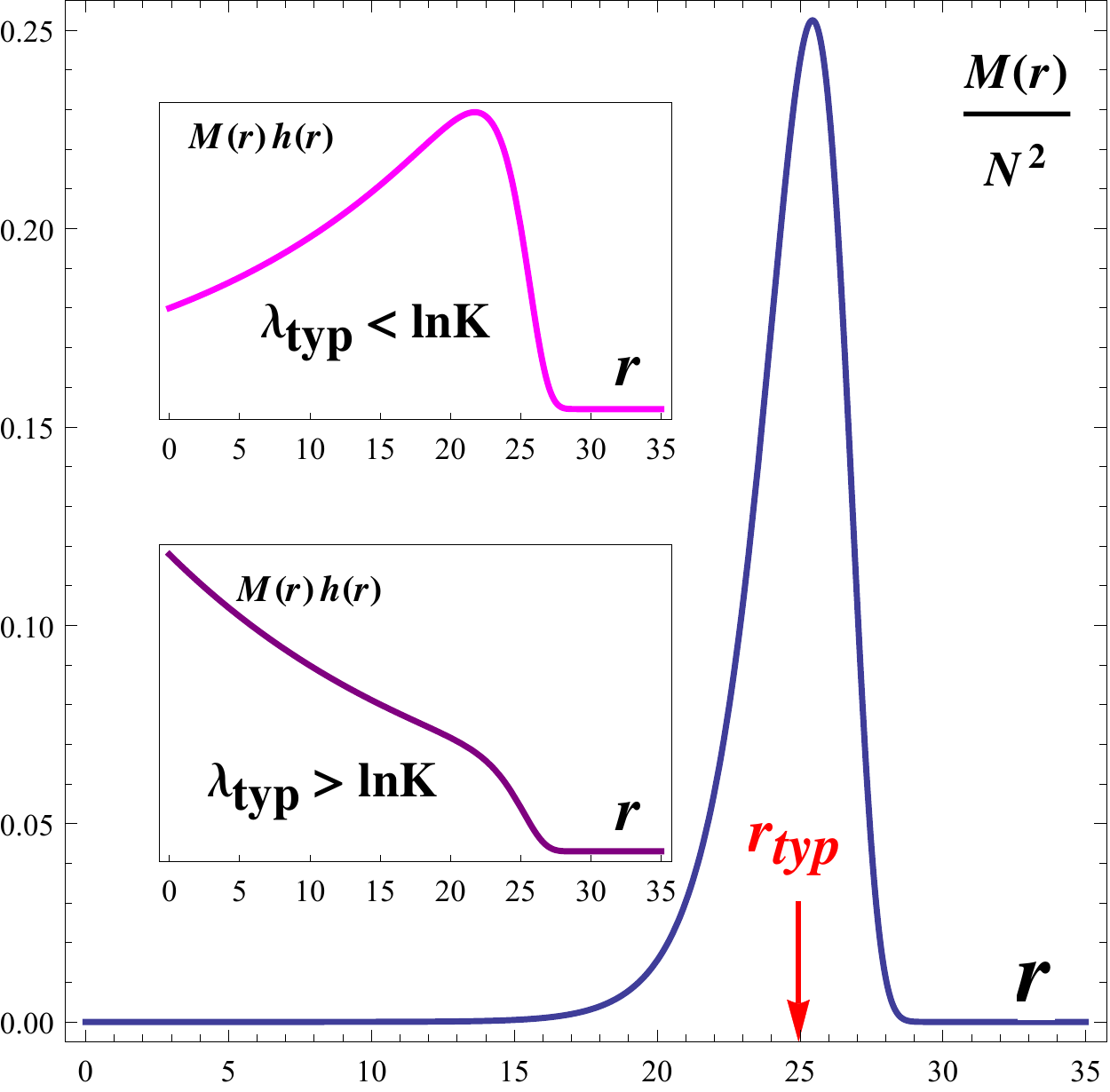}
\caption{(Color online) (Left panel): Sparse matrix of short-range hopping
between neighboring sites $i,j$ ($r\equiv||i-j||=1$) on RRG. It
contains $K+1=3$ non-zero off-diagonal matrix elements $h_{ij}=1$
(shown by green squares) in each row or column. (Middle panel): Transition
amplitudes $h_{ij}(r_{{\rm typ}})=e^{-\lambda_{{\rm typ}}\,r_{{\rm typ}}}$
(shown by green squares) between sites $i,j$ at a typical distance
$||i-j||=r_{{\rm typ}}$ between two sites on RRG. At $N\rightarrow\infty$
the typical distance is approximately equal to the graph diameter
$r_{{\rm typ}}\approx d_{{\rm RRG}}$. The distribution of matrix
elements $h_{ij}(r_{{\rm typ}})$ corresponds to \textit{macroscopically
large} effective connectivity $K_{{\rm eff}}\sim N/2$; (Right panel):
Distribution of distances $r$ between sites on RRG at $N=10^{8}$.
A typical distance is $r_{{\rm typ}}\approx d_{{\rm RRG}}\approx\ln N/\ln K$.
In the limit $N\rightarrow\infty$ the distribution shrinks to a delta-function
$\delta(r-d_{{\rm RRG}})$. Inset: Optimal hopping distances found
by maximization of the product $M(r)\,h(r)$ of the number of pairs
of sites $M(r)\sim N\,K^{r}\,e^{-c\,N^{-1}\,K^{r}}$ at a distance
$r$ on RRG and the absolute value $h(r)=e^{-\lambda_{{\rm typ}}\,r}$
of the corresponding transition amplitude. Optimal hopping is at short
distance $r=1$ for $\lambda_{{\rm typ}}>\ln K$ and at large distances
$r\approx d_{{\rm RRG}}$ for $\lambda_{{\rm typ}}<\ln K$. In the
first regime RRG is equivalent to BL, while in the second regime it
is reminiscent of the Rosenzweig-Porter random matrix model. In the
limit $N\rightarrow\infty$ the switching from one regime to the other
happens abruptly at $\lambda_{{\rm typ}}=\ln K$ which corresponds
to $W_{0}\approx10$ at $K=2$ and $E=0$. At small disorder $W<W_{0}$
equivalence between RRG and BL breaks down and a transition from NEE
to EE phase \cite{AltshulerCuevasIoffeKravtsov2016} is not excluded.
\label{Fig:matrices} }
\end{figure*}

At large distances a typical off-diagonal Green's function becomes
exponentially small, expanding the recursion (\ref{eq:G_ij^(l+1)})
in $\mathfrak{g}_{2\ell}(E)$ gives precisely (\ref{eq:lambda_typ-def-mf}-\ref{eq:lambda_av-def-mf}).
At any finite distance, however, rare fluctuations might lead to an
arbitrary large value of the product $\prod_{P}|G_{i_{P}}|$. Physically,
this corresponds to the resonance between sites $i$ and $j$. In
contrast to the product $\prod_{P}|G_{i_{P}}|$ the real Green's function
given by the off-diagonal part of $\mathcal{G}_{ij}^{(\ell+1)}(E)$
remains finite even when one $G_{i}$ in the product is infinite.
This makes the use of the matrix recursion (\ref{eq:G_ij^(l+1)})
more reliable at moderate sizes.

The linear RSB theory makes two approximations: it replaces the actual
correlations between real parts of $G_{i}$ by the effective distribution
of single site energies and it neglects the imaginary part of the
Green's function. Both approximations work well at large $W$ leading
to the good agreement between population dynamics and RSB results
at $W\gtrsim12$. However, at small disorder the essential difference
emerges. The Lyapunov exponents $\lambda_{\text{typ}}$ and $\lambda_{\text{av}}$
obtained by equilibrium population dynamics are always \textit{above}
the clean limit provided by Bethe lattice without the disorder, $(1/2)\,\ln K$,
they approach this level only as $W\rightarrow0$. In contrast, the
RSB theory predicts exponents below this level that signals that at
small $W$ the non-linear in $\Im G$ terms neglected in the theory
become relevant. Notice that the equilibrium population dynamics corresponds
to the ATL when $N\rightarrow\infty$ first, whilst the linear RSB
theory corresponds to IOTL when the limit $\eta\rightarrow0$ is taken
first. Therefore, the difference between the results is not surprising.

The Anderson transition, $W=W_{c}$, coincides well with the point
where $\lambda_{\text{av}}(W)=\ln K$. Indeed, the average Green's
function is dominated by rare fluctuations and thus occasionally may
decay slowly along an atypical path. In the localized phase these
pathes are rare. Exactly at the delocalization transition such a
path connects a typical site with sites far away. The contribution
of this single path to the \textit{average} Green's function (which
is a sum of the total of $K^{\ell}$ paths) is proportional to $K^{-\ell}=e^{-\ln K\,\ell}$.
This is equivalent to $\lambda_{av}=\ln K$. In the fully ergodic
phase, the particle density is spread over the whole lattice which
corresponds to $\mathfrak{g}_{\ell}=|G_{C}|^{\ell}=K^{-\ell/2}$,
so in this case $\lambda_{\text{typ}}=\lambda_{\text{av}}=\ln K/2$.

However, this latter limit is not reached in the population dynamics
even at $W=3$. This is in agreement with the conclusion of section
\ref{sec:Analytical-results-for-D(W)} that ergodicity is fully restored
only in the limit $W\rightarrow0$, although the numerical accuracy
does not allow us to exclude the transition in which the ergodicity
is fully restored at some very small $W\lesssim3$.

The decay of the Green's function between different sites allows us
to check the consistency of the Bethe lattice approximation for finite
random regular graphs. On Bethe lattice, the site at distance $\ell$
from the origin can be reached by one and only one path. On random
regular graphs this remains correct if the distance between the sites
is less than the graph diameter, $\ell<d_{\text{RRG}}=\ln N/\ln K$
(see Fig.~\ref{Fig:matrices}). Large loops appearing at distances
$d_{\text{RRG}}$ imply that the RRG becomes multiply connected at
these scales. The typical transition amplitude (effective hopping
matrix element in Fig.~\ref{Fig:matrices}) at these distances is
given by $\exp\left[-\lambda_{\text{typ}}d_{\text{RRG}}\right]$.
One can identify two different regimes by studying the optimal hopping
distance found by maximization of the product $M(r)\,h(r)$, where
$M(r)$ is the number of pairs of sites at a distance $r$ and $h(r)=e^{-\lambda_{{\rm typ}}}\,r$
is the typical transition amplitude between them. On BL $M(r)\sim K^{r/2}$
for $r<d_{{\rm BL}}\approx2\ln N/\ln K$ and the maximum of $M(r)\,h(r)$
is reached at the shortest distance $r=1$ for all $\lambda_{{\rm typ}}>(1/2)\,\ln K$.
In contrast, on RRG $M(r)\sim K^{r}$ for $r<d_{{\rm RRG}}\approx\ln N/\ln K$,
and the optimal hopping distance $r=1$ only for $\lambda_{{\rm typ}}>\ln K$.
Under this condition RRG is equivalent to BL. With decreasing disorder
below the Anderson transition point $\lambda_{{\rm typ}}(W)$ decreases
from the value $\lambda_{{\rm typ}}(W_{c})>\lambda_{{\rm av}}(W_{c})=\ln K$
and at some point $W=W_{0}<W_{c}$ reaches the value $\ln K$:
\begin{equation}
\lambda_{{\rm typ}}(W_{0})=\ln K.\label{W0}
\end{equation}
For $W<W_{0}$ ($W_{0}\approx10$ for $K=2$) typical Lyapunov exponent
$\lambda_{{\rm typ}}<\ln K$ , so that the optimal hopping distance
changes from short-range $r=1$ to long-range $r=d_{{\rm RRG}}\approx\ln N/\ln K$.
At lower $W$ it remains almost constant until $\lambda_{{\rm typ}}$
reaches its minimal value $(1/2)\ln K$. The distribution of transition
amplitudes in this regime is nearly bi-modal with optimal hopping
matrix elements shown by green squares in the middle panel of Fig.~\ref{Fig:matrices}.
These matrix elements $h_{ij}$ are distributed almost homogeneously
in the corresponding effective Hamiltonian which corresponds to macroscopically
large effective connectivity $K_{{\rm eff}}\sim N/2$. Thus, in this
regime the model becomes similar to Rosenzweig-Porter random matrix
model discussed in Sec.\ref{sec:Application-to-Rosenzweig-Porter}
where ergodic extended phase is present \cite{KravtsovKhaymovichCuevas2015}.

Note that the switching from the short-range to long-range effective
hopping is abrupt in the limit $N\rightarrow\infty$, so that the
equivalence of RR and BL may also break down abruptly at $W=W_{0}$.
This may lead to a discontinuous transition reported in work\cite{AltshulerCuevasIoffeKravtsov2016}.

\section{Phase diagram\label{sec:Phase-diagram}}

\begin{figure*}[th]
\centering{ \includegraphics[width=0.45\linewidth]{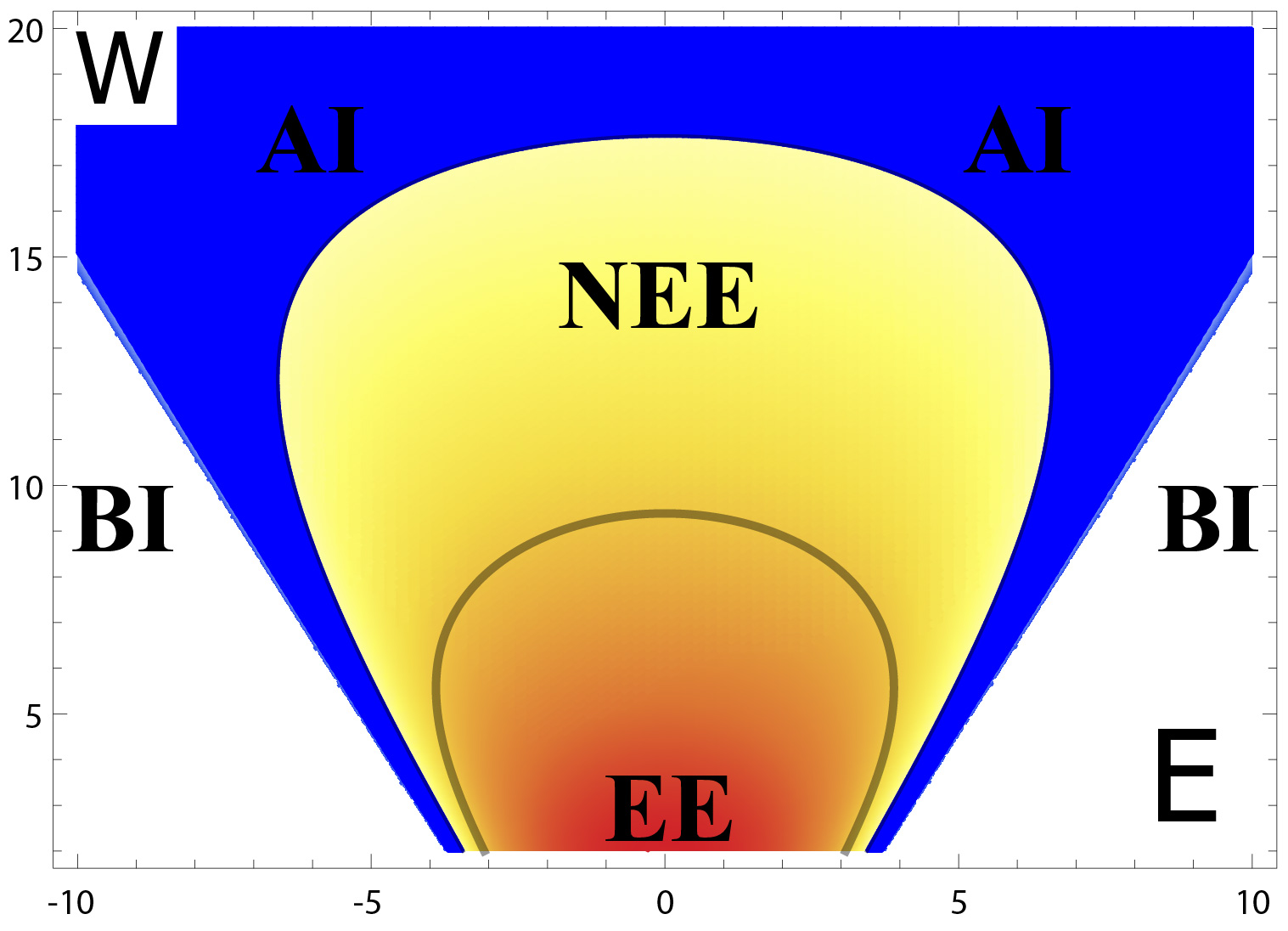}
\includegraphics[width=0.45\linewidth]{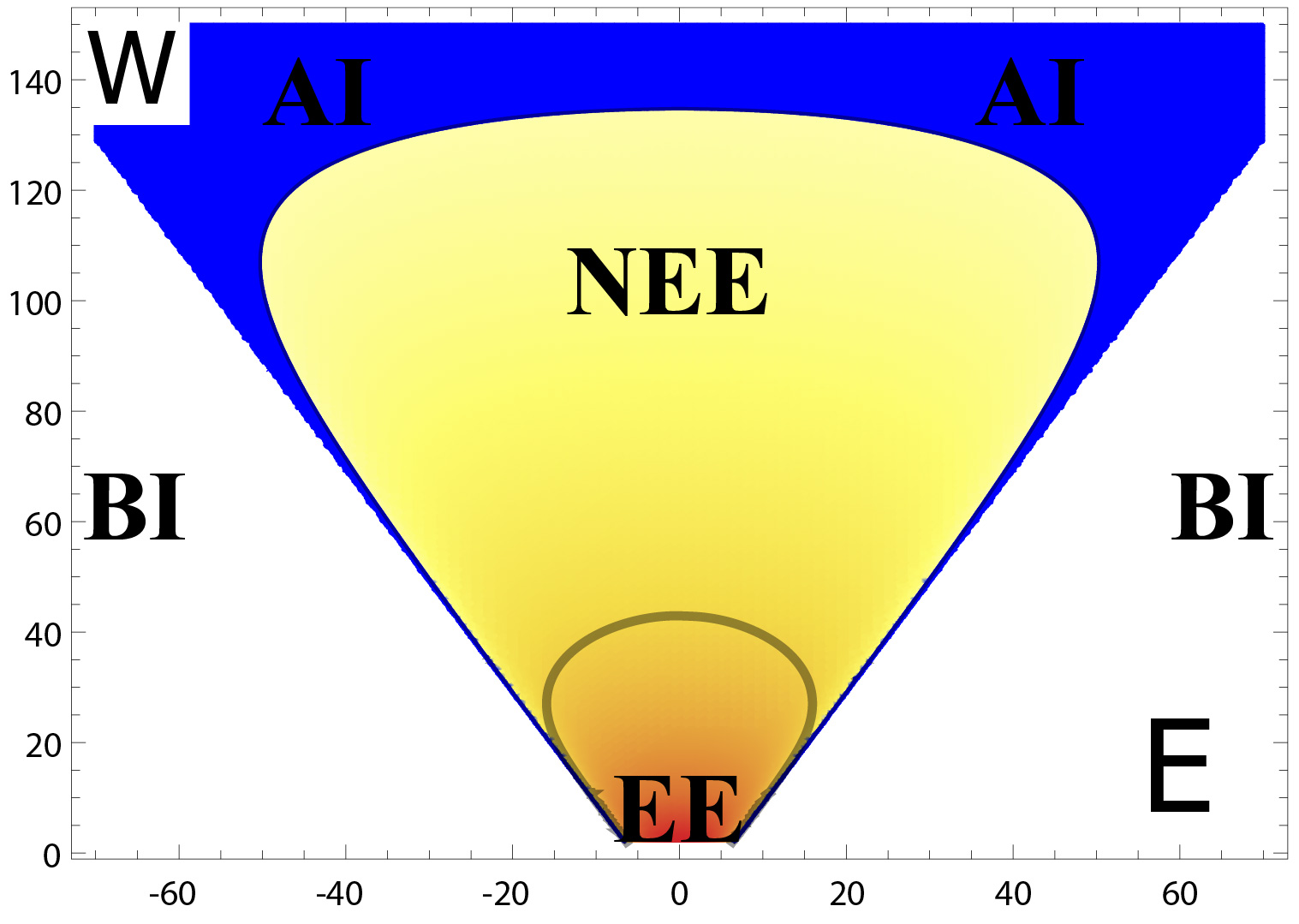} }
\caption{(Color online) Sketch of the phase diagram in the $E-W$ plane for
small $K$ ($K=2$, left panel) and for medium $K$ ($K=8$, right
panel). The blue area indicates Anderson insulator (AI) with non-zero
density of states and localized eigenfunctions, it is separated by
a well defined transition line from the non-ergodic extended state
(NEE) shown in yellow. At smaller disorder $W$ and/or enegy $E$
the anomalous dimension characterizing the wave functions in the non-ergodic
state grows. For Bethe lattice the full ergocity, $D=1$ (red area)
is restored only at $W\rightarrow0.$ In contrast, for RRG a transition
form non-ergodic (NEE) to ergodic (EE) phase is likely to happen at
the line $\lambda_{{\rm typ}}(W,E)=\ln K$ (shown on a plot) or inside
the area bounded by it. The white area corresponds to the band insulator
(BI) beyond the spectral edge where there are no states. \label{Fig:phase-diagram} }
\end{figure*}

In order to determine the phase diagram in the plane $(W-E)$ we use
the equations (\ref{eq:lambda_av(E,W_c)},\ref{eq:lambda_typ(E,W_E)})
that determine the positions of the Anderson transition and ergodicity
restoration. We compute the values of $\lambda_{\text{av}}$ and $\lambda_{\text{typ}}$
that enter these equations in RSB theory developed in Sections \ref{sec:Large-connectivity-approximation},\ref{sec:Minimal-account-for-real-part}
and \ref{sec:Analytical-results-for-D(W)} that we generalize here
to the case of $E\neq0$. For this purpose we employ the ansatz for
$F_{{\rm eff}}(\epsilon)$ which extends (\ref{low-cutoff}) to the
case of $E\neq0$:

\begin{widetext}
\begin{equation}
F_{{\rm eff}}(\epsilon;E,W)=C(E,W)\,\left[\theta\left(\frac{W}{2}-\epsilon\right)\theta\left(\epsilon-E-\frac{1}{\frac{W}{2}-E}\right)+\theta\left(\frac{W}{2}+\epsilon\right)\theta\left(E-\epsilon-\frac{1}{\frac{W}{2}+E}\right)\right],\label{F-eps-omega}
\end{equation}
where $C(E,W)$ is the normalization constant. The corresponding $\Lambda(E,m)$
is given by $\Lambda(E,m)=\frac{1}{m}\,\ln(K\,\tilde{I}_{m})$, where:
\begin{equation}
\tilde{I}_{m}=\frac{1}{(1-2m)\left(W-\frac{4W}{W^{2}-4E^{2}}\right)}\,\left[\left(\frac{W}{2}-|E|\right)^{1-2m}-\left(\frac{W}{2}-|E|\right)^{2m-1}+\left(\frac{W}{2}+|E|\right)^{1-2m}-\left(\frac{W}{2}+|E|\right)^{2m-1}\right].\label{Im-anal-cont}
\end{equation}
\end{widetext}

Equations (\ref{F-eps-omega},\ref{Im-anal-cont}) are a good approximation
only for $W/2-|E|\gg\sqrt{K}$, and they are only qualitatively valid
close to the edge of the spectrum. However, it gives correctly the
main characteristic features of the phase diagram which is presented
in Fig.~\ref{Fig:phase-diagram}.

In this approximation Lyapunov exponents become:
\begin{equation}
\lambda_{\text{typ}}^{\text{(RSB)}}=\int F_{{\rm eff}}(\epsilon;E,W)\,\ln|\epsilon-E|\,d\epsilon,\label{eq:lambda_typ^(RSB)}
\end{equation}

\begin{equation}
\lambda_{\text{av}}^{\text{(RSB)}}=-\ln\int F_{{\rm eff}}(\epsilon;E,W)\,\frac{d\epsilon}{|E-\epsilon|}.\label{eq:lambda_av^(RSB)}
\end{equation}

Together with the equations (\ref{eq:lambda_av(E,W_c)},\ref{eq:lambda_typ(E,W_E)})
they give two lines in $W-E$ plane. The first line corresponds to
the transition from non-ergodic to ergodic state, the second from
non-ergodic to localized. Evaluation of the integrals gives the position
of transition lines.

The computation outlined above was done in the framework of the RSB
theory that treats only approximately the correlations between $G_{i}$
that enter in the products along the paths, such as (\ref{eq:y}).
The population dynamics that evaluates these products exactly gives
results very close to the RSB theory for the Anderson transition but
does not show a transition into the fully ergodic state at finite
$W.$ We thus believe that the true behavior in the bulk of a large  Bethe lattice
corresponds to a gradual crossover to $D=1$ as $W\rightarrow0$ but
$D$ becomes very close to 1 at $W$ at which the condition $\lambda_{\text{typ}}^{\text{(RSB)}}=\ln K/2$
together with RSB equation (\ref{eq:lambda_typ^(RSB)}) predicts the
phase transition. Both a transition predicted by RSB theory and a crossover
to the ergodic phase predicted by population dynamics for infinite
Bethe lattice might be preempted by a sharp transition on random regular
graphs when large loops become relevant. The arguments of section
\ref{sec:Population-dynamics-Lyapunov} and Eq.(\ref{W0}) show that
the equivalence of BL and RRG breaks down when $\lambda_{\text{typ}}<\ln K$.
Inside this area ( which borderline is shown in Fig.~\ref{Fig:phase-diagram}
) the behavior on RRG is largely unknown. However it is very likely
that a transition to the EE phase is happening on RRG at or below
$W_{0}$ defined in Eq.(\ref{W0}).

All these results are summarized by the phase diagrams shown in Fig.~\ref{Fig:phase-diagram}.

Concluding this Section we would like to emphasize the difference
between the phase diagram for $K=2$ and $K=8$. At large $K$ the
crossover into fully ergodic phase happens at $W_{E}\propto\sqrt{K}$,
while the critical $W_{c}$ of Anderson transition scales like $K\ln K$.
Furthermore, the part of the phase diagram ($W_{c}>W>W_{0}\sim K$
at $E=0$) that is correctly described by the RSB theory grows with
$K$ as discussed in section \ref{sec:Analytical-results-for-D(W)}.
As the result, the relative area of the fully ergodic phase shrinks
as $K$ increases.

This can be interpreted as a relative insignificance of the fully
ergodic and fully localized regime in the classical limit. Indeed,
the parameter, $r_{s}$, that quantifies the quantum-to-classical
crossover is the ratio of the typical potential energy, the on-site
energy fluctuations, to the typical kinetic energy, the bandwidth
computed at $W=0$. In our case it is $r_{s}=W/\sqrt{K}$. The Anderson
transition corresponds to $r_{s}=W_{c}/\sqrt{K}\sim\sqrt{K}\ln K$,
so that the classical limit is the limit of large $K$. Our results
show that in this limit the non-ergodic phase, which in many respects
is similar to glass, is occupying the lion share of the phase diagram
corresponding to extended states, while the insulating phase is pushed
to very strong disorder. This is to be expected because full localization
is impossible in the classical theory.

\section{Discussion\label{sec:Discussion}}

The main conclusions of the paper summarized in Fig.~\ref{Fig:Intro}(a,b)
differ from the results of earlier works. There are two striking differences.
\begin{enumerate}
\item A number of works predict different scaling dependence for the characteristic
critical volume $N_{c}$ associated with the localization transition
on the Bethe lattice.\cite{Zirn-BL,MirlinFyodorov1991,MF-ergodic-theory,Efetov-book}
In particular, these works predict asymmetric dependence as a function
of $W-W_{c}$ below and above the Anderson transition. Whilst the
scaling in the localized phase is similar to the one obtained here,
$N_{c}\sim{\rm exp}(a/|W-W_{c}|)$ for $W>W_{c}$, the scaling in
the extended phase for $W<W_{c}$ was predicted to obey a different
dependence, $N_{c}\sim{\rm exp}(b/\sqrt{W_{c}-W})$. In contrast,
the scaling obtained in this work using the identification $\rho_{\text{typ}}\sim N_{c}^{-1}$
is symmetric: on both sides of the transition $N_{c}\sim{\rm exp}(a/\left|W-W_{c}\right|)$
(see Fig.~\ref{Fig:Intro}(b)). As explained in sections \ref{sec:RSB-results-for-rho_typ}
and \ref{sec:Population-dynamics-for-ImG} we have checked this analytical
result by the extensive population dynamics and can rule out square
root behavior $N_{c}\sim{\rm exp}(a/\sqrt{W_{c}-W})$ in the delocalized
phase. Below we discuss in detail the origin for this discrepancy.
Here we only note that most of the works \cite{Zirn-BL,MirlinFyodorov1991,MF-ergodic-theory,Efetov-book}
solved the supersymmetric \textit{sigma model} on the Bethe lattice
which might correspond to a different physical problem.
\item The conclusion of the symmetric behavior of $N_{c}$ on both sides
of Anderson transition agrees qualitatively with another conclusion
of this work that the fractal dimension $D_{1}(W)$ is a \textit{continuous}
function of disorder at the localization transition on Bethe lattice.
The fractal dimension $D_{1}=0$ for $W\geq W_{c}$ and it continuously
grows from 0 to 1 throughout the non-ergodic phase as disorder decreases
below $W_{c}$, see Fig.~\ref{Fig:Intro}(a). It is hard to imagine
a continuous transition with the asymmetric exponents of the critical
volume. Correspondingly, no evidence of non-ergodic extended phase
was reported in the earlier works \cite{Zirn-BL,MirlinFyodorov1991,MF-ergodic-theory,Efetov-book}
that predicted asymmetric critical behavior. The conclusion that $D=1$
in the delocalized regime was reached in recent mostly numerical studies
\cite{TMS,Lemarie} that we discuss in detail below.
However, very recently two papers \cite{TihMir} reported the existence of the non-ergodic
extended, multifractal phase found in the framework of the non-linear
sigma-model on a finite Cayley tree.
While the parameters of multifractality depend on the distance $r_{i}=s\,\ln N/\ln K$ from the  observation point to the root of the Cayley tree,
the ergodic transition happens at the same disorder $W_{E}\approx 5.7$ for all $s<1$.  The authors of this work argued
that these results are not applicable to the Random Regular Graph
(RRG) for which all fractal dimensions jump from $0$ at $W>W_{c}$
to 1 at $W<W_{c}$ in agreement with earlier results \cite{MF-ergodic-theory}.

We discuss the reasons for the applicability of the population dynamics and RSB theory to RRG in detail below.
Here we note that  the result of the recent work\cite{TihMir} directly contradicts the ones of the population dynamics.
Indeed, the inflationary population dynamics of Sec.~\ref{sec:Analytical-results-for-D(W)} shows that $1-D_{1}\propto W^{4}$
at least down to $W=4.5$, in contrast with the prediction\cite{TihMir} of the ergodic phase with $D_{1}=1$ at all $W<W_{E}=5.7$ for $K=2$.
Furthermore, the Lyapunov exponents obtained numerically in Sec.~\ref{sec:Population-dynamics-Lyapunov} do not reach the ergodic limit $\lambda=\frac{1}{2}\ln K$ even at $W=3$.
This discrepancy raises doubts on the applicability of results of Ref.\cite{TihMir} to the Cayley tree with one orbital per site, because there is no doubt that population dynamics describes the bulk of a large Cayley tree.

\end{enumerate}

\subsubsection*{Bethe lattice vs. RRG. }

It is indeed possible that a finite Cayley tree and RRG might be very
different. This possibility was extensively discussed in the context
of the theory of spin glasses (see Ref.\cite{Mezard-Parisi}). In
the context of localization problem a finite Cayley tree is a statistically
inhomogeneous system in which typical wave function amplitudes decreases
exponentially from a certain center (the \char`\"{}head\char`\"{}),
the states are localized or extended depending only on the relationship
between the Lyapunov exponent $\lambda_{\text{av}}$ that characterizes
the exponential decay of a wave function and the increment $\ln K$
in the exponential growth of the phase volume. This behavior is reminiscent
of the localization problem in finite dimensional systems with power-law
dependence of hopping integrals on the distance \cite{Shlyapnikov}.
In this case a typical wave function also decreases from a single
center. This decrease is power-law that competes with the polynomial
growth of the phase volume, similarly to the competition between exponential
decrease of the wave functions and exponential increase of the phase
volume on Bethe lattice. In contrast to a finite Cayley tree, in RRG
a typical wave function is \char`\"{}multi-headed\char`\"{}. The eigenstates
may be ergodic or non-ergodic depending on the distribution of \char`\"{}heads\char`\"{}
in space. Thus the difference between the finite Cayley tree and RRG
might be really dramatic.

\subsubsection*{Symmetry of the correlation volume dependence on $W-W_{c}$ in different
models}

The first work\cite{Zirn-BL} predicting asymmetric behavior of the
correlation volume performed a careful analysis of the non-linear
supersymmetric sigma model on the Bethe lattice. A rigorous derivation
of this model by Wegner \cite{Wegner} starts from the $\mathfrak{N}$-orbital
model, a tight-binding model with $\mathfrak{N}\gg1$ states per site.
All states of neighboring sites are connected by a random hopping
and they also have random on-site energies. Physically, this model
corresponds to large quantum dots connected by tunnel junctions. In
contrast, the model considered in this paper has $\mathfrak{N}=1$
states per site. Qualitatively one expects very different regimes
in the system of quantum dots depending whether the width of each
level, $\Gamma$, is larger or smaller than the level spacing, $\delta$
inside each dot. Supersymmetric sigma model can be derived in the
limit $\delta\sim1/\mathfrak{N}\rightarrow0$. Thus the results obtained
in this framework correspond to $\Gamma\gg\delta$ whereas single
particle localization problem on Bethe lattice considered here in
the Inverse-Order Thermodynamic Limit (IOTL) corresponds to the opposite
limit, $\Gamma\ll\delta$. Formally, one can argue that 'large dot'
model introduces one more parameter, $\delta\sim1/\mathfrak{N}\rightarrow0$,
so that the results depend on the order of limits, $\delta\rightarrow0$,
$\eta\sim\Gamma\rightarrow0$ and $N\rightarrow\infty$,
the "sigma-model limit" corresponding to $\mathfrak{N}\rightarrow\infty$ before taking any other limit.

In this respect it would be interesting to study the models with finite
but large $\mathfrak{N}\gg1$. We believe that these models will show
the crossover from the regime equivalent to that described by the
non-linear sigma model at sufficiently large $1-W/W_{c}$ to the regime
equivalent to $\mathfrak{N}$=1 in the vicinity of $W_{c}$. We expect
this crossover to happen happen when $\rho_{\text{typ }}\sim1/\mathfrak{N}$.
At sufficiently large $1-W/W_{c}$ the level width, $\rho_{\text{typ }}$,
becomes much larger than the distance between them, $\delta\sim1/\mathfrak{N},$
so the individual level structure inside the dot is not resolvable
and the system should be equivalent to the dots with $\mathfrak{N}\rightarrow\infty$.
Conversely, at $\rho_{\text{typ }}\ll1/\mathfrak{N}$ levels inside
the dot are not hybridized, so the system should be equivalent to
$\mathfrak{N}=1$ studied in this work with the modified density of
states. In the former regime the equations similar to those derived
in section \ref{sec:RSB-results-for-rho_typ} should produce $\rho_{\text{typ }}\sim{\rm exp}(-a/\sqrt{W_{c}-W})$.
This is likely to be due to the $\Xi(\rho,W,m)\sim\ln\rho$ for $\rho_{\text{typ }}<\rho<\rho_{c}$
in this regime leading to $\ln^{2}(\rho_{c}/\rho_{\text{typ }})$
contribution to the integral (\ref{eq:rho_0}).

A distinct set of works summarized in \cite{Efetov-book} used supersymmetric
formulation to develop the\emph{ }\textit{\emph{effective medium approximation}}
in high but finite dimension $d$ to the Anderson localization problem
as a zero-order approximation in $1/d$ expansion. As one might expect,
because in any finite dimension the phase volume grows polynomially
with the length, the symmetry of the characteristic length below and
above the transition was restored \cite{Efetov-book} and exponential
dependence disappeared both below and above the transition: $N_{c}\sim|W-W_{c}|^{-d/2}$.

The only work that studied the same model of single particle localization
on Bethe lattice with $\mathfrak{N}=1$ is the work \cite{MirlinFyodorov1991}
that used the supersymmetry method but does not employ the mapping
to the supersymmetric sigma model. We believe that it is very difficult
to get the correct results for $N_{c}$ in delocalized regime by this
method due to a lack of physical transparency of intermediate results
within the supersymmetric formalism. In particular, the account of
the symmetry which we discuss in section \ref{sec:RSB-results-for-rho_typ}
and Appendix \ref{app:Proof-of-the-symmetry}, in \emph{both} linear
and non-linear equation is crucial. Whilst the symmetry of the linear
equations was embedded in the solution \cite{MirlinFyodorov1991},
it is not clear to us if this symmetry was properly accounted for
in the non-linear equations.

Generally we believe that a difference between the results of the
large dot model and the single level model studied in this paper implies
that localization in infinite dimensional space may belong to several
different universality classes. It is very likely that these classes
translate into different physical properties of the many body localization
of physically relevant systems.

A possible way to identify these classes is to characterize the geometrical
properties of graphs representing a many-body Hamiltonian via connections
(\char`\"{}edges\char`\"{}) between the single-particle states (\char`\"{}sites\char`\"{})
provided by interaction. Once geometry of graphs is established, the
next step is to study single-particle localization in the Anderson
model on such \textit{random} graphs with \textit{random} on-site
energies. Random Regular Graphs and Random Tree Graphs discussed in
this paper are only two important examples of such an approach.

\subsubsection*{Evidence for direct transition from ergodic to localized states.}

We now to turn to the works that claim to observe direct transition
from localized to ergodic states. We start with the very recent work
\cite{Lemarie}. This work studied the model in the regimes were non-ergodic
phase is expected to be narrow or disappear, so such conclusion is
to be expected. Indeed, the model studied by the extensive numerical
simulations in this work corresponds to the average branching number
$K$ close to 1. As we show above, the non-ergodic regime becomes
wide at large $K$, it is obviously absent at $K=1,$ so we expect
it to be very narrow or even absent at $K\rightarrow1$ which makes
it hard to detect numerically.

The absence of non-ergodic extended state on RRG in work \cite{MetzCastillo}
is based on an analysis of the level compressibility $\chi$ \cite{AltShkl,ChalKravLer,Bogomol-Chi}
\begin{equation}
\chi_{\eta}(L)\equiv\frac{\langle(\delta n)^{2}\rangle}{\langle n\rangle},\label{Chi-def}
\end{equation}
where $n$ is the fluctuating number of energy levels in an energy
window $L=s/N$, and $s=\langle n\rangle$ is the average number of
levels in this window. Using the combination of analytical theory
of \cite{MetzCastillo-PRL} of the spectra of random matrices with
local tree structure and population dynamics this work shows that
$\chi_{\eta}(L\rightarrow0)\rightarrow0$ at sufficiently small $L$
and makes the conclusion that the entire extended phase is ergodic.
However, the population dynamics of this work used fixed $\eta=10^{-6}$
while it corresponds to $N\rightarrow\infty$. In this situation the
width of energy levels $\eta$ becomes much larger than their mean
spacing $\delta$. The individual states then completely lose their
meaning giving way to the \textit{wave packets} for which the theory
of level compressibility \cite{AltShkl,ChalKravLer,Bogomol-Chi} is
inapplicable.

\subsubsection*{Comparison with RSB and supersymmetric approach}

Despite the obviously different conclusions and completely different
methods many equations derived in the present work have their counterparts
in the supersymmetric treatment of Ref.\cite{TihMir}. This by no
means is accidental and reflects a deep connection between the linear
one-step RSB and the corresponding supersymmetric (SUSY) formalism.

In particular, the equations for $\Lambda(m)$ and $D_{1}(W)$ have
their equivalents in \cite{TihMir}. The difference arises only when
non-linear in $\Im G$ terms in RSB and non-linearity of self-consistency
equations in SUSY formalism become important. Indeed, one can make
a dictionary that establishes a direct correspondence between the
quantities and equations in this paper and in work \cite{TihMir}.
Consider (\ref{eq:Lambda-averaging}) that describes the dependence
of the increment $\Lambda(m)$ on the parameter $m$ of the one-step
RSB theory. Its counterpart in Ref.\cite{TihMir} is Eq. (20) for
the velocity $v(\beta)$ of the kink soliton. The quantity $\epsilon_{\beta}$
is analogous to $\tilde{I}_{m}$ of this work and possesses the same
symmetry with respect to $\beta\rightarrow1-\beta$. Both equations
are crucial for the respective theories and in both cases one has
to optimize with respect to $m$ or $\beta$.

Furthermore, in both theories it is necessary to terminate the processes
(\char`\"{}inflation\char`\"{} of $\Im G$ in this paper and motion
of a soliton in \cite{TihMir}) described by $\Lambda_{m}$ and $v(\beta)$.
In the linear case this termination is due to a finite system size,
it occurs before a large $\Im G$ is developed that corresponds to
the equilibrium. This termination is described by the similar equations
(\ref{ell-t}) in the present work and Eq.(8) in Ref.\cite{TihMir},
where $s_{0}$ stands for $\ell_{t}$ and $m$ stands for $K$.

The statement in Ref.\cite{TihMir} that $D_{1}=1$ in the entire
extended phase on RRG is equivalent to a statement that the distances
$\ell$ larger than the graph diameter $d_{\text{RRG}}$ are relevant
for localization problem on a graph and $\ell_{t}>d$. An example
of the Rosenzweig-Porter RMT (section \ref{sec:Application-to-Rosenzweig-Porter})
shows that this is not the case.

The most important coincidence is the key equation (\ref{Lambda-D})
for the fractal dimension $D_{1}$ (first published in \cite{AltshulerCuevasIoffeKravtsov2016})
and the corresponding (24) in \cite{TihMir}. Given the above correspondence
between the notations they are just \textit{identical}.

The main difference between the results of this work in the linear
regime and those of work \cite{TihMir} is the issue of applicabilty
of the results to finite systems. We do believe that these results apply
to RRG with a few states per site, $\mathfrak{N}\sim1$.
The behavior at finite $\mathfrak{N}\gg1$ might be very different and deserves
future studies as explained above.

\subsubsection*{Statistical accuracy of dimension computation}

Many works (e.g. \cite{TMS}) derive fractal dimensions $D_{1}$ and
$D_{1}$ from exact diagonalization of the Anderson model on RRG by
computing corresponding moments of wave function amplitude $\langle|\psi|^{2q}\rangle$
($q=1,2$). The inherent problem with these computation is that these
moments (and all other moments with $q>1/2$) are dominated by rare
events and need enormous statistics to get them accurately. This is
a difficult problem, especially at large $N$ when the cost of diagonalization
increases as $N^{3}$. The situation is much better if instead of
$D_{1}$ one studies $2-\alpha_{0}$, where $\alpha_{0}$ characterizes
the most abandoned, \textit{typical value} of $|\psi|^{2}$ (see sec.
\ref{sec:Distribution-of-LDoS} and Appendices \ref{app:-Relation-between-alpha-and-D},\ref{app:Extraction-of-D(W)}).
The exponent $2-\alpha_{0}$ bears the same information as $D_{1}$
or $D_{2}$ but it requires much less disorder realizations to get
a satisfactory statistics.

It is exactly because of this issue of statistics that we employed
population dynamics to compute a \textit{typical} value of $\Im G$
rather than, say, its second moment $\langle(\Im G)^{2}\rangle$.
Even within the PD recursive algorithm that is much less expensive
than exact diagonalization in terms of CPU and allows for population
sizes as large as $10^{8}$ it appeared to be an almost impossible
task to find the second moment of $\Im G$ with the required accuracy.

Thus we believe that the results based on calculation of moments $\langle|\psi|^{2q}\rangle$
with $q=1,2$ are not reliable for large $N$ when computation cost
allows only for relatively small number of disorder realizations.
Much more promising seem to be computational schemes, like funding
$\alpha_{0}$, that employ the typical averages.

\subsubsection*{Distibution of $|\psi|^{2}$ vs. that of $\Im G$ }

Finally, we notice the important difference between the distribution
of wave functions amplitudes $|\psi|^{2}$ and that of $\Im G$ or
LDoS. In the NEE phase both of them are given by the \textit{large-deviation
ansatz} $P(\ln Z)=A\exp\left[\ln\mathfrak{M}\,f\left(\frac{\ln Z}{\ln\mathfrak{M}}\right)\right]$,
cf. (\ref{eq:P(lnx)-general}). This is a very general type of distribution
function characterized by a function $f(\alpha)$ and a large parameter
$\mathfrak{M}\rightarrow\infty$. As is shown above, the distribution
function of $|\psi|^{2}$ and that of LDoS $\rho$ are both of this
type. Moreover, the function $f(\alpha)$, which is very different
for these two distributions, nonetheless in both cases obeys the Mirlin-Fyodorov
symmetry (\ref{MF-sym}). However, in order to describe \textit{multifractality}
as a certain scaling with $N$, the large parameter $\mathfrak{M}$
in the large-deviation ansatz must be proportional to the system size $N$.
In the NEE phase this is the case for the distribution
of $|\psi|^{2}$ \textit{at any} sufficiently large system size $N$.
In contrast, distribution of LDoS exhibits a crossover: the parameter
$\mathfrak{M}$ is proportional to the system size $N$ for modestly
large $N\ll N_{c}$ and ceases to depend on $N$ in the limit $N\rightarrow\infty$:
for $N\gg N_{c}$ the parameter $\mathfrak{M}\propto N_{c}$ freezes
at the correlation volume $N_{c}$. Thus the distribution of LDoS
in the ATL is no longer multifractal in a sense of power-law dependence
of its moments with $N$, though its shape is highly asymmetric close
to the Anderson transition.

This principle difference between the distribution of wave function
amplitudes and that of the LDoS implies that it is the structure of
\textit{local spectrum} that shows a crossover as $N$ increases above
$N_{c}$, whereas statistics of wave functions remain multifractal
in NEE phase at all system sizes and does not feel the scale $N_{c}$
whatsoever.

\section{Conclusion\label{sec:Conclusion}}

In this paper we show the existence of the extended non-ergodic (NEE)
phase using the one-step Replica Symmetry Breaking (RSB) approach
and the novel \char`\"{}Inflationary Population Dynamics\char`\"{}
numerical algorithm for both Bethe lattice (BL) and Random Regular
Graphs (RRG). This algorithm allows to implement the unusual (\char`\"{}inverted-order\char`\"{})
thermodynamic limit (when the level width $\eta\rightarrow0$ prior
to the system size $N\rightarrow\infty$) that gives an access to
the statistics of single-wavefunctions. We use a standard population
dynamics algorithm to compute in the Anderson thermodynamic limit
($N\rightarrow\infty$ prior to $\eta\rightarrow0$) various quantities
such as the typical imaginary part of a single-site Green's function,
dynamical correlation function of the local density of states and
the Lyapunov exponents.

We show that the predictions of RSB approach are reliable and coincide
with the ones of the inflationary population dynamics at moderately
large disorder $W$ that constitute main part of the phase diagram.
At relatively small $W$ the RSB predicts the continuous transition
into the fully ergodic (EE) state while IPD predicts smooth crossover
in which $D\rightarrow1$ as $W\rightarrow0$. In the regime of small
$W$ where RSB results become not reliable, the results of the population
dynamics for Bethe lattice show the fractal dimension $D_{1}$ to
behave as $1-D_{1}\sim W^{4}$.

We argue that the Anderson localization model on random regular graph
is equivalent to that on BL as long as the \textit{typical} Lyapunov
exponent $\lambda_{{\rm typ}}>\ln K$ and thus the NEE phase exists
on RRG at least as long as the above inequality holds true and until
the localization transition that happens when the \textit{average}
Lyapunov exponent $\lambda_{{\rm av}}=\ln K$. In other words the
existence of the NEE phase is related with the lack of self-averaging
for the Lyapunov exponent which results in a gap between $\lambda_{{\rm typ}}$
and $\lambda_{{\rm av}}$.

We argue that the continuous behavior of the fractal dimension $D_{1}(W,E)$
on disorder and energy is a specific property of the Anderson localization
model on the locally tree-like graphs with \textit{a few orbitals}
per site. The non-linear sigma-model on such graphs may show a different
(e.g. discontinuous) behavior as it corresponds to \textit{an infinite
number of orbitals} per site.

The same is true for the behavior of the correlation volume $N_{c}$
near the localization transition. We show that for BL and RRG the
characteristic length $L_{c}=\ln N_{c}\propto1/|W-W_{c}|$ has a \textit{symmetric}
behavior above and below the Anderson transition, in contrast to the
asymmetric behavior ($L_{c}\propto1/\sqrt{W_{c}-W}$ for $W<W_{c}$
and $L_{c}\propto1/(W-W_{c}$) for $W>W_{c}$) derived from the nonlinear
sigma-model on BL and RRG.

These results have important implications for the physical properties
of the disordered physical systems with interaction. Generally, knowing
the Hamiltonian of an interacting system and applying interaction
as perturbation progressively one can restore the corresponding graphs.
We expect the physical properties to be very different for the graphs
with the number of \char`\"{}orbitals\char`\"{} $\mathfrak{N}\sim1$
and $\mathfrak{N}\gg1$. In particular, the corresponding universality
classes of interacting Hamiltonians will or will not exhibit the intermediate
non-ergodic states and transitions between non-ergodic to ergodic
phases. Different universality classes of many body systems might
also translate into the different properties of local spectrum.\cite{PinoKrav}
Classifying interacting systems by the corresponding graphs and studying
localization properties on such graphs is a challenging project for
future studies.

The developed theory relies on the exact symmetry (\ref{symmetry-omega}).
Using it we suggested the approximation for the critical disorder
$W_{c}$ for the Anderson model on Bethe lattice with the box probability
distribution of random on-site energies. This approximation appeared
to be the best available so far. It also allows us to obtain a phase
diagram in the disorder-energy plane for the disordered Bethe lattice
and express the phase boundary in terms of the disorder- and energy-dependent
Lyapunov exponents.

\section{Acknowledgment}

We appreciate discussions with I. Aleiner, G. Biroli, J. T. Chalker,
E. Cuevas, M. Feigelman, I. Khaymovich, A. Yu. Kitaev, G. Parisi,
A. Scardicchio, M. Tarzia, K. Tikhonov. We are especially grateful
to E. Bogomolny and S. Warzel who helped us to understand importance
and pertinence of the recent rigorous mathematical results on localization
on BL. A support from LPTMS of University of Paris-Sud at Orsay, from
LPTHE of the Paris University Pierre and Marie Curie, and from College
de France and ICTP (Trieste) where important part of this work was
done, is gratefully appreciated. This research was partially supported
by the Russian Science Foundation grant No. 14-42-00044.

\appendix

\section{Relation between $\alpha_{0}$ and $D_{1}$\label{app:-Relation-between-alpha-and-D}}

Here we derive the relation between the exponents $\alpha_{0}$ and
$D_{1}$ assuming a scale invariant distribution of the wave function
amplitudes and duality between small and large local densities of
states established in many different models.

A broad class of the distributions, $P(x)$, of the normalized wave
function amplitudes, $x=N\psi^{2}$, is given by the \textit{\emph{generic
multifractal ansatz introduced in section \ref{sec:Analytical-results-for-D(W)}
(see also}} \cite{Our-BL}):
\begin{equation}
P(x)=\frac{A}{x}\,N^{f(\alpha)-1}\text{, where }\alpha=1-\frac{\ln x}{\ln N}.\label{multifractal-ansatz}
\end{equation}
Here the function $f(\alpha)$ is defined for $\alpha\geq0$. The
normalization condition for the probability distribution $\int P(x)dx=1$
implies that it has a maximum at some point $\alpha=\alpha_{0}$,
and its value at the maximum is $f(\alpha_{0})=1$, provided that
the normalization constant $A$ depends only logarithmically on $N$.
The position of the maximum, $\alpha_{0}$, determines the most abundant,
\textit{typical} value of $x=N|\psi|^{2}$:
\begin{equation}
|\psi|_{\text{typ}}^{2}={\rm exp}\left[\int\ln x\,P(x)\,dx\right]\sim N^{-\alpha_{0}}.\label{g-typ}
\end{equation}
Normalization of the wave function $\sum_{i}|\psi(i)|^{2}=1$ and
the Jensen inequality \cite{Jensen} $|\psi|_{{\rm typ}}^{2}\leq\langle|\psi|^{2}\rangle$
requires $|\psi|_{{\rm typ}}^{2}\leq N^{-1}$, i.e. $\alpha_{0}\geq1$,
with the equality attained in the ergodic state. Thus $\alpha_{0}>1$
implies fractality of the wave function and a non-ergodic state. Furthermore,
$\alpha_{0}$ is directly related to the fractal dimension $D_{1}$.

The relation between $\alpha_{0}$ and $D_{1}$ is due to the symmetry
of the function $f(\alpha)$. The latter follows from the symmetry
of the distribution function, $P(\tilde{\rho})$, of the reduced local
density of states, $\tilde{\rho}=\rho/\langle\rho\rangle,$ defined
in the Anderson thermodynamic limit:
\begin{equation}
P(1/\tilde{\rho})=\tilde{\rho}^{3}\,P(\tilde{\rho}).\label{eq:duality}
\end{equation}

\begin{figure}
\includegraphics[width=0.9\columnwidth]{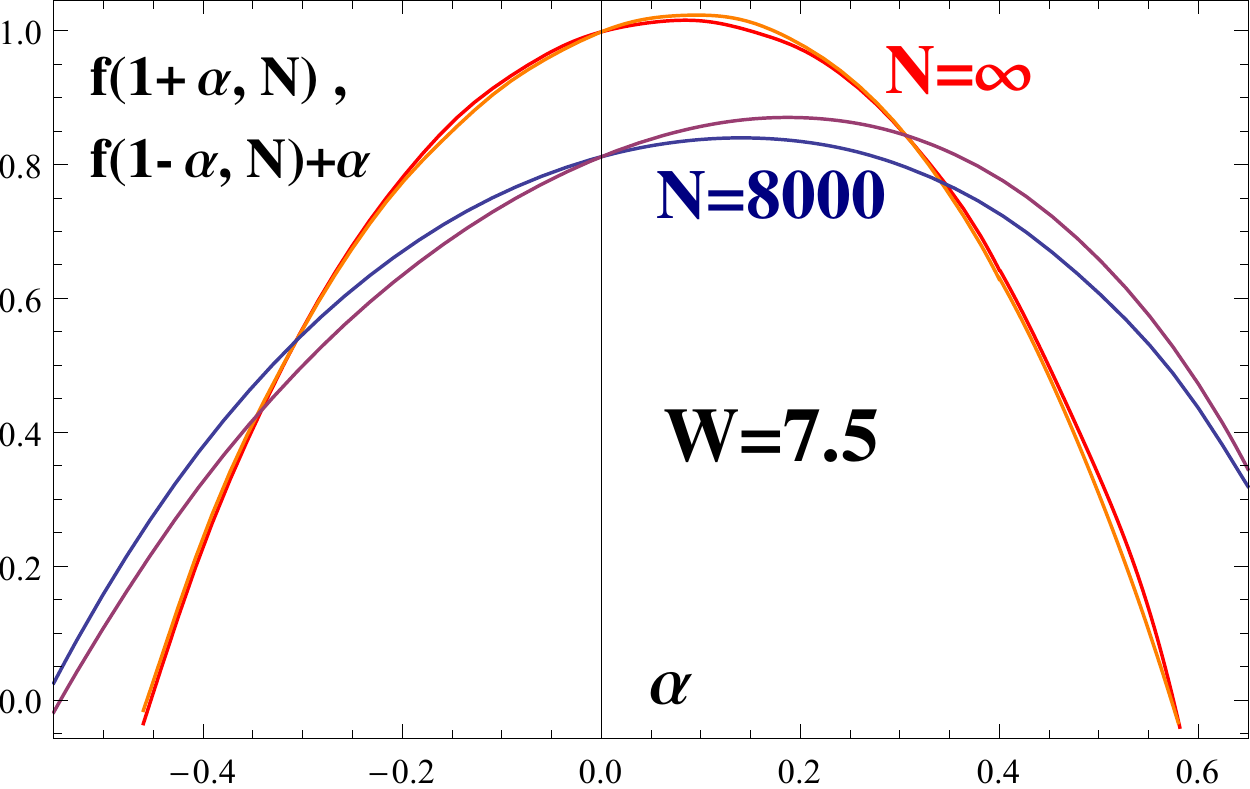}

\caption{The symmetry of the distribution of wave function amplitudes expressed
in terms of $f(\alpha)$ for the actual data at $W=7.5$ and $N=8000$
(blue and purple curves) and for the data extrapolated to $N=\infty$
(red and orange curves).\label{fig:The-symmetry-of-wave-f}}
\end{figure}

The symmetry (\ref{eq:duality}) seems to be a very general property
of the disordered systems. It was discovered first in\cite{Alt-Prig}
for the LDoS distribution in strictly one-dimensional systems using
the Berezinskii technique \cite{Ber}, it was later proved for systems
of \textit{any} dimensions in the framework of the nonlinear supersymmetric
sigma model \cite{MF}. Later works \cite{Savin-Fyod} showed that
(\ref{eq:duality}) is valid under very general conditions in both
localized and extended phases. It is generally believed that the symmetry
(\ref{eq:duality}) is exact in disordered (chaotic) systems where
the phase of wave function is completely random.

The symmetry (\ref{eq:duality}) translates into the symmetry of the
wave function distribution. For sufficiently small sizes the main
contribution to the local density of states comes from a few correlated
wave functions, so $\rho\sim\psi^{2}$. In this situation the distribution
function of $\rho$ coincides with the distribution function of individual
wave functions. At larger sizes these distributions might become different
because $\rho$ involves averaging over many states belonging to the
energy interval $\rho_{\text{typ}}$. When the mean level spacing
$\delta\sim1/N$ becomes smaller than this interval, the distribution
of $\rho$ ceases to depend on $N$. The crossover volume $N_{c}\sim\rho_{\text{typ}}^{-1}$
diverges as $W$ approaches $W_{c}$ and remains numerically large
in a wide range of $W$. In contrast, the distribution of $\psi^{2}$
does not experience a similar crossover as is evidenced by the absence
of a well defined crossover scale in IOTL (we remind that $\rho_{{\rm typ}}\propto\eta\rightarrow0$
in this limit). Thus, the symmetry (\ref{eq:duality}) should hold,
at least approximately for all $W$ for which $N_{c}\gg1$. This conclusion
is corroborated by the analysis of the data of direct diagonalization
of work\cite{Our-BL}.

The symmetry expressed in terms of function $f(\alpha)$ becomes
\begin{equation}
f(\alpha+1)=f(1-\alpha)+\alpha.\label{MF-sym}
\end{equation}

We re-plot the data from the work\cite{Our-BL} for $W=7.5$ in Fig.~\ref{fig:The-symmetry-of-wave-f}
that shows a fairly good symmetry at $N=8000$ and perfect symmetry
of the curve extrapolated to $N=\infty$. For $W=7.5$ the typical
imaginary part of the Green's function is $\rho_{\text{typ}}\approx0.055$
that translates into $N_{c}\approx200.$ The data shown in Fig.~\ref{fig:The-symmetry-of-wave-f}
for $N=8000$ prove that at large $N$ when the distributions of $\rho$
and $\psi^{2}$ differ, the symmetry still holds.

Using $f(\alpha_{0})=1$ one immediately finds from (\ref{MF-sym}):
\begin{equation}
2-\alpha_{0}=f(2-\alpha_{0}).\label{eq:2-alpha}
\end{equation}
Generally, the dimensions of the wave function moments, $D_{q},$
can be related to $\alpha_{q}$ defined as the root of the equation
$f'(\alpha_{q})=q$:
\begin{equation}
D_{q}=\frac{q\alpha_{q}-f(\alpha_{q})}{q-1}\label{eq:D_q}
\end{equation}

The normalization condition $\sum_{i}|\psi(i)|^{2}=1$ implies that
\begin{equation}
f(\alpha_{1})=\alpha_{1},\label{eq:norm_condition}
\end{equation}
so that the dimension $D_{1}$ should be understood as a limit $D_{1}=\lim_{q\rightarrow1}D_{q}$.
Using the definition of $\alpha_{q}$ we get

\[
D_{1}=\alpha_{1}+\frac{\partial\alpha_{q}}{\partial q}\,\left[q-f'(\alpha_{q})\right]_{q=1}=\alpha_{1}
\]

On the other hand, the equation $f(x)=x$ has only one solution, so
comparing (\ref{eq:2-alpha}) and (\ref{eq:norm_condition}) we conclude
\begin{equation}
2-\alpha_{0}=\alpha_{1}=D_{1}.\label{D-D1}
\end{equation}

\section{$m_{0}$, $m_{1}$ and the power-law distribution of $|\psi|^{2}$.\label{app:m_0-m_1-and-the-power}}

\begin{figure}[h]
\includegraphics[width=0.9\linewidth]{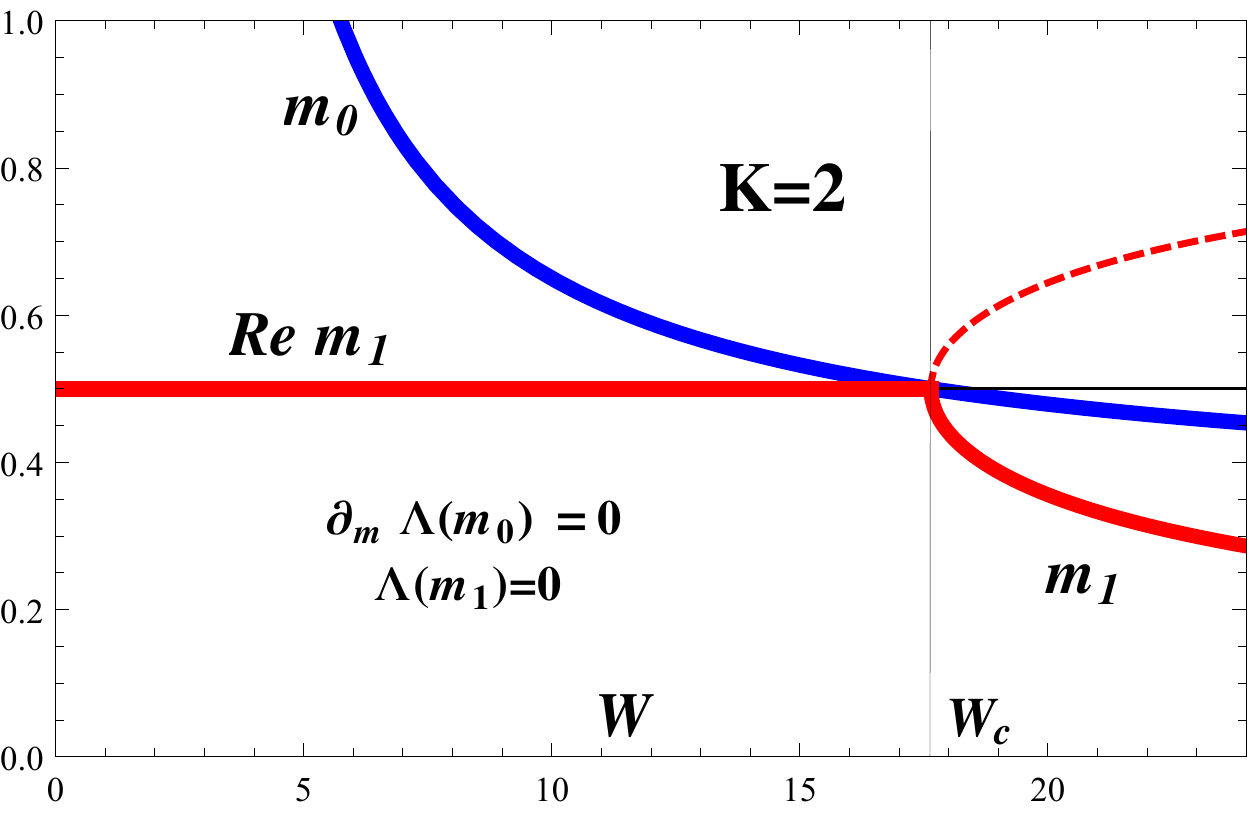}
\caption{(Color online) The solutions to the equations $\partial_{m}\Lambda(m_{0})=0$
(blue) and $\Lambda(m_{1})=0$ (red) obtained for $E=0$ using $F_{{\rm eff}}(\epsilon)$
from Eq.(\ref{low-cutoff}). The termination point $W=W_{E}$ of the
RSB solution corresponds to $m_{0}=1$. At the AT point $W=W_{c}$
$m_{0}=m_{1}=1/2$. In the limit $1/\ln K\rightarrow0$ where Eq.(\ref{low-cutoff})
is exact the solution $m_{0}\rightarrow1/2$ for all $W>W_{E}$. For
$W>W_{c}$ the equation $\Lambda(m_{1})=0$ has two roots, the smaller
of them describes the exponent of the power-law distribution of wave
function amplitudes. At $W<W_{0}$ the solution $m_{1}=1/2\pm i\;\Im m_{1}$
is a complex number with the real part equal to $1/2$ for \textit{all}
$W<W_{c}$. }
\label{Fig:m0-m1}
\end{figure}

\begin{figure}[t]
\includegraphics[width=0.9\columnwidth]{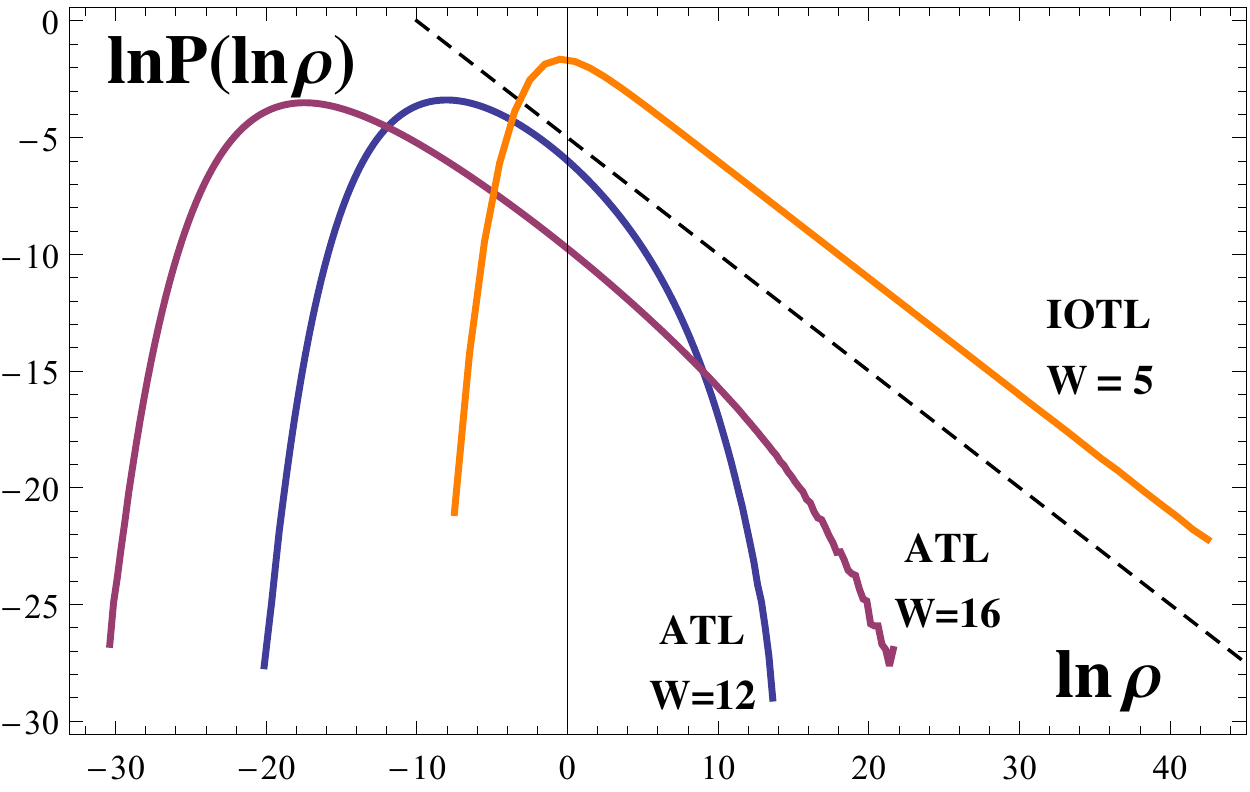}
\caption{Distribution functions of $\rho$ obtained by stationary population
dynamics converge to the power law as $W\rightarrow W_{c}.$ These
distribution functions correspond to the behavior in Anderson thermodynamic
limit (ATL). Distribution functions for $\rho$ in the inverted order
thermodynamic limit (IOTL) are power law with $m=1/2$ for all $W<W_{c}$.
The dashed line shows power law $m=1/2.$ \label{fig:Distribution-functions-of-rho}}
\end{figure}

In this section we establish the exact correspondence between RSB
parameter $m$ and distribution function of the imaginary part of
self energy, $P(\mathfrak{S})$ in the limit when $\eta\rightarrow0$
prior to $N\rightarrow\infty$ (IOTL). In this \char`\"{}inflationary\char`\"{}
regime the typical $\mathfrak{S}$ grows with generation number $\ell$,
however the distribution function of $\mathfrak{S}/\mathfrak{S}_{\text{typ}}$
acquires a stationary form. To find it we factor out the uniformly
growing factor, thus defining $\mathfrak{r}_{i}^{(\ell)}$ by
\begin{equation}
\mathfrak{S}_{i}^{(\ell)}=\mathfrak{r}_{i}^{(\ell)}\,e^{\Lambda\,\ell},\label{inflation}
\end{equation}
which acquires a stationary (independent of the generation $\ell$)
distribution function, $P_{0}(\mathfrak{r})$. Substituting (\ref{inflation})
into the \textit{linearized} equation (\ref{eq:ImG_lin}) we obtain:
\begin{equation}
\mathfrak{r}_{i}^{(\ell+1)}=\sum_{j(i)}\frac{\mathfrak{r}_{j}^{(\ell)}\,e^{-\Lambda}}{\epsilon_{j}^{2}}.\label{Eq-infl}
\end{equation}
Next we exponentiate $e^{-s\,\mathfrak{r}_{i}}$ and average over
disorder to obtain the Laplace transform ${\cal P}(s)$ of the distribution
function $P_{0}(\mathfrak{r})$. We obtain (cf. Eq.(29) in Ref. \cite{FeigelmanIoffeMezard}):

\begin{equation}
{\cal P}(s)=\left[\int F_{{\rm eff}}(\epsilon)\,{\cal P}(s\,e^{-\Lambda}/\epsilon^{2})\,d\epsilon\right]^{K},\label{eq:P(s)-recursion-a}
\end{equation}
where we took into account the correlations in $\Re G$ at different
sites by introducing the \textit{effective} distribution $F_{{\rm eff}}(\epsilon)$
of $\epsilon=\Re G^{-1}$ obeying (at $E=0$) the symmetry Eq.(\ref{symmetry-omega}).

Next we establish the condition when $P_{0}(\mathfrak{S})$ (and thus
also $P_{0}(\mathfrak{r})$) has a power-law tail at large $\mathfrak{S}$
(or large $\mathfrak{r}$). To this end we look for a solution to
Eq.(\ref{eq:P(s)-recursion-a}) in the form \cite{AbouChacAnd,FeigelmanIoffeMezard}:
\begin{equation}
{\cal P}(s)=1-A\,s^{m}.\label{eq:P(s)-expansion}
\end{equation}
One can easily see that at $s\ll1$ this form is consistent with Eq.(\ref{eq:P(s)-recursion-a})
provided that:
\begin{equation}
K\int\frac{F_{{\rm eff}}(\epsilon)}{\epsilon^{2m}}\,d\epsilon=e^{\Lambda\,m},\label{cond-power-law}
\end{equation}
where $\Lambda=\Lambda(m_{0})$ is given by (\ref{eq:Lambda-averaging},\ref{Imm}).

Comparing the left hand side of this equation with (\ref{eq:Lambda-averaging},\ref{Imm})
we conclude that Eq.(\ref{cond-power-law}) is equivalent to:
\begin{equation}
\Lambda(m)=\Lambda(m_{0}),\Rightarrow m=m_{0}.\label{Lambda(m)}
\end{equation}
Thus the distribution function $P_{0}(\mathfrak{S})$ is a power law:
\begin{equation}
P_{0}(\mathfrak{S})\sim\frac{\mathfrak{S}_{0}^{m_{0}}}{\mathfrak{S}^{1+m_{0}}}.\label{PL}
\end{equation}
The parameter $\mathfrak{S}_{0}$ in Eq.(\ref{PL}) has a meaning
of the lower cutoff of the power-law of the order of the typical value
of $\mathfrak{S}_{\text{typ}}$.

The result Eq.(\ref{PL}) is derived in the inflationary regime of
IOTL. The presence of the factor $e^{\Lambda(m_{0})\,m}$ in the right
hand side of (\ref{cond-power-law}) is crucial for the conclusion
that exponent $m=m_{0}$. In contrast, in the standard ATL the distribution
function $P(\mathfrak{S})$ at $W<W_{c}$ is determined by the equilibration
due to the non-linearity of (\ref{eq:ImG-recursion}) and it is power-law
only in the vicinity of the AT point $W\approx W_{c}$ (see Fig.~\ref{fig:Distribution-functions-of-rho}).

The situation is completely different in the insulating phase $W>W_{c}$.
In this regime the instability is absent and the linearized (\ref{eq:ImG_lin})
always applies at sufficiently small but finite $\eta$. However,
in this case one cannot neglect the source term $\eta$ in the linearized
equations:
\[
\Im G_{i}^{(\ell)}(E)=\frac{\sum_{j(i)}\Im G_{j}^{(\ell-1)}}{\left(E-\epsilon_{i}-\Re\Sigma_{i}^{(\ell)}\right)^{2}}+\eta
\]
that translate into
\begin{equation}
{\cal P}(s)=e^{-s\eta}\left[\int F_{{\rm eff}}(\epsilon)\,{\cal P}(s/\epsilon^{2})\,d\epsilon\right]^{K}\label{eq:P(s)-recursion-b}
\end{equation}
for the Laplace transforms. In contrast to (\ref{eq:P(s)-recursion-a}),
the equation (\ref{eq:P(s)-recursion-b}) does not contain a free
parameter, $\Lambda$. Its solution corresponds to the stationary
distribution in the presence of the source term $\eta$. Assuming
that at small $s$ ${\cal P}(s)$ has the expansion (\ref{eq:P(s)-expansion})
we conclude that $m$ satisfies the equation
\[
K\int\frac{F_{{\rm eff}}(\epsilon)}{\epsilon^{2m}}\,d\epsilon=1
\]
which is equivalent to:
\begin{equation}
\Lambda(m)=0\label{m1}
\end{equation}
in the RSB theory. In the following discussion we denote the smaller
root of this equation as $m_{1}$.

In this case of $W>W_{c}$ (insulator phase) $P(\mathfrak{S}/\eta)$
in the ATL essentially describes the distribution of the wave function
amplitudes $|\psi(i)|^{2}$. The corresponding power law is
\begin{equation}
P(|\psi|^{2})\sim\frac{1}{(|\psi|^{2})^{1+m_{1}}}\label{PL-psi}
\end{equation}
Note that real solutions to Eq.(\ref{Lambda(m)}) at $E=0$ exist
only for $W>W_{c}$ (see Fig.~\ref{Fig:m0-m1}), in the Anderson
insulator phase. For $W<W_{c}$ the two solutions to Eq.(\ref{Lambda(m)})
with $F_{{\rm eff}}(\epsilon)$ given by Eq.(\ref{low-cutoff}) are
complex conjugated numbers which real part is $1/2$.

There are two real solutions to Eq.(\ref{Lambda(m)}) at $W>W_{c}$,
and it is the smaller of them (which decreases to zero in the limit
$W\rightarrow\infty$) that determines the power law Eq.(\ref{PL-psi}).
One can check that the exponent of the power-law distribution of $|\psi|^{2}$
in the Anderson insulator on BL computed within the \char`\"{}directed
polymer\char`\"{} approximation in Ref.\cite{Our-BL} obeys the equation
identical to (\ref{m1}) for the box-shaped $F_{{\rm eff}}(\epsilon)=W^{-1}\,\theta(W/2-|\epsilon|)$.

\begin{figure}[h!]
\includegraphics[width=0.9\linewidth]{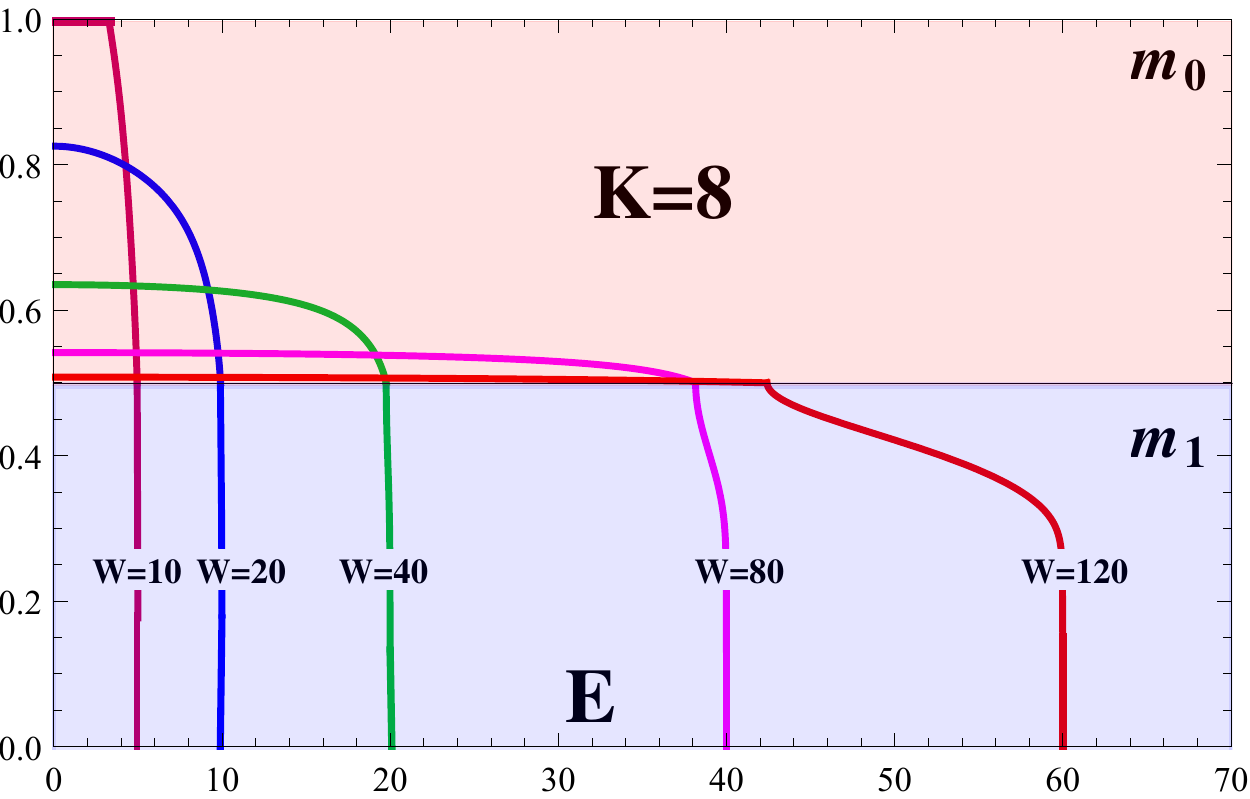}
\caption{(Color online) The solution $m_{1}$ to the equation $\Lambda(m_{1})=0$
in the region of localized states (blue) and the solution $m_{0}$
to the equation $\partial_{n}\Lambda(m_{0})=0$ in the region of extended
states (rose) as functions of energy $E$ at fixed disorder $W=10,20,40,80,120$
for $K=8$. For $W=10,20,40$ the drop of $m_{1}$ from $m_{1}=1/2$
to $m_{1}=0$ is practically vertical in our large-$K$ approximation
Eq.(\ref{Im-anal-cont}). Thus at $W<50$ the energy window for the
Anderson insulator phase is exponentially narrow in the parameter
$1/\ln K$ that controls the approximation.}
\label{Fig:band-edge}
\end{figure}

As the localization becomes stronger with increasing disorder or approaching
the band-edge, the exponent $m_{1}$ in Eq.(\ref{PL-psi}) decreases.
At $E=0$ and increasing disorder $m_{1}\approx\ln K/\ln[(W/2)^{2}]\rightarrow0$.
It also decreases logarithmically $m_{1}\approx\ln K/\ln[1/(E_{g}-E)^{2}]\rightarrow0$
as the observation energy $E$ approaches the band-edge $E_{g}$.

The spectral band-edge can be found from the condition:
\begin{equation}
m_{1}(E=E_{g})=0.
\end{equation}
One can see from Fig.~\ref{Fig:band-edge} that for small enough
$W$ and at large branching number $K$ the energy window $0<m_{1}<1/2$
close to the band-edge where the Anderson insulator phase exists,
is exponentially narrow in the control parameter parameter $1/\ln K$
of our approximation or may be absent completely \cite{Warzel}.

\section{Termination point of RSB solution\label{app:Termination-point-of-RSB-solution}}

The goal of this Appendix is to prove that for \textit{any} distribution
$F_{\text{eff }}(\epsilon)$ one-step RSB solution terminates at a
finite disorder $W=W_{E}$ and that at the termination point

\begin{subequations}
\begin{align}
m_{0}(W_{E}) & =1,\\
D(W_{E}) & =1,\\
dD(W)/dW|_{W_{E}} & =0
\end{align}
\label{eq:termination-equations}\end{subequations}

This implies that it is not the particular form (\ref{low-cutoff})
of $F_{{\rm eff}}(\epsilon)$ which is responsible for the absence
of ergodic transition at a finite $W_{E}$ on BL (as evidenced by
inflationary population dynamics) but it is rather a failure of the
one-step RSB ansatz at small $W$ due to non-local correlations in
$\Re G$ along a path.

We start by proving that $m_{0}(W)$ reaches 1 at a finite disorder.
To prove it we show that in the limit $W\rightarrow0$ the formal
solution to $\partial_{m}\Lambda(m)=0$ diverges $m_{0}(W)\rightarrow\infty$.
Because at the Anderson transition point $m_{0}(W_{c})=1/2$ by continuity
it implies that $m_{0}(W)=1$ for some $W_{E}>0$. Indeed, in the
limit $W\rightarrow0$ the function $F_{{\rm eff}}(\epsilon)$ in
the integral
\begin{equation}
\tilde{I}_{m}=\int_{0}^{\infty}F_{{\rm eff}}(\epsilon)\,\epsilon^{-2m}\,d\epsilon\label{IIm}
\end{equation}
shrinks to a delta-function $F_{{\rm eff}}(\epsilon)=\delta(\epsilon-1)$,
and thus \textit{for all} $m$ we have:
\begin{equation}
\tilde{I}_{m}\rightarrow1.\label{Im1}
\end{equation}
Using the expression for $\Lambda(m)$:
\begin{equation}
\Lambda(m)=\frac{1}{m}\,\ln\left(K\,\tilde{I}_{m}\right)\label{Lam}
\end{equation}
we rewrite the equation for $m_{0}$ as
\begin{equation}
\partial_{m}\ln{I}_{m}=\Lambda(m).\label{derLam}
\end{equation}
Because in the limit $W\rightarrow0$ we have $\tilde{I}_{m}\rightarrow1$
for all $m$, $\partial_{m}I_{m}\rightarrow0$ and Eq.(\ref{derLam})
is reduced to: $\Lambda(m_{0})\rightarrow0$. Then it immediately
follows from (\ref{Lam}) that in this limit $m_{0}\rightarrow\infty$.

In Sec.\ref{sec:Analytical-results-for-D(W)} we have already proven
that $m_{0}=1$ implies $D(W_{E})=1$. Now we prove that $\partial_{W}D(W_{E})=0$.
Indeed,
\[
d\Lambda/dW=\partial_{W}\Lambda(m_{0},W)+\partial_{m}\Lambda(m_{0},W)\,dm_{0}/dW.
\]
Since $\partial_{m}\Lambda(m_{0},W)=0$ and $dm_{0}/dW$ at $W=W_{E}$
is finite one obtains:
\begin{equation}
\frac{dD}{dW}=\frac{\partial_{W}(\ln\tilde{I}_{m_{0}})}{\ln K\,m_{0}}.\label{dD-dW}
\end{equation}
We conclude from Eq.(\ref{dD-dW}) that $dD/dW=0$ at $W=W_{E}$,
because $m_{0}=1$ at the termination point, and $\tilde{I}_{1}=\tilde{I}_{0}=1$
is independent of $W$ due to the symmetry of $F_{{\rm eff}}(\epsilon)=F_{{\rm eff}}(\epsilon^{-1})$.

This proof remains valid as long as the derivative $\partial_{m}\tilde{I}_{m}$
at $m=1$ (which is equal to $-\partial_{m}\tilde{I}_{1-m}$) is finite,
i.e. for $F_{{\rm eff}}(\epsilon)$ decreasing at large $\epsilon$
faster than $\epsilon^{-1}\,\ln^{-2}(\epsilon)$.

\section{Proof of the symmetry and its numerical verification.\label{app:Proof-of-the-symmetry}}

In this Appendix we give a proof of the symmetry of the distribution
function for the Green's functions products along a path that justifies
the requirement (\ref{symmetry-omega}) and approximation (\ref{low-cutoff}).
We also derive the general properties of RSB solutions that follow
from this symmetry and do not rely on the particular approximation
such as (\ref{low-cutoff}). Our first goal is to prove that for long
paths ($\ell\gg1)$ the distribution function of the product

\begin{equation}
y=\prod_{k=1}^{\ell}|\Re G_{k}(i_{k})|^{-1}\label{eq:y_def}
\end{equation}
obeys the symmetry

\begin{equation}
{\cal P}(y)={\cal P}(1/y).\label{eq:P(y)_symm}
\end{equation}
It is important that this symmetry holds only for the distribution
of the \textit{real} part of $G^{-1}$ in the situation when $\Im G$
can be neglected, it is thus directly applicable to the study of IOTL
in the whole range of $W$ and to ATL at $W\rightarrow W_{c}$.

We distinguish the Green's functions associated with a given path
and all others in the recursion (\ref{eq:G_i^(l+1)}) and rewrite
it as
\begin{equation}
G_{k-1}^{-1}(i_{k-1})=E-E_{i_{k-1}}-G_{k}(i_{k}),\label{G-eq}
\end{equation}
where $G_{k}(i_{k})$ is the Green's function in a point $i_{k}$
of the $k^{\text{th}}$ generation, and we introduced the notation:
\begin{equation}
E_{i_{k-1}}=\varepsilon_{i_{k-1}}+\sum_{j(i_{k-1}),j\neq i_{k}}G_{k}(j).\label{E}
\end{equation}
\begin{figure}[t]
\includegraphics[width=1.1\linewidth]{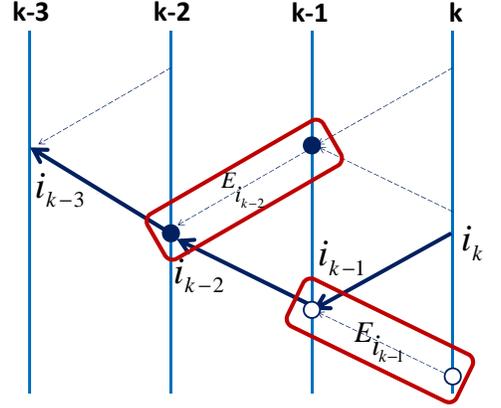} \caption{(Color online) Sites involved in $E_{i_{k-1}}$ (open circles) and
$E_{i_{k-2}}$ (full circles) belong to different branches of the
tree and thus $E_{i_{k-1}}$ and $E_{i_{k-2}}$ are statistically
independent. Vertical lines are generations, the fat solid arrows
denote the path, the dashed arrows denote links other than those belonging
to the path. }
\label{Fig:path}
\end{figure}

We notice that the energies $E_{i_{k}}$contains the contributions
of different side branches and thus are statistically independent
for different sites of the path (see Fig.~\ref{Fig:path}). Then
the measure $d\mu=\prod_{k=1}^{\ell}dE_{i_{k}}F_{0}(E_{i_{k}})$ along
a path is given by
\begin{equation}
d\mu=\prod_{k=1}^{\ell}dx_{k}\,F_{0}\left(E-x_{k}-\frac{1}{x_{k+1}}\right),\label{measure}
\end{equation}
where $x_{k}=G_{k}^{-1}(i_{k})$ and $F_{0}(E_{i_{k}})$ is the probability
distribution function of $E_{i_{k}}$ which is independent of $i_{k}$.

Consider a slightly modified measure in which $x_{\ell+1}=x_{1}$.
Then the ratio $d\mu/\Pi$, where $\Pi=\prod_{i}x_{i}$ is invariant
under the transformation:
\begin{equation}
x_{k}\rightarrow x_{\ell-k}^{-1}\label{eq:x_k_symmetry_transf}
\end{equation}
that inverts the order of variables along the loop and inverts each
of them. This transformation changes $y\rightarrow1/y$.

Then applying this transformation to: ${\cal P}(y^{-1})=\int d\mu\,\delta(y^{-1}-\Pi)=\int(d\mu/\Pi)\,\Pi\,\delta(y^{-1}-\Pi)$
one obtains Eq.(\ref{eq:P(y)_symm}):
\begin{eqnarray}
{\cal P}(y^{-1}) & = & \int\frac{d\mu}{\Pi}\;\Pi^{-1}\,\delta(y^{-1}-\Pi^{-1})\nonumber \\
 & = & \int\frac{d\mu}{y}\;y^{-1}\delta(y-\Pi)\,y^{2}={\cal P}(y).
\end{eqnarray}
The actual measure (\ref{measure}) differs from the closed loop measure
only by the end point that introduces corrections that become irrelevant
in the limit of $\ell\gg1$, we shall return to this justification
below.

We now use the exact symmetry to prove general relations of RSB theory
that do not depend on a particular approximation. Using (\ref{all-pathes})
we can express $\Lambda(E,m)$ in terms of ${\cal P}(y)\equiv{\cal P}_{\ell\rightarrow\infty}(y)$,
instead of $F_{{\rm eff}}(\epsilon)$:
\begin{align}
\Lambda(E,m) & =\frac{1}{m}\ln\left[K\left\langle \prod_{k=1}^{\ell}|G_{k}(i_{k})|^{2m}\right\rangle ^{\frac{1}{\ell}}\right]\label{RSB-eq}\\
 & =\frac{1}{m}\ln\left\{ K\left[\int\frac{dy}{y^{2m}}\,{\cal P}(y)\right]^{\frac{1}{\ell}}\right\} .\nonumber
\end{align}
Notice that any finite factor $A$ in ${\cal P}(y)$ would drop out
of this equation, as it would enter as $A^{\frac{1}{\ell}}\rightarrow1$.
This justifies the neglect of the effect of the boundary conditions
on the symmetry reasoning above.

\begin{figure}[t]
\includegraphics[width=0.9\linewidth]{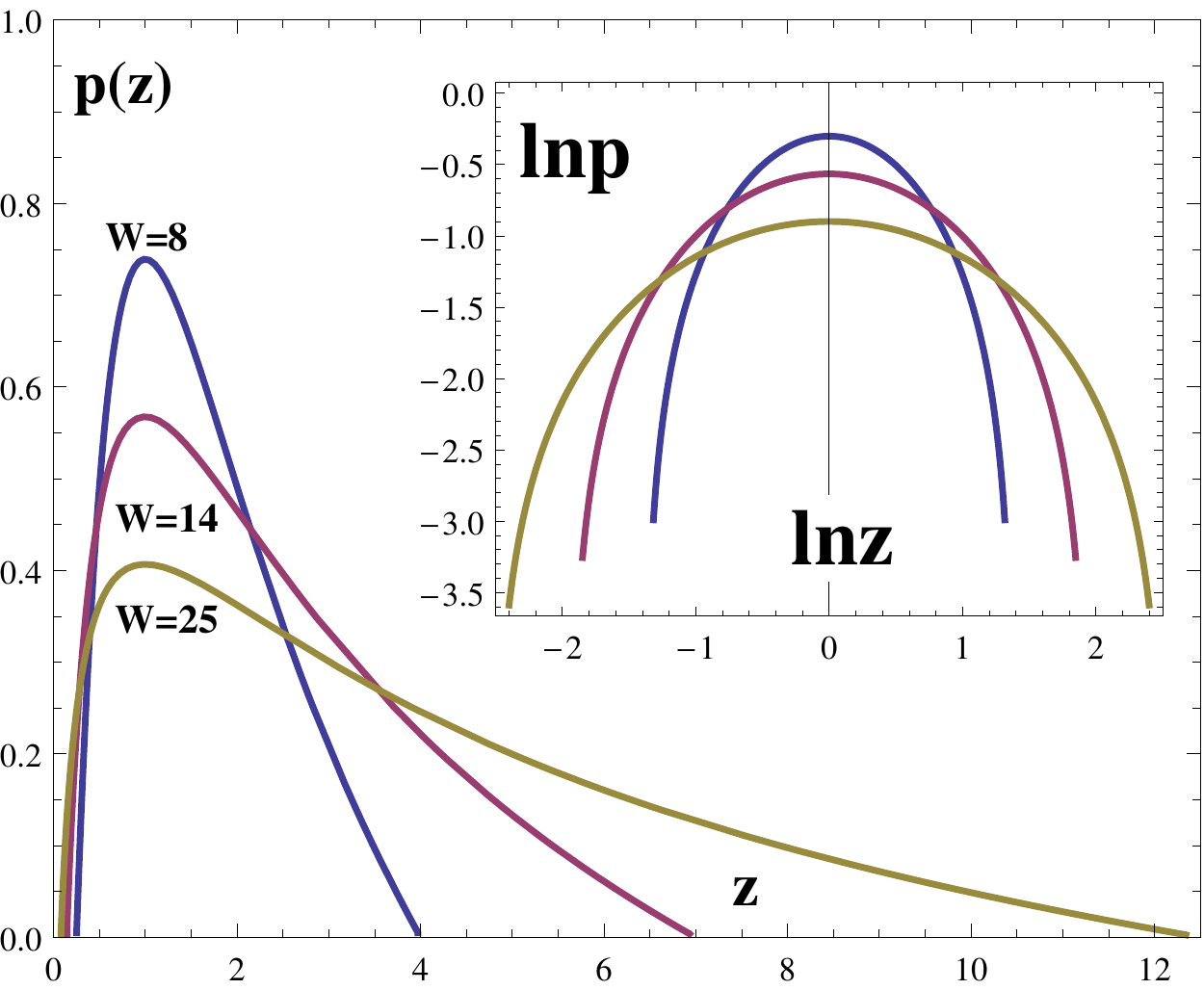}
\caption{(Color online) Functions $p(z)$ given by (\ref{eq:p(z)-approx})
for $W=8,14,25$ corresponding to the approximation Eq.(\ref{low-cutoff})
for $F_{{\rm eff}}(\epsilon)$. \label{Fig:approx} }
\end{figure}

At large $\ell\rightarrow\infty$ it is convenient to introduce the
function $p_{\ell}(z)$, with $p(z)=\lim_{\ell\rightarrow\infty}p_{\ell}(z)$,
such that
\begin{equation}
[p_{\ell}(z)]^{\ell}={\cal P}_{\ell}(z^{\ell}),\;\Rightarrow p(z)=p(z^{-1}).\label{eq:p(z)-definition}
\end{equation}

and evaluate the integral in (\ref{RSB-eq}) in the saddle-point approximation.
Then we obtain in the limit $\ell\rightarrow\infty$:
\begin{equation}
\left[\int\frac{dy}{y^{2m}}\,{\cal P}(y)\right]^{\frac{1}{\ell}}=p(z_{m})\,z_{m}^{1-2m},\label{saddle}
\end{equation}
where the saddle-point $z_{m}$ is the solution to:
\begin{equation}
\partial_{z}\ln p(z)+\frac{(1-2m)}{z}=0.\label{stat}
\end{equation}
The normalization of $\int{\cal P}(y)\,dy=1$ then imposes the normalization
of $p(z)$:
\begin{equation}
p(z_{0})\,z_{0}=1.\label{norm-p}
\end{equation}
Comparing (\ref{RSB-eq}) and (\ref{saddle}) with (\ref{eq:Lambda-averaging})
one concludes that:
\begin{equation}
\tilde{I}_{m}=\int F_{{\rm eff}}(\epsilon+E)\,\frac{d\epsilon}{|\epsilon|^{2m}}\Rightarrow I_{m}=p(z_{m})\,z_{m}^{1-2m}.\label{correspon}
\end{equation}
One can represent (\ref{saddle}), (\ref{stat}), (\ref{correspon})
in a more elegant form of the Legendre transformations. We denote

\begin{eqnarray*}
x & = & \ln z,\\
f(x) & = & \ln p(z),\\
q & = & 2m-1,\\
\tau(q) & = & -\ln I_{m}.
\end{eqnarray*}
Then (\ref{saddle}, \ref{correspon}) take a standard form of the
Legendre transform:

\begin{subequations}
\begin{align}
\partial_{x}f(x_{q}) & =q,\\
\tau(q) & =-f(x_{q})+qx_{q}.
\end{align}

\label{eq:Legendre}\end{subequations}

Equations (\ref{eq:Legendre}) can be inverted:

\begin{subequations}
\begin{align}
\partial_{q}\tau(q_{x}) & =x,\\
f(x) & =-\tau(q_{x})+x\,q_{x}.
\end{align}

\label{eq:inverted-Legendre}\end{subequations}

This allows to compute the function $p(z)$ that corresponds to the
approximation for $F_{{\rm eff}}(E)$ given by (\ref{low-cutoff}).
It can be obtained in the following parametric form:
\begin{eqnarray}
z & = & e^{\left(\frac{1}{u}-\coth(u)\right)\,\ln(W/2)},\nonumber \\
p(z) & = & A(W)\,e^{\left[1-u\coth(u)+\ln\left(\frac{\sinh(u)}{u}\right)\right]},\label{eq:p(z)-approx}
\end{eqnarray}
where $A(W)=\frac{2\ln(W/2)}{W/2-2/W}$ and $u\in[-\infty,+\infty]$.
The plot of this function at different values of disorder $W$ is
given in Fig.~\ref{Fig:approx}. Note that $p(z)/A(W)$ is a universal
function of $\ln z/\ln(W/2)$.

The results of this Appendix demonstrate that the notion of \char`\"{}effective
distribution of on-site energies\char`\"{} $F_{{\rm eff}}(\epsilon)$
is convenient for presentation but it is not necessary to obtain the
main results of the paper. They can be formulated entirely in terms
of $I_{m}$ given by Eq.(\ref{correspon}). In particular, the symmetry
$p(z)=p(1/z)$ is sufficient to prove the symmetry:

\begin{equation}
z_{m}=z_{1-m},\Rightarrow I_{m}=I_{1-m},\label{m-symm}
\end{equation}
which plays the same role as the symmetry $F_{{\rm eff}}(\epsilon)=F_{{\rm eff}}(1/\epsilon)$.
Notice that this symmetry is equivalent to the symmetry with respect
to $\beta\rightarrow1-\beta$ in the original work \cite{AbouChacAnd},
cf. equation (6.8) of Ref.\cite{AbouChacAnd}.

\begin{figure}[t!]
\includegraphics[width=1\linewidth]{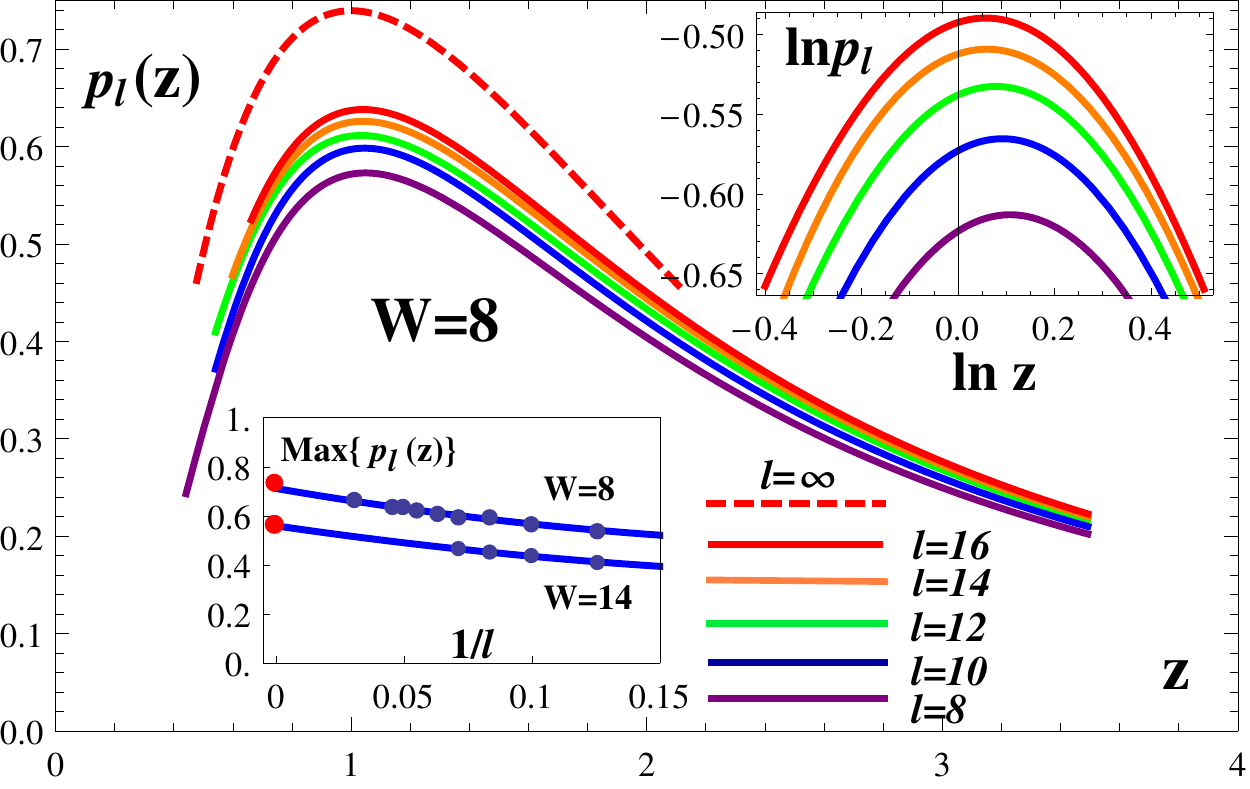}
\caption{(Color online) The function $p_{\ell}(z)$ for the infinite Cayley
tree with the branching number $K=2$ and disorder strength $W=8$
obtained by averaging over $5\times10^{9}$ pathes of the length $\ell=8-32$.
The red dashed line shows the $\ell=\infty$ limit given by (\ref{eq:p(z)-approx}).
The upper inset: the corresponding $\ln p_{\ell}(z)$ vs. $\ln z$
demonstrates restoration of symmetry $p_{\ell}(z)=p_{\ell}(z^{-1})$
as $\ell$ increases. The lower inset shows extrapolation of the maximum
of $p_{\ell}(z)$ to $\ell\rightarrow\infty$ from $\ell=8-32$ for
$W=8$ and from $\ell=8-14$ for $W=14$ using the second order polynomial
fit in $1/\ell$. The red spots are the maxima ${\rm Max}_{z}p(z)=A(W)$
of $p(z)$ given by (\ref{eq:p(z)-approx}). \label{Fig:p_z_num} }
\end{figure}

It follows immediately from (\ref{m-symm}) that $I_{1}=I_{0}=1$,
which is what we need to prove that $D=1$ at $m=1$, i.e. the existence
of the replica-symmetric solution. It is also sufficient to prove
Eq.(\ref{eq:lambda_av(E,W_c)}). Another useful relation that follows
from Eq.(\ref{m-symm}) is
\begin{equation}
\partial_{m}I_{m}=-\partial_{m}I_{1-m}.\label{m-der-id}
\end{equation}
Equation (\ref{m-der-id}) is sufficient to prove that $m=1/2$ at
the AT point.

The Lyapunov exponents can be expressed in terms of $I_{m}$ and its
derivative:
\begin{align}
\lambda & =-\ln I_{\frac{1}{2}},\label{I-lambda}\\
\lambda_{{\rm typ}} & =-\frac{1}{2}\,\partial_{m}I_{m}|_{m=0}.
\end{align}

We conclude this Appendix by presenting the results of numerical evaluation
of the function $p_{\ell}(z)$ by the PD with branching number $K=2$
and disorder strength $W=8$ and $W=14$ averaging over $5\times10^{9}$
pathes of the length $\ell=8-32$. In these calculations the typical
imaginary part of $G$ was held a small constant by decreasing $\eta$
at each run of iteration, so that this was a \textit{non-equilibrium}
PD, similar to the inflationary one. The anomalously large and anomalously
small contributions of end points were suppressed by multiplying the
product of Green's functions along the path by $(G_{\ell}+G_{1}^{-1})^{-1}\,(G_{1}+G_{\ell}^{-1})^{-1}$.

The result is shown in Fig.~\ref{Fig:p_z_num}. It is clearly seen
that as $\ell$ increases the symmetry $p_{\ell}(z)=p_{\ell}(z^{-1})$
becomes more and more explicit and the form of the function $p_{\ell}(z)$
approaches the one given by (\ref{eq:p(z)-approx}).

\section{Applicability of power-law distribution Eq.(\ref{P-power}).\label{app:power-law-distribution-function}}

It was shown in Appendix \ref{app:m_0-m_1-and-the-power} that the
distribution of $\Im G$ obtained in the IOTL by RSB formalism has
a power-law tail in non-ergodic extended phase. In section \ref{sec:RSB-results-for-rho_typ}
we assumed that this power-law dependence remains valid in the conventional
ATL where non-linear in $\Im G$ terms are important to reach the
equilibrium distribution. In this Appendix we justify this assumption
and show that the power-law distribution (\ref{P-power}) is a very
good approximation close to Anderson transition at small $1-W/W_{c}$.
Surprisingly, this approximation remains reasonably good at small
$W$ because in this limit only $\rho$ in the small interval are
relevant.

We assume the distribution function of $\rho$ in the ATL in a general
form of \textit{large deviation ansatz} that is consistent with a
well-defined limit at $N\gg N_{c}$(see sections \ref{sec:Analytical-results-for-D(W)},\ref{sec:Discussion}
and Appendix \ref{app:-Relation-between-alpha-and-D}):
\begin{equation}
\tilde{P_{0}}(\ln\rho)=A_{\rho}\,N_{c}^{f_{\rho}(\beta)}.\label{eq:P(ln_rho)}
\end{equation}
Here $\ln N_{c}\sim\ln\rho_{\text{typ}}^{-1}\gg1$ is the characteristic
critical length, $\beta=-\ln\rho/\ln N_{c}$, $A_{\rho}$ is the normalization
constant and $f_{\rho}(\beta)$ is a certain function with a maximum
at $\beta=\beta_{0}$. Normalization condition implies that at its
maximum $\beta=\beta_{0}$ the function $f_{\rho}(\beta)$ is zero:
\begin{equation}
f_{\rho}(\beta_{0})=0.\label{A}
\end{equation}
At small sizes $N\lesssim N_{c}$ the distribution (\ref{eq:P(ln_rho)})
crosses over to
\[
\tilde{P_{0}}(\ln\rho)=\tilde{A}_{\rho}\,N^{\tilde{f}_{\rho}(\beta)}
\]
with a \textit{different} function $\tilde{f}_{\rho}(\beta)$ and
$\beta=-\ln\rho/\ln N$ that coincides with the distribution of $N\,\psi^{2}$
as discussed in section \ref{sec:Analytical-results-for-D(W)} and
Appendix {\ref{app:-Relation-between-alpha-and-D}}. Thus in the
limit $1\ll N\ll N_{c}$ the function $\tilde{f}_{\rho}(\beta)$ is
approximately the same as $f(\alpha)-1$ in Appendix
\ref{app:-Relation-between-alpha-and-D}.

Let us expand $\ln\tilde{P}_{0}(\ln\rho)$ up to the second order
in $x=\ln(\rho/\rho_{m})$ near the point $\rho_{m}=N_{c}^{-\beta_{m}}$
where $f_{\rho}'(\beta_{m})=m$. We obtain:
\begin{eqnarray}
\ln\tilde{P}_{0}(\ln\rho) & \approx & \ln\tilde{P}_{0}(\ln\rho_{m})-m\,x+\frac{1}{2}f_{\rho}''(\beta_{m})\,\frac{x^{2}}{\ln N_{c}}\nonumber \\
P_{0}(\rho) & \approx & \frac{C_{m}}{\rho^{1+m}}\,{\rm exp}\left[-\frac{1}{2}|f_{\rho}''(\beta_{m})|\,\frac{\ln^{2}(\rho/\rho_{m})}{\ln N_{c}}\right],\label{eq:P_0(rho)}
\end{eqnarray}
where $\ln C_{m}=\ln A_{\rho}+f_{\rho}(\beta_{m})\,\ln N_{c}+m\,\ln\rho_{m}$.

Eq.(\ref{eq:P_0(rho)}) shows that \textit{any} distribution of the
type Eq.(\ref{eq:P(ln_rho)}) is \textit{locally power-law} with the
power $m$ depending on the point $\rho\approx\rho_{m}$. The log-normal
\textit{correction factor} to this power-law is controlled by the
large parameter $\ln N_{c}$ and depends on the curvature $|f_{\rho}''(\beta_{m})|$.
In order for the derivation in section \ref{sec:RSB-results-for-rho_typ}
to be valid, this factor should not be small for all $\rho_{\text{typ}}<\rho<\rho_{m}$:
\begin{equation}
\frac{|f_{\rho}''(\beta_{m})|}{2\ln N_{c}}\;\ln^{2}\left(\frac{\rho_{m}}{\rho_{\text{typ}}}\right)\lesssim1.
\end{equation}
Recalling that $\rho_{m}=N_{c}^{-\beta_{m}}$ and $\rho_{\text{typ}}=N_{c}^{-\beta_{0}}$
we obtain a criterion of validity of ansatz Eq.(\ref{P-power}):
\begin{equation}
\frac{1}{2}\,|f''(\beta_{m})|\,(\beta_{0}-\beta_{m})^{2}\,\ln N_{c}\lesssim1.\label{criterion-P-pow}
\end{equation}
To further simplify Eq.(\ref{criterion-P-pow}) we employ the symmetry
(\ref{eq:duality}) for $P_{0}(\rho)$ which in terms of $f_{\rho}(\beta)$
reads:
\begin{equation}
f_{\rho}(\beta)=f_{\rho}(-\beta)+\beta.
\end{equation}
From this symmetry one immediately finds:
\begin{equation}
f_{\rho}'(0)=1/2,\;\Rightarrow\;\beta_{1/2}=0.
\end{equation}
Since $m=m_{0}=1/2$ at the Anderson transition $W=W_{c}$, we obtain
in the vicinity of this transition:
\begin{equation}
\beta_{m}\approx0.
\end{equation}
Near the Anderson transition there is only one divergent scale $N_{c}\sim\rho_{\text{typ}}^{-1}$.
This corresponds to $\beta_{0}=1$ at the transition.

Thus near the localization transition the condition Eq.(\ref{criterion-P-pow})
is simplified:
\begin{equation}
\frac{1}{2}\,|f_{\rho}''(0)|\,\ln N_{c}\lesssim1.\label{cr-near-AT}
\end{equation}
Since $\ln N_{c}\approx\ln\rho_{\text{typ}}^{-1}\sim(W_{c}-W)^{-1}$,
we conclude that in order to satisfy the condition (\ref{cr-near-AT})
of validity of ansatz (\ref{P-power}), the curvature $|f_{\rho}''(0)|$
should vanish faster than $|W-W_{c}|$ near the Anderson transition.
This fast decrease is evident in the numerical data, see Fig.~\ref{fig:Distribution-functions-of-rho}.

We now turn to the regime of weak multifractality. In this regime
$P_{0}(\rho)$ is expected to be log-normal and given by (\ref{P_0-AKL}).
It translates into

\begin{subequations}
\begin{align}
f_{\rho}(\beta) & =1-\frac{\beta_{0}}{4}\,\left(1-\frac{\beta}{\beta_{0}}\right)^{2},\\
\beta_{0}= & u\,/\ln N_{c}\ll1,
\end{align}
\end{subequations}which gives

\begin{subequations}
\begin{align}
|f_{\rho}''| & =\frac{1}{2\beta_{0}},\\
\beta_{m} & =-(2m-1)\,\beta_{0}.
\end{align}
\end{subequations}

We apply these equations to the regime of moderately small $W\approx W_{E}$
where RSB theory predicts a transition and population dynamics predicts
crossover into ergodic state. In RSB theory in this regime we have
$m\approx1$, so that the condition Eq.(\ref{criterion-P-pow}) becomes
\begin{equation}
\beta_{0}\ln N_{c}=u\lesssim1.
\end{equation}
We see that the ansatz (\ref{P-power}) works both near the localization
and near the ergodic transition expected in RSB theory.

\section{Extraction of D(W) by extrapolation of exact diagonalization data\label{app:Extraction-of-D(W)}}

\begin{figure}[t]
\includegraphics[width=0.9\columnwidth]{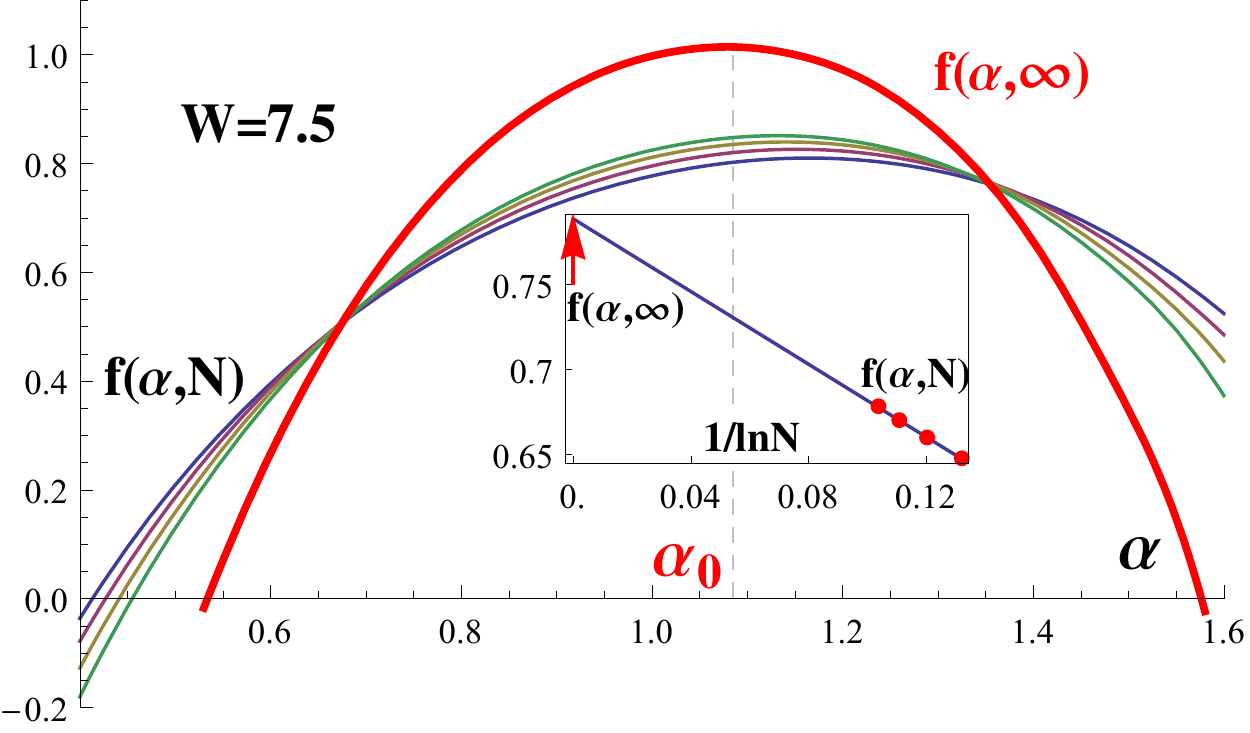}
\caption{Extrapolation of $f(\alpha,N)$ for $W=7.5$ and extraction of $D=2-\alpha_{0}$.\label{fig:Extrapolation-of-f(alpha)}}
\end{figure}
In order to compare the results of population dynamics numerics and
analytical RSB calculations (which both correspond to an infinite
BL), with the results of exact diagonalization of the Anderson model
on finite RRG we employ the procedure developed in \cite{Our-BL}.
It consists of few steps. First, we obtain the distribution function
of $|\psi(i)|^{2}$ by numerical diagonalization of the Anderson model
on RRG of modestly large sizes $N=2000,\,4000,\,8000,\,16000$. Second
we extract the distribution function of the wave function envelope.
This is an important step, because generally the wave function at a
given site can be small due to two reasons: it might be localized
or fractal and because the given site is close to its node. The latter
effect is not relevant and has to be de-convoluted from the raw data.
We define the envelope wave function by $\psi(i)=\psi_{\text{env}}(i)\,\psi_{\text{PT}}(i)$
where $\psi_{PT}(i)$ is the Porter-Thomas wave function describing
random de Broglie oscillations of unit amplitude. We assume the Porter-Thomas
distribution of the latter $P_{\text{PT}}(x)=(2\pi x)^{-\frac{1}{2}}\;e^{-x/2}$,
where $x=|\psi_{PT}|^{2}$, and use Laplace transform to extract the
distribution function of the envelope $P(\ln|\psi_{\text{env}}|^{2})$
and the finite-size spectrum of fractal dimensions:
\begin{align*}
f(\alpha,N) & =\ln[N\,P(\ln|\psi_{\text{env}}|^{2})]/\ln N\\
\alpha & =-\ln|\psi_{\text{env}}|^{2}/\ln N
\end{align*}
Finally, we use the linear in $\ln N$ extrapolation
\[
f(\alpha,\infty)=f(\alpha,N)+c(\alpha)/\ln N
\]
to find the value of $f(\alpha,\infty)$ in the thermodynamic limit.

We find $\alpha_{0}$ as the point of maximum of the \textit{extrapolated}
$f(\alpha,\infty)$ as shown in Fig.~\ref{fig:Extrapolation-of-f(alpha)}.
We repeat this procedure for several values $W$ of the disorder to
find $D(W)=2-\alpha_{0}(W)$.

\end{document}